\documentclass[12pt]{elsarticle}                                          
\usepackage{graphicx}                                                         
\usepackage{a41}                                                                
\usepackage{xcolor}                     
\usepackage[rflt]{floatflt}
\usepackage{float}
\usepackage{lscape}
\usepackage{breakurl} 
\usepackage{times}
\usepackage{array}
\usepackage[english]{babel}
\usepackage[T1]{fontenc}
\usepackage{ae}
\usepackage{url}
\usepackage{amsmath, amsthm, amssymb}
\usepackage{slashed}

\usepackage{rotating}
\usepackage{graphicx}
\usepackage{comment}
\newcounter{mmacnt}
\def\restartmma{\setcounter{mmacnt}{0}}
\restartmma \catcode`|=\active
\def|#1|{\mathrm{#1}}
\catcode`|=12
\newenvironment{mma}{
\par\smallskip
\catcode`|=\active
\parskip=0pt\parindent=0pt 
\small
\def\In##1\\{%
\def\linebreak{\hfill\break\null\qquad}%
\refstepcounter{mmacnt}
\hangindent=2.5em\hangafter=0
\leavevmode
\llap{\tiny\sffamily In[\arabic{mmacnt}]:=\kern.5em}%
\mathversion{bold}\footnotesize$
\displaystyle##1$\normalsize
\mathversion{normal}\par
 }%
\def\Print##1\\{%
\def\linebreak{\hfill\break}%
\hangindent=2.5em\hangafter=0
\leavevmode ##1\par}%
\def\Out##1\\{%
\def\linebreak{$\hfill\break\null\hfill$}%
\kern\abovedisplayskip\par
\hangindent=2.5em\hangafter=0
\leavevmode
\llap{\tiny\sffamily Out[\arabic{mmacnt}]=\kern.5em}
\footnotesize$\displaystyle##1$
\normalsize\hfill\null\par
\kern\belowdisplayskip
}%
\def\Warning##1##2\\{%
\def\linebreak{\hfill\break}%
\hangindent=2.5em\hangafter=0
\leavevmode
{\scriptsize##1 : ##2}\par}%
}{%
\par\smallskip
}

\usepackage{color}
\newenvironment{fshaded}{%
\MakeFramed {\FrameRestore}
}%
{\endMakeFramed}

\makeatletter
\def\ps@pprintTitle{%
\let\@oddhead\@empty
\let\@evenhead\@empty
\def\@oddfoot{\reset@font\hfil\thepage\hfil}
\let\@evenfoot\@oddfoot
}
\makeatother
\usepackage{tikz}
\usetikzlibrary{matrix}
\allowdisplaybreaks[4]

\begin{document}
\begin{frontmatter}
\title{\Large
\textbf{
Probing one--loop--induced decay
channel $H^\pm \to W^\pm\gamma$ in
the Two Higgs Doublet Models
at muon--TeV colliders}}
\author[1,2]{Dzung Tri Tran}
\author[3]{Quang Hoang-Minh Pham}
\author[3]{Khoa Ngo-Thanh Ho}
\author[1,2]{Khiem Hong Phan}
\ead{phanhongkhiem@duytan.edu.vn}
\address[1]{\it Institute of Fundamental
and Applied Sciences, Duy Tan University,
Ho Chi Minh City $70000$, Vietnam}
\address[2]{Faculty of Natural Sciences,
Duy Tan University, Da Nang City $50000$,
Vietnam}
\address[3]
{\it VNUHCM-University of Science,
$227$ Nguyen Van Cu, District $5$,
Ho Chi Minh City $700000$, Vietnam}
\pagestyle{myheadings}
\markright{}
\begin{abstract} 
In this work, we study the one--loop--induced decay channel
$H^{\pm} \rightarrow W^{\pm}\gamma$ in the general
$\mathcal{R}_{\xi}$ gauge within Two Higgs Doublet Models.
We analytically verify the gauge invariance ($\xi$-independence),
ultraviolet finiteness, and renormalization-scale
independence of the one-loop form factors,
thereby confirming the consistency of our calculations. On the
phenomenological side, we perform a parameter scan of the Type-I
THDM and, based on the viable parameter space, evaluate the
branching ratios of this decay process. Furthermore,
we investigate charged Higgs pair production
at muon--TeV colliders through the representative processes
$\mu^+\mu^- \rightarrow H^{+}H^{-} \rightarrow W^+W^- h\gamma$
and
$\mu^+\mu^- \rightarrow \gamma\gamma
\rightarrow H^{+}H^{-} \rightarrow W^+W^- h\gamma$,
as typical applications of our results.
The events for the processes are computed
within the allowed parameter regions
of the Type-I THDM. The corresponding
signal significances are evaluated, including the relevant
Standard Model backgrounds, at a center-of-mass energy
of $\sqrt{s} = 3$~TeV.
 With the high integrated
luminosity expected at muon--TeV colliders, reaching up to
$\mathcal{L} = 3000~\text{fb}^{-1}$, our analysis indicates
that the signals can be detected with a statistical
significance of $5\sigma$ for several benchmark scenarios
in the viable parameter space of the Type-I THDM.
\end{abstract}
\end{frontmatter}
\section{Introduction}
In addition to the high-precision measurements
of the properties of the observed Standard Model
like Higgs boson at the Large Hadron Collider (LHC), the
exploration of additional scalar states predicted
in extensions of the Standard Model (SM) is also a
central objective for future collider programs.
The discovery of such non-standard Higgs bosons
would yield critical insights into the structure
of the scalar potential and provide a deeper
understanding of the mechanisms underlying
the electroweak symmetry breaking (EWSB).
Among the possible production modes of
additional Higgs states, the production
of charged Higgs bosons is of particular
interest.

From the experimental perspective,
searches for light charged Higgs production in
association with top-quark decays at
$\sqrt{s}=7$~TeV and $\sqrt{s}=8$~TeV at the
LHC have been reported
in~\cite{CMS:2012fgz,CMS:2015lsf,
ATLAS:2023bzb,ATLAS:2024oqu}.
Additional searches for
$H^{+} \to \tau \nu$~\cite{ATLAS:2012nhc}
and $H^+ \to c\bar{s}$~\cite{ATLAS:2013uxj}
in top-quark pair production have also been
performed with the ATLAS collaboration.
For heavy charged Higgs bosons, searches
based on the decay channels
$H^{\pm} \to tb$~\cite{ATLAS:2015nkq,
ATLAS:2021upq,CMS:2020imj} and
$H^{\pm} \to W^{\pm}Z$~\cite{ATLAS:2015edr,
ATLAS:2018iui} at $\sqrt{s}=8$~TeV have
been conducted by the ATLAS and CMS collaborations.
The CMS Collaboration has carried out a search
for charged Higgs bosons produced through
vector boson fusion in proton--proton collisions at
$\sqrt{s}=13$~TeV~\cite{CMS:2021wlt}.
Updated results on $H^\pm \rightarrow cb$
and $H^\pm \rightarrow cs$ at $\sqrt{s}=8$~TeV
have been released in Refs.~\cite{CMS:2018dzl,CMS:2020osd}.
The decay of a charged Higgs boson into a heavy neutral
Higgs boson associated with a $W$ boson
has been investigated at $\sqrt{s}=13$~TeV at the
LHC~\cite{CMS:2022jqc,ATLAS:2024rcu}.
More recently, the ATLAS and CMS Collaborations
have focused on searches for charged
Higgs bosons produced in top-quark decays
as well as in association with top quarks,
with subsequent decays through
$H^{\pm} \to \tau^{\pm}\nu_{\tau}$
\cite{ATLAS:2018gfm,ATLAS:2024hya,CMS:2019bfg}.

From the theoretical perspective, studies of
the charged Higgs signal via $pp \to tH^-
\to t W^- b\bar{b}$ have been carried out within the
THDM~\cite{Arhrib:2017veb} at LHC Run~II,
including the effects of top-quark polarization as
discussed in Ref.~\cite{Arhrib:2018bxc}.
Signals for $H^{+} \to t\overline{b}$
within MSSM scenarios have also been investigated
at the LHC~\cite{Arhrib:2019ykh}.
In addition, double production of singly charged
Higgs bosons at the HL-LHC and
HE-LHC has been analyzed in the bosonic decay
channels~\cite{Arhrib:2019ywg}.
Within the THDM framework, the discovery
prospects of a light charged Higgs boson
decaying into vector bosons have been
analyzed using LHC Run~III
data, as reported in~\cite{Arhrib:2020tqk}.
Probing for a light charged Higgs boson
through the $H^{\pm}h/A$ production
and $pp \to H^{+}H^{-}$ channels,
with subsequent decays
$H^{\pm} \to W^{\pm}\phi_j$ ($\phi_j = h,\,A$),
has been investigated within the THDM at
the LHC~\cite{Arhrib:2021xmc}.
Further analyses of charged Higgs
searches at the LHC can be found
in Refs.~\cite{Wang:2021pxc,
Krab:2022lih, Arhrib:2024sfg},
together with analyses of charged
Higgs bosons produced via vectorlike
top-quark pairs~\cite{Arhrib:2024nbj}.
The $W\gamma$ decay mode of a charged
Higgs boson has also been investigated
at the LHC~\cite{Logan:2018wtm}.
Relevant studies for future lepton
colliders include the production of
charged Higgs bosons within the THDM framework
at both $\mu^{+}\mu^{-}$ and $e^{+}e^{-}$ colliders,
as explored in Refs.~\cite{Ouazghour:2023plc,
Ouazghour:2024twx, Ouazghour:2025owf,
BrahimAit-Ouazghour:2025mhy}.
Probing heavy charged Higgs bosons at
$\gamma\gamma$ colliders using multivariate
techniques, as well as the decay
$H^{\pm} \to W^{\pm}H$ at high-energy lepton
colliders, has been studied in
Refs.~\cite{Ahmed:2024oxg, Hashemi:2023osd}.

Among the possible decay processes of the
charged Higgs boson, the channel
$H^{\pm} \to W^{\pm}\gamma$ is of particular
interest for several reasons:
(i) its loop--induced nature makes it highly sensitive to
new-physics effects, and
(ii) a precise study of this channel could
help discriminate between different THDM scenarios.
Given its importance, theoretical calculations of
loop-induced charged Higgs decays have been
extensively pursued.  The one-loop decay
$H^{\pm} \to W^{\pm}\gamma$ was first computed in
Ref.~\cite{Arhrib:2006wd}, with further
developments including studies of the loop--induced
decay $H^{\pm}\to W^{\mp}Z$ in the THDM with CP
violation~\cite{Kanemura:2024ezz}, together
with one-loop corrections to charged Higgs
two-body decays within the THDM~\cite{Aiko:2021can},
analyses within a nonlinear
gauge~\cite{Hernandez-Sanchez:2004yid},
and investigations of Yukawa textures in
the Type-III THDM~\cite{BarradasGuevara:2010xs}.

In this work, we present alternative analytic
results for the one--loop--induced decay channel
$H^{\pm} \rightarrow W^{\pm}\gamma$ in the general
$\mathcal{R}_{\xi}$ gauge. In contrast to previous
studies, we explicitly verify gauge invariance
(i.e., $\xi$-independence and the validity of
the Ward identity for the one-loop amplitude
with an external on-shell photon), together with
ultraviolet finiteness and renormalization-scale
independence of the one-loop form factors.
These checks confirm the overall consistency
of our calculations. In the phenomenological
analysis, we perform a parameter scan of the
Type-I THDM and, based on the allowed parameter
space, evaluate the branching ratios of the
decay process. Furthermore, we investigate charged
Higgs pair production at muon--TeV colliders
through the representative processes
$\mu^+\mu^- \rightarrow H^{+}H^{-}
\rightarrow W^+W^- h\gamma$
and $\mu^+\mu^- \rightarrow \gamma\gamma
\rightarrow H^{+}H^{-}
\rightarrow W^+W^- h\gamma$,
as typical applications of our results.
The event rates of the considered channels
are evaluated within the viable
parameter space of the Type-I THDM.
The corresponding signal significances,
including the relevant Standard Model
backgrounds, are computed at a
center-of-mass energy of $\sqrt{s}=3$~TeV.
With the high integrated luminosity anticipated
at muon--TeV colliders, up to
$\mathcal{L}=3000~\text{fb}^{-1}$, we find that
the signals can be observed with a $5\sigma$
significance for several benchmark points
in the viable Type-I THDM parameter space.

Our work is organized as follows. Section~2
provides a review of the Two Higgs Doublet Model
together with a summary of the relevant theoretical
and experimental constraints. In Section~3, we derive
the one-loop expressions for the decay
$H^{\pm} \to W^{\pm}\gamma$ in the general
$\mathcal{R}_{\xi}$ gauge. Section~4 presents
the simulation of singly charged Higgs pair production at
muon--TeV colliders. Conclusions and outlook are
given in Section~5. The Appendices contain
detailed analytic expressions for the one-loop
form factors. In particular, Appendix~D provides
explicit checks of $\xi$-independence and
the Ward identity for the one-loop form factors.
\section{Review of Two Higgs Doublet
Models, theoretical and experimental
constraints}
\subsection{Review of THDM}%
We begin with a brief review of the THDM;
for an extensive overview, see Ref.~\cite{Branco:2011iw}.
In this framework, the fermion content and gauge sector
remain identical to those of the Standard Model, while
an additional scalar doublet $\Phi_2$ with
hypercharge $Y = +1$ is introduced into the scalar sector.
The CP-conserving scalar potential of the THDM,
in its most general form consistent with gauge
invariance and renormalizability, can be written as
follows:
\begin{eqnarray}
\label{V2HDM}
\mathcal{V}(\Phi_1,\Phi_2)
&=&
m_{11}^2\Phi_1^\dagger \Phi_1
+
m_{22}^2\Phi_2^\dagger \Phi_2
-
\Big[
m_{12}^2\Phi_1^\dagger \Phi_2
+{\rm H.c.}
\Big]
\\
&&
+
\frac{\lambda_1}{2}
(\Phi_1^\dagger \Phi_1)^2
+
\frac{\lambda_2}{2}
(\Phi_2^\dagger \Phi_2)^2
+ \lambda_3(\Phi_1^\dagger \Phi_1)
(\Phi_2^\dagger \Phi_2)
\nonumber
\\
&&
+
\lambda_4(\Phi_1^\dagger \Phi_2)
(\Phi_2^\dagger \Phi_1)
+
\Big[
\frac{\lambda_5}{2}
~(\Phi_1^\dagger \Phi_2)^2
+ ~{\rm H.c.}
\Big].
\nonumber
\end{eqnarray}
As mentioned, this corresponds to the CP-conserving version of the THDM,
in which all fundamental parameters of the scalar potential are real.
To prevent flavor-changing neutral currents (FCNCs) at tree level,
a $Z_2$ symmetry is applied for the scalar potential, defined by the
transformations $\Phi_1 \to \Phi_1$ and $\Phi_2 \to -\Phi_2$.
A soft-breaking term, $m_{12}^2 \Phi_1^\dagger \Phi_2 + \text{H.c.}$,
is also allowed, with $m_{12}^2$ serving as the breaking scale of the
$Z_2$ symmetry.

After electroweak symmetry breaking (EWSB),
the two scalar doublets are parametrized
around their vacuum expectation values
(VEVs) as
\begin{eqnarray}
\label{representa-htm}
\Phi_k &=&
\begin{bmatrix}
\phi_k^+ \\
(v_k
+
\phi^0_k
+
i
\psi^0_k)
/\sqrt{2}
\end{bmatrix}
\quad \textrm{for}
\quad k=1,2.
\end{eqnarray}
Here $v_k$ ($k=1,2$) denotes the vacuum
expectation values (VEVs) of the two
scalar doublets. The combined VEV is given by
$v = \sqrt{v_1^2 + v_2^2}$ and is fixed at
$v \simeq 246~\text{GeV}$, consistent with the
Standard Model case.

The physical spectrum of the THDM after EWSB consists of two CP-even Higgs bosons:
the lighter state $h$, identified with the SM-like Higgs boson observed at the LHC,
and the heavier state $H$. In addition, the model predicts a CP-odd Higgs boson ($A$)
and a pair of charged Higgs bosons ($H^\pm$).
The physical masses of these scalar states are obtained by diagonalizing
the corresponding mass matrices derived from the scalar potential
in the original Lagrangian.
The associated rotation matrices are given by
\begin{eqnarray}
\begin{pmatrix}
\phi^0_1\\
\phi^0_2
\end{pmatrix}
&=&
\begin{pmatrix}
c_{\alpha}
&
-s_{\alpha}
\\
s_{\alpha}
&
c_{\alpha}
\end{pmatrix}
\begin{pmatrix}
H\\
h
\end{pmatrix},
\\
\begin{pmatrix}
\phi_1^{\pm}\\
\phi_2^{\pm}
\end{pmatrix}
&=&
\begin{pmatrix}
c_{\beta}
&
-s_{\beta}
\\
s_{\beta}
&
c_{\beta}
\end{pmatrix}
\begin{pmatrix}
G^{\pm}\\
H^{\pm}
\end{pmatrix},
\\
\begin{pmatrix}
\psi^0_1\\
\psi^0_2
\end{pmatrix}
&=&
\begin{pmatrix}
c_{\beta}
&
-s_{\beta}
\\
s_{\beta}
&
c_{\beta}
\end{pmatrix}
\begin{pmatrix}
G^{0}\\
A
\end{pmatrix}.
\end{eqnarray}
In these rotations, the unphysical fields
$G^0$ and $G^{\pm}$ are interpreted as
the neutral and charged Goldstone bosons,
which are absorbed by the $Z$ and $W^{\pm}$
gauge bosons to provide their longitudinal
components and thereby generate their masses.
The mixing angle between the charged Higgs
bosons and the charged Goldstone bosons
is denoted by $\beta$, defined through
$t_{\beta} \equiv \tan\beta = v_2/v_1$.
The physical masses of the remaining
scalar states can then be written
in terms of the bare parameters as
\begin{eqnarray}
M_{H^{\pm}}^{2}
&=&
M^{2}-
\frac{1}{2}
\lambda_{45}
v^{2},
\quad
M_{A}^{2}  =
M^{2}-\lambda_{5}v^{2},
\\
M_{h}^{2} &=& (M_{11}^{\phi})^{2}
s_{\beta-\alpha}^{2}
+ (M_{22}^{\phi})^{2}c_{\beta-\alpha}^{2}
+ (M_{12}^{\phi})^{2}s_{2(\beta-\alpha)},
\\
M_{H}^{2} &=& (M_{11}^{\phi})^{2}c_{\beta-\alpha}^{2}
+ (M_{22}^{\phi})^{2}s_{\beta-\alpha}^{2}
- (M_{12}^{\phi})^{2}s_{2(\beta-\alpha)},
\end{eqnarray}
where $
m_{12}^{2}=M^{2}s_{\beta}c_{\beta}$,
and
\begin{eqnarray}
(M_{11}^{\phi})^{2}&=&
(\lambda_{1}c_{\beta}^{4}
+\lambda_{2}s_{\beta}^{4})v^{2}
+\frac{v^{2}}{2}\;
\lambda_{345}
\; s_{2\beta}^{2}, \\
(M_{22}^{\phi})^{2}
&=&
M^{2}
+ \frac{v^{2}}{4}
\Big[
\lambda_{12}
-2\lambda_{345}
\Big]
s_{2\beta}^{2}, \\
(M_{12}^{\phi})^{2} &=&
(M_{21}^{\phi})^{2}  =
-\frac{v^{2}}{2}
\Big[
\lambda_{1}c_{\beta}^{2}
-\lambda_{2}s_{\beta}^{2}
-
\lambda_{345}\; c_{2\beta}
\Big]
s_{2\beta}.
\end{eqnarray}
For brevity, we employ the shorthand notation
$\lambda_{ij\cdots} \equiv
\lambda_i + \lambda_j + \cdots$.

Finally, we conclude the review of the model
by considering the Yukawa sector of the THDM.
To eliminate tree-level flavor-changing neutral
currents, a discrete $Z_2$ symmetry is imposed,
as discussed above. Depending on the $Z_2$
charge assignments of the fermions and scalar doublets,
the Yukawa interactions are categorized into four distinct types,
commonly denoted as Type-I, Type-II, Type-X, and Type-Y,
as summarized in Table~\ref{Z2-assignment}
(see also Ref.~\cite{Aoki:2009ha}).
Within this framework, the Yukawa interactions
can be written in the general form:
\begin{eqnarray}
{\mathcal L}_\text{Y}
&=&
-\sum_{f=u,d,\ell}
\left(
\sum_{\phi_j=h, H}
g_{\phi_j ff}\cdot
\phi_j{\overline f}f
+
g_{Aff}\cdot
A
{\overline f}
\gamma_5f
\right)
+\cdots
\\
&=&
-\sum_{f=u,d,\ell}
\left(
\sum_{\phi_j=h, H}
\frac{m_f}{v}\xi_{\phi_j}^f
\phi_j {\overline f}f
-i\frac{m_f}{v}\xi_A^f
{\overline f}
\gamma_5fA
\right)
\nonumber\\
&&
+
\frac{
\sqrt{2}
}{v}
\left[
\bar{u}_{i}
V_{ij}\left(
m_{u_i}
\xi^{u}_A P_L
+
\xi^{d}_A
m_{d_j} P_R \right)d_{j} H^+
\right]
\nonumber\\
&&
+ \frac{\sqrt{2}}{v}
\bar{\nu}_L
\xi^{\ell}_A
m_\ell \ell_R H^+
+ \textrm{H.c}.
\end{eqnarray}
Here, $V_{ij}$ denotes the $(i,j)$-th
element of the CKM matrix,
and $\ell_{L/R}$ represent the left- and
right-handed lepton fields, respectively.
The projection operators
$P_{L/R} = \frac{1 \mp \gamma_5}{2}$
are included in the Lagrangian above.
\begin{table}[H]
\begin{center}
\begin{tabular}
{|c|ccccccc|cccccc|}
\hline\hline
Types & $\Phi_{1}$ & $\Phi_{2}$ &$Q_L$
&$L_L$&$u_R$&$d_R$&$e_R$&$\xi^u_A$
&$\xi^d_A$&$\xi^{\ell}_A$
&$\xi_h^u$
&$\xi_h^d $
&$\xi_h^{\ell}$
\\\hline\hline
I & $+$ & $-$
& $+$ & $+$ &
$-$
& $-$ & $-$ &
$\cot\beta$
& $-\cot\beta$
& $\cot\beta$
&
$\frac{c_\alpha}{
s_\beta}$
&
$
\frac{c_\alpha}{
s_\beta}$
&
$
\frac{c_\alpha}{
s_\beta}$
\\ \hline
II & $+$ & $-$ & $+$ & $+$
& $-$ & $+$ & $+$ &
$-\cot\beta$
&$-\tan\beta$
&$-\tan\beta$
&
$
\frac{c_\alpha}{
s_\beta}$
&
$-
\frac{s_\alpha}{
c_\beta}$
&
$-
\frac{s_\alpha}{
c_\beta}
$
\\ \hline
X & $+$ & $-$ & $+$
& $+$ & $-$ & $-$ & $+$
&
$-\cot\beta$
& $\cot\beta$
& $-\tan\beta$
&
$
\frac{c_\alpha}
{s_\beta}$
&
$
\frac{c_\alpha}{
s_\beta}$
&
$-
\frac{s_\alpha}
{c_\beta}$
\\ \hline
Y & $+$ & $-$
& $+$ & $+$ & $-$
& $+$ & $-$ &
$-\cot\beta$
&$-\tan\beta$
&$\cot\beta$
&
$
\frac{c_\alpha}
{s_\beta}$
&
$
-
\frac{s_\alpha}
{c_\beta}$
&
$
\frac{c_\alpha}
{s_\beta}$
\\
\hline
\hline
\end{tabular}
\caption{
\label{Z2-assignment}
The Table summarizes the $Z_2$ charge assignments
and the factors $\xi^f_{A(h)}$
($f = u, d, \ell$) for the four THDM types.
The Yukawa couplings of the CP-even Higgs
boson $H$ to fermion pairs ($\xi^f_H$)
can be obtained by exchanging $c_\alpha$
and $s_\alpha$ in the corresponding
expressions for $\xi^f_h$.
}
\end{center}
\end{table}
The following relations are important for
understanding the behavior of fermionic
loop contributions to charged Higgs decay
channels. The relations are
\begin{eqnarray}
\frac{c_{\alpha}}{s_{\beta}}
&=& s_{\beta -\alpha} +\cot\beta \; c_{\beta -\alpha}, \\
\frac{s_{\alpha}}{c_{\beta}}
&=& -s_{\beta -\alpha} + t_{\beta}\;  c_{\beta -\alpha}, \\
\frac{s_{\alpha}}{s_{\beta}}
&=& c_{\beta -\alpha}  -\cot\beta \;  s_{\beta -\alpha}, \\
\frac{c_{\alpha}}{c_{\beta}}
&=& c_{\beta -\alpha} + t_{\beta} \; s_{\beta -\alpha}.
\end{eqnarray}
It is noted that the Yukawa coupling of the
charged Higgs to the top and bottom quarks
is proportional to $\cot\beta$ in the Type-I THDM.
Consequently, the fermionic loop contributions
to the one-loop–induced decays of the singly
charged Higgs are suppressed in the large
$t_{\beta}$ region, providing a representative
example.
\subsection{Theoretical and
experimental constraints}
This subsection focuses on the theoretical
and experimental constraints on the THDM.
Theoretical constraints arise from the
perturbative unitarity, perturbativity,
and vacuum stability of the model,
while experimental constraints stem
from measurements of Higgs boson properties,
flavor-changing neutral current processes,
and electroweak precision observables.
In the following, we discuss each of
these constraints in detail.
\begin{itemize}
\item
{\bf Unitarity:}
Particle physics models must satisfy
fundamental principles such as perturbative unitarity.
This condition can be examined by analyzing
scattering processes involving scalar and gauge bosons,
as discussed in detail in Refs.~\cite{Kanemura:1993hm}.
\item {\bf Vacuum stability:}
This requirement implies that the scalar potential
must be bounded from below. Consequently, the following
conditions are imposed:
\begin{eqnarray}
\label{vacumm-stability}
\lambda_{1,2}>0,  \quad
\lambda_3>- \sqrt{\lambda_1\lambda_2},
\quad
\lambda_3+\lambda_4-|\lambda_5|
> - \sqrt{\lambda_1\lambda_2}.
\end{eqnarray}
\item {\bf Perturbativity:}
These constraints require that
the quartic couplings
of the scalar potential satisfy
\begin{equation}
|\lambda_i| < 8\pi, \quad (i = 1, \dots, 5),
\end{equation}
as discussed in Ref.~\cite{Branco:2011iw}.
\item {\bf The EW precision
observables (EWPOs):}
The electroweak oblique parameters,
$S$, $T$, and $U$~\cite{Peskin:1991sw},
which are related to the $W$ boson mass
$M_W$, are also used as constraints.
Theoretical predictions for these
parameters within the THDM must lie within
the 95\% confidence level (C.L.)
of their experimentally determined values,
as measured by CDF-II~\cite{ParticleDataGroup:2024cfk}.
The latest global fit results
are reported in Refs.~\cite{ParticleDataGroup:2024cfk,
BrahimAit-Ouazghour:2025mhy}
as follows:
\begin{eqnarray}
S_0 = -0.05 \pm 0.07, \quad
T_0 = 0.00 \pm 0.06, \quad
\rho_{ST} = 0.93
\end{eqnarray}
where \(\rho_{ST}\)
denote the correlation
coefficients among
the $S$,~$T$ oblique parameters.
For all the aforementioned constraints,
we employ the program
{\tt 2HDMC-1.8.0}~\cite{Eriksson:2009ws},
in which both the theoretical
conditions and the EWPOs
have been implemented.
In addition, {\tt 2HDMC} provides built-in
interfaces to {\tt HiggsBounds-5.10.1}
~\cite{Bechtle:2020pkv} and
{\tt HiggsSignals-2.6.1}~\cite{Bechtle:2020uwn},
as outlined below
\item \textbf{BSM Higgs Boson Exclusions:}
To ensure that the theoretical predictions
obtained in this study are consistent
with existing exclusion limits on
BSM Higgs bosons, we utilize the program
{\tt HiggsBounds-5.10.1}~\cite{Bechtle:2020pkv}.
The program evaluates our model predictions
in comparison with the exclusion limits established
at the 95\% confidence level, based on Higgs
boson searches performed at LEP,
the LHC, and the Tevatron.
\item \textbf{SM-like Higgs Boson Properties:}
In order to guarantee compatibility between
the model and the observed properties of the SM-like
Higgs boson measured at the LHC, we utilize
the program
{\tt HiggsSignals-2.6.1}~\cite{Bechtle:2020uwn}.
This tool assesses the consistency of
the predicted signal rates with experimental data
from various Higgs boson searches,
including the most recent results
at $\sqrt{s} = 13~\text{TeV}$.
\item \textbf{Flavor Constraints:}
Theoretical predictions from the THDM must also
comply with constraints from flavor physics,
particularly observables in $B$-meson decays.
To evaluate these constraints, we utilize
the \textbf{SuperIso v4.1}
package~\cite{Mahmoudi:2008tp}.
In particular, consistency with experimental
measurements is required at the $2\sigma$
confidence level, as summarized in Table~\ref{FLPhys}.
\begin{table}[H]
\centering
\setlength{\tabcolsep}{7pt}
\renewcommand{\arraystretch}{1.2} %
\begin{tabular}{|l||c|c|}
\hline
Observable&Experimental result&95\% C.L. Bounds\\\hline
BR($B_{\mu}\to \tau\nu$)\cite{Haller:2018nnx}
&$(1.06 \pm 0.19) \times 10^{-4}$
&$ [0.68\times 10^{-4} , 1.44\times 10^{-4} ]$\\\hline
BR($B_{s}^{0}\to \mu^{+}\mu^{-}$)\cite{Haller:2018nnx}
&$(2.8 \pm 0.7) \times 10^{-9}$
&$[1.4 \times 10^{-9}, 4.2\times 10^{-9}]$\\\hline
BR($B_{d}^{0}\to \mu^{+}\mu^{-}$)\cite{Mahmoudi:2008tp}
&$(3.9\pm 1.5)\times10^{-10}$
&$[0.9\times 10^{-10}, 6.9\times10^{-10}$]\\\hline
BR($\bar{B}\to X_{s}\gamma$)\cite{Haller:2018nnx,
HFLAV:2016hnz,HFLAV:2022esi}
&$(3.49\pm 0.19)\times10^{-4}$
&$[3.11\times 10^{-4} , 3.87\times 10^{-4}]$\\\hline
\end{tabular}
\caption{\label{FLPhys}
Experimental results of
$B_{\mu}\to \tau\nu$, $B_{s,d}^{0}
\to \mu^{+}\mu^{-}$ and $\bar{B}
\to X_{s}\gamma$ at 95$\%$ C.L.}
\end{table}
\end{itemize}
\section{One-loop induced formulas       
for $H^{\pm}\rightarrow W^{\pm}\gamma$}  
In this section, we revisit the calculation of
the one--loop--induced decay process
$H^{\pm} \rightarrow W^{\pm} \gamma$, performed
in the general $R_{\xi}$ gauge within the THDM.
We place particular emphasis on verifying the
gauge independence of the process, i.e.,
explicitly demonstrating the analytical cancellation
of the $\xi$-dependence and verifying the Ward identity
for the one-loop amplitude. Furthermore, we confirm
the ultraviolet divergence cancellation
and the renormalization-scale
independence of the one-loop form factors,
thereby validating the consistency of
our analytical results.

In this paper, we consider the CP-conserving
formulation of the THDM. Consequently, the decay
rates for $H^{\pm}(p) \rightarrow W^{\pm}_{\mu}(k_1)\,
\gamma_{\nu}(k_2)$ are identical. In the general
$R_{\xi}$ gauge, all one-loop Feynman diagrams
are presented in Appendix~B.
The one-loop amplitude is conveniently
parameterized by the following Lorentz
structures:
\begin{eqnarray}
\label{ampHWgam}
\mathcal{A}_{H^{\pm}
\rightarrow W^{\pm} \gamma}
=
\mathcal{A}^{\mu\nu}
\varepsilon^{*}_{\mu} (k_1)
\varepsilon^{*}_{\nu} (k_2)
=
\Big[
F_1
\,
g^{\mu\nu}
+
F_2
\,
k_2^{\mu}
k_1^{\nu}
+
F_3
\times
\big(
i
\epsilon^{\mu\nu\rho\sigma}
\,
k_{1, \rho}
\,
k_{2, \sigma}
\big)
\Big]
\varepsilon^{*}_{\mu} (k_1)
\varepsilon^{*}_{\nu} (k_2).
\end{eqnarray}
In the amplitude, $\epsilon^{\mu\nu\rho\sigma}$
denotes the totally antisymmetric tensor; $p$ is
the incoming momentum, while $k_{1}$ and $k_{2}$
represent the outgoing momenta; and
$\varepsilon^{*}_{\mu}$ ($\varepsilon^{*}_{\nu}$)
denote the polarization vectors of the external
vector bosons $W^{\pm}$ and $\gamma$, respectively.
Owing to the presence of the final on-shell photon,
the amplitude is required to satisfy the Ward identity.
That is, the amplitude vanishes when the photon
polarization vector is replaced as
$\varepsilon^{*}_{\nu} \rightarrow k_{2\nu}$.
Consequently, one obtains
the following relations between
the one-loop form factors:
\begin{eqnarray}
\label{wardiden}
&&
F_2
=
-
\dfrac{F_1}{(k_1 \cdot k_2)}
=
\dfrac{-2}{\big(M_{H^\pm}^2
- M_W^2\big)}
\,
F_1.
\end{eqnarray}
It is important to emphasize that all form
factors in Eq.~\ref{ampHWgam} are computed
independently by extracting the coefficients
associated with the tensors
$g^{\mu\nu}$, $k_2^{\mu}k_1^{\nu}$, and so forth.
The relation given in Eq.~\ref{wardiden} is
then verified analytically. Details of the
derivation are provided in Appendix~C.

The one-loop form factors $F_{i}$ ($i = 1, 2, 3$)
are computed from the complete set of
one-loop Feynman diagrams presented in Appendix~B
and can be expressed in the following form:
\begin{eqnarray}
F_i
&=&
F^{f}_{i}
+
F^{W^\pm H^\pm}_{i}.
\end{eqnarray}
Each form factor $F_i$ is decomposed into
fermionic, $F^{f}_{i}$, and bosonic,
$F^{W^\pm H^\pm}_{i}$, contributions,
which are given by
\begin{eqnarray}
F^{P}_{i}
&=&
F^{P}_{i, \Delta}
+
F^{P}_{i, \Pi}
+
F^{P}_{i, T},
\end{eqnarray}
for $P = f, \{W^\pm H^\pm\}$.
In this equation, the symbols
$\Delta$, $\Pi$, and $T$ denote
the one-loop triangle, self-energy,
and tadpole Feynman diagrams,
respectively.

Since the one-loop form factor $F_{1}$
can be replaced by $F_{2}$ using
the Ward identity in Eq.~\ref{wardiden},
the decay rate can be expressed
in terms of $F_{2}$ and $F_{3}$ as follows:
\begin{eqnarray}
\Gamma_{H^{\pm}\rightarrow W^{\pm}\gamma}
&=&
\frac{M_{H^\pm}^3}{32 \pi}
\Big(
1
-
\frac{M_{W}^2}{M_{H^\pm}^2}
\Big)^3
\Big(
\,
\big|
F_2
\big|^2
+
\big|
F_3
\big|^2
\,
\Big).
\end{eqnarray}
This implies that two one-loop form factors,
$F_2$ and $F_3$, are required to evaluate
the decay width.
The form factors can be formulated using
scalar one-loop one-, two-, and three-point
functions as follows.
First, the contributions arising from the
mixing between the vector boson $W$ and the
charged Higgs boson inside the loop
are given by
\begin{eqnarray}
\label{masterFWH}
F^{W^\pm H^\pm}_{2}
&=&
-
\dfrac{e}{(4\pi)^2}
\dfrac{1}{
M_{H^\pm}^2
M_{W}^2 \big(M_{H^\pm}^2-M_{W}^2\big)^2}
\sum
\limits_{\phi = h, H}
g_{\phi \, H^- W^+}
\times
\\
&& \times
\Big[
c^{0}_{\phi}
+
c^{1}_{\phi}
\,
A_0(M_{\phi}^2)
+
c^{2}_{\phi}
\,
A_0(M_{H^\pm}^2)
+
c^{3}_{\phi}
\,
A_0(M_{W}^2)
+
c^{4}_{\phi}
\,
B_0(M_{H^\pm}^2,M_{\phi}^2,M_{H^\pm}^2)
\nonumber \\
&&\hspace{0.0cm}
+
c^{5}_{\phi}
\,
B_0(M_{H^\pm}^2,M_{\phi}^2,M_{W}^2)
+
c^{6}_{\phi}
\,
B_0(M_{W}^2,M_{\phi}^2,M_{H^\pm}^2)
+
c^{7}_{\phi}
\,
B_0(M_{W}^2,M_{\phi}^2,M_{W}^2)
\nonumber \\
&&\hspace{0.0cm}
+
c^{8}_{\phi}
\,
C_0(M_{W}^2,0,M_{H^\pm}^2,
M_{\phi}^2,M_{W}^2,M_{W}^2)
+
c^{9}_{\phi}
\,
C_0(M_{W}^2,0,M_{H^\pm}^2,
M_{\phi}^2,M_{H^\pm}^2,M_{H^\pm}^2)
\Big].
\nonumber
\end{eqnarray}
Here, we unify the coefficients
$c^{i}_{\phi}$ for $i = 0, \ldots, 9$,
which are represented in terms of
the two general couplings
$g_{\phi W^\pm W^\mp}$ and
$g_{\phi H^\pm H^\mp}$, as follows:
\begin{eqnarray} 
c^{0}_{\phi}
&=&
-
M_{H^\pm}^2
\big(M_{H^\pm}^2-M_{W}^2\big)
\Big[
M_{\phi}^2
\,
g_{\phi W^\pm W^\mp}
-
(2 M_{W}^2)
\,
g_{\phi H^\pm H^\mp}
\Big],
\\
c^{1}_{\phi}
&=&
\big(M_{H^\pm}^2-M_{W}^2\big)
\Big[
\big(2 M_{W}^2\big)
\,
g_{\phi H^\pm H^\mp}
-
\big(
M_{\phi}^2
-
M_{H^\pm}^2
-
M_{W}^2
\big)
\,
g_{\phi W^\pm W^\mp}
\Big],
\\
c^{2}_{\phi}
&=&
-
2
M_{W}^2
\big(M_{H^\pm}^2-M_{W}^2\big)
g_{\phi H^\pm H^\mp},
\\
c^{3}_{\phi}
&=&
\big(M_{H^\pm}^2-M_{W}^2\big)
\,
M_{\phi}^2
\,
g_{\phi W^\pm W^\mp},
\\
c^{4}_{\phi}
&=&
2 M_{W}^2
\Big[
M_{\phi}^2
M_{W}^2
-
2
M_{H^\pm}^2
\big(
M_{\phi}^2
-
M_{H^\pm}^2
\big)
\Big]
\,
g_{\phi H^\pm H^\mp},
\\
c^{5}_{\phi}
&=&
\Big\{
\big(M_{H^\pm}^2-M_{W}^2\big)
\Big[
\big(
M_{\phi}^2
-
M_{H^\pm}^2
-
M_{W}^2
\big)^2
-
4
M_{H^\pm}^2
M_{W}^2
\Big]
\nonumber \\
&&
-
M_{H^\pm}^2
\big(M_{H^\pm}^2-M_{\phi}^2+M_{W}^2\big)
\big(M_{\phi}^2+M_{H^\pm}^2-3 M_{W}^2\big)
\Big\}
\,
g_{\phi W^\pm W^\mp},
\\
c^{6}_{\phi}
&=&
2
M_{H^\pm}^2
M_{W}^2
\big(
M_{\phi}^2-M_{H^\pm}^2-M_{W}^2
\big)
\,
g_{\phi H^\pm H^\mp},
\\
c^{7}_{\phi}
&=&
-
M_{H^\pm}^2
\Big[
M_{\phi}^2
(M_{\phi}^2-M_{H^\pm}^2-3 M_{W}^2)
-
2 M_{W}^2
(M_{H^\pm}^2-3 M_{W}^2)
\Big]
g_{\phi W^\pm W^\mp},
\\
c^{8}_{\phi}
&=&
2
M_{H^\pm}^2
M_{W}^2
\big(M_{H^\pm}^2-M_{W}^2\big)
\big(
3 M_{W}^2
-
M_{\phi}^2
-
M_{H^\pm}^2
\big)
\,
g_{\phi W^\pm W^\mp},
\\
c^{9}_{\phi}
&=&
4
M_{H^\pm}^4
M_{W}^2
\big(M_{H^\pm}^2-M_{W}^2\big)
\,
g_{\phi H^\pm H^\mp}.
\end{eqnarray}
For fermionic contributions,
we have the following form factors:
\begin{eqnarray}
F^{f}_{2}
&=&
-
\dfrac{e}{(4\pi)^2}
\sum
\limits_{ \{ f, f' \} }
\dfrac{2 N^C_{f f'}
\,
g_{W^+ \bar{f} \, f'}}
{M_{H^\pm}^2
(M_{H^\pm}^2-M_{W}^2)^2}
\Big\{
c_{2f}^0
+
c_{2f}^1
[
A_{0}(m_{f'}^2)
-
A_{0}(m_{f}^2)
]
\nonumber\\
&&
+
c_{2f}^2
B_{0}(M_{W}^2,m_{f'}^2,m_{f}^2)
+
c_{2f}^3
B_{0}(M_{H^\pm}^2,m_{f'}^2,m_{f}^2)
\\
&&
+
c_{2f}^4
C_{0}(0,M_{H^\pm}^2,M_{W}^2,
m_{f'}^2,m_{f'}^2,m_{f}^2)
+
c_{2f}^5
C_{0}(M_{W}^2,0,M_{H^\pm}^2,
m_{f'}^2,m_{f}^2,m_{f}^2)
\Big\}.
\nonumber
\end{eqnarray}
Where the corresponding coefficients
are listed as follows:
\begin{eqnarray}
c_{2f}^0 &=& M_{H^\pm}^2
(M_{W}^2 - M_{H^\pm}^2)
\big(
Q_{f}
+
Q_{f'}
\big)
\Big(
g^{R}_{H^- f \bar{f'}}
\,
m_{f}^2
+
g^{L}_{H^- f \bar{f'}}
\,
m_{f'}^2
\Big),
\\
c_{2f}^1 &=&
\Delta Q_{ff'}
\big( M_{H^\pm}^2-M_{W}^2 \big)
\Big(
g^{L}_{H^- f \bar{f'}}
\,
m_{f'}^2
+
g^{R}_{H^- f \bar{f'}}
\,
m_{f}^2
\Big),
\\
c_{2f}^2 &=&
M_{H^\pm}^2
\Big\{
\Delta Q_{ff'}
\Big[
g^{L}_{H^- f \bar{f'}}
\,
m_{f'}^2
(m_{f'}^2-m_{f}^2-M_{H^\pm}^2)
\nonumber\\
&&
\hspace{2.5cm}
+
g^{R}_{H^- f \bar{f'}}
\,
m_{f}^2
(m_{f'}^2-m_{f}^2+M_{H^\pm}^2)
\Big]
\nonumber \\
&&
+
2 M_{W}^2
\Big[
\big( Q_{f'} m_{f'}^2 \big)
\,
g^{L}_{H^- f \bar{f'}}
+
\big( Q_{f} m_{f}^2 \big)
\,
g^{R}_{H^- f \bar{f'}}
\Big]
\Big\},
\\
c_{2f}^3 &=&
g^{L}_{H^- f \bar{f'}}
\,
m_{f'}^2
\Big\{
M_{H^\pm}^2
\Delta Q_{ff'}
(2 m_{f}^2-2 m_{f'}^2+M_{H^\pm}^2)
\nonumber \\
&&
+
M_{W}^2
\Big[
Q_{f'}
(m_{f'}^2-m_{f}^2-2 M_{H^\pm}^2)
+
Q_{f}
(m_{f}^2-m_{f'}^2)
\Big]
\Big\}
\nonumber\\
&& +
g^{R}_{H^- f \bar{f'}}
m_{f}^2
\Big\{
Q_{f}
\Big[
M_{H^\pm}^2
\big(
2 m_{f'}^2
-
2 m_{f}^2
+
M_{H^\pm}^2
-
2 M_{W}^2
\big)
+
M_{W}^2
\big(
m_{f}^2
-
m_{f'}^2
\big)
\Big]
\nonumber \\
&&
-
Q_{f'}
\Big[
M_{H^\pm}^2
\big(
2 m_{f'}^2
-
2 m_{f}^2
+
M_{H^\pm}^2
\big)
+
M_{W}^2
\big(
m_{f}^2
-
m_{f'}^2
\big)
\Big]
\Big\},
\\
c_{2f}^4 &=&
M_{H^\pm}^2
\big( M_{W}^2 - M_{H^\pm}^2 \big)
\big( Q_{f'} m_{f'}^2 \big)
\Big[
g^{L}_{H^- f \bar{f'}}
(2 m_{f'}^2-M_{H^\pm}^2+M_{W}^2)
+
2 g^{R}_{H^- f \bar{f'}} m_{f}^2
\Big],
\\
c_{2f}^5 &=& M_{H^\pm}^2
\big( M_{W}^2 - M_{H^\pm}^2 \big)
\big( Q_{f} m_{f}^2 \big)
\Big[
2 g^{L}_{H^- f \bar{f'}}
m_{f'}^2
+
g^{R}_{H^- f \bar{f'}}
(2 m_{f}^2-M_{H^\pm}^2+M_{W}^2)
\Big].
\end{eqnarray}
The last form factor is given by
\begin{eqnarray}
F^{f}_{3}
&=&
-
\dfrac{e}{(4\pi)^2}
\sum
\limits_{ \{ f, f' \} }
\dfrac{2 N^C_{f f'}
\,
g_{W^+ \bar{f} \, f'}
}{
\big( M_{H^\pm}^2-M_{W}^2 \big)}
\Big\{
c_{3f}^1
\Big[
B_{0}(M_{H^\pm}^2,m_{f'}^2,m_{f}^2)
-
B_{0}(M_{W}^2,m_{f'}^2,m_{f}^2)
\Big]
\nonumber \\
&&
+
c_{3f}^2
C_{0}(0,M_{H^\pm}^2,M_{W}^2,
m_{f'}^2,m_{f'}^2,m_{f}^2)
+
c_{3f}^3
C_{0}(M_{W}^2,0,
M_{H^\pm}^2,m_{f'}^2,m_{f}^2,m_{f}^2)
\Big\}.
\end{eqnarray}
The corresponding coefficients are given by
\begin{eqnarray}
c_{3f}^1 &=&
\Delta Q_{ff'}
\Big(
g^{L}_{H^- f \bar{f'}}
\,
m_{f'}^2
+
g^{R}_{H^- f \bar{f'}}
\,
m_{f}^2
\Big),
\\
c_{3f}^2 &=&\big( M_{H^\pm}^2-M_{W}^2 \big)
g^{L}_{H^\pm f \bar{f'}}
\big( Q_{f'} m_{f'}^2 \big),
\\
c_{3f}^3 &=&-\big( M_{H^\pm}^2-M_{W}^2 \big)
g^{R}_{H^- f \bar{f'}}
\big( Q_{f} m_{f}^2 \big).
\end{eqnarray}
In all the above form factors, the color factor
$N^C_{ff'}$ appears for quarks (leptons) in the loop,
with a value of $3$ $(1)$, respectively.
We define $\Delta Q_{ff'} = Q_{f'} - Q_{f}$, and
$\{ f, f' \}$ denotes the fermion doublets:
$\{ t, b \}$, $\{ c, s \}$, $\{ u, d \}$ for quarks,
and $\{ \ell, \nu_\ell \}$ for leptons,
with $\ell = e, \mu, \tau$.
The corresponding general couplings for the vector
boson--fermion vertices are given by
$g^{L}_{\gamma f \bar{f}} = g^{R}_{\gamma f \bar{f}}
\equiv g_{\gamma f \bar{f}},
\;
g^{R}_{W^\pm f \bar{f'}} = 0,
\;
g^{L}_{W^\pm f \bar{f'}} \equiv g_{W^\pm f \bar{f'}}
= \frac{e}{\sqrt{2}\, s_W}.$

In order to verify the UV-finiteness and
$\mu^2$-independence of all form factors,
we first expand the scalar one-loop
functions $A_0$ and $B_0$ in
terms of the UV-divergent parameter
$\Delta = \tfrac{1}{\varepsilon}
- \gamma_E + \ln(4\pi)$
and the renormalization
scale $\mu^2$, as
follows~\cite{Denner:1991kt,
Hahn:1998yk}:
\begin{eqnarray}
A_{0}(m^2)
&=&
m^2
\Big[
\Delta
+
\ln
\Big(
\dfrac{\mu^2}{m^2}
\Big)
\Big]
+
\text{finite terms},
\\
B_{0}(p^2,m_{1}^2,m_{2}^2)
&=&
\Delta
+
\ln
\Big(
\dfrac{\mu^2}{m_{2}^2}
\Big)
+
\text{finite terms}.
\end{eqnarray}
While the scalar one-loop
function $C_0$
is itself UV-finite and
independent
of the scale $\mu^2$.
We consequently confirm
the following relation:
\begin{eqnarray}
\Big[
B_{0}(M_{H^\pm}^2,m_{f'}^2,m_{f}^2)
-
B_{0}(M_{W}^2,m_{f'}^2,m_{f}^2)
\Big]
\Big|_{\Delta, \mu^2 - \text{term}}
&=&
0
\end{eqnarray}
As a result, we verify that
the form factor $F^{f}_{3}$ is
both UV-finite and independent
of the scale $\mu^2$.
In fact, one can check that
\begin{eqnarray}
F^{f}_{3}|_{\Delta,~\mu^2-\text{term}}
= 0.
\end{eqnarray}
Similarly,
the form factor $F^{f}_{2}$,
can be factored of
$\Delta$ and $\ln \, (\mu^2)$
as follows:
\begin{eqnarray}
F^{f}_{2}|_{\Delta,~\mu^2 - \text{term}}
&=&
-
\dfrac{e}{(4\pi)^2}
\sum
\limits_{ \{ f, f' \} }
\dfrac{2 N^C_{f f'}
\,
g_{W^+ \bar{f} \, f'}}
{M_{H^\pm}^2 (M_{H^\pm}^2-M_{W}^2)^2}
\Big[
\big(
m_{f'}^2
-
m_{f}^2
\big)
c_{2f}^1
+
c_{2f}^2
+
c_{2f}^3
\Big]
\\
&=&
-
\dfrac{e}{(4\pi)^2}
\sum
\limits_{ \{ f, f' \} }
\dfrac{2 N^C_{f f'}
\,
g_{W^+ \bar{f} \, f'}}
{M_{H^\pm}^2
(M_{H^\pm}^2-M_{W}^2)^2}
\Big[
c_{L}
\,
m_{f'}^2
\,
g^{L}_{H^- f \bar{f'}}
+
c_{R}
\,
m_{f}^2
\,
g^{R}_{H^- f \bar{f'}}
\Big].
\nonumber
\end{eqnarray}
Both coefficients $c_{L}$
and $c_{R}$ vanish,
which can be verified
analytically as
\begin{eqnarray}
c_{L}
&=&
\big(
m_{f'}^2
-
m_{f}^2
\big)
\Delta Q_{ff'}
\big( M_{H^\pm}^2-M_{W}^2 \big)
+
M_{H^\pm}^2
\Big[
\Delta Q_{ff'}
\big(
m_{f'}^2-m_{f}^2-M_{H^\pm}^2
\big)
+
2 Q_{f'}
M_{W}^2
\Big]
\nonumber \\
&&
+
M_{H^\pm}^2
\Delta Q_{ff'}
(2 m_{f}^2-2 m_{f'}^2+M_{H^\pm}^2)
+
M_{W}^2
\Big[
Q_{f'}
(m_{f'}^2-m_{f}^2-2 M_{H^\pm}^2)
+
Q_{f}
(m_{f}^2-m_{f'}^2)
\Big]
\nonumber \\
&=&
M_{H^\pm}^2
\Delta Q_{ff'}
\Big[
\big(
m_{f'}^2
-
m_{f}^2
\big)
+
\big(
m_{f'}^2-m_{f}^2-M_{H^\pm}^2
\big)
+
(2 m_{f}^2-2 m_{f'}^2+M_{H^\pm}^2)
\Big]
\nonumber \\
&&
+
M_{W}^2
\Big[
Q_{f'}
(m_{f'}^2-m_{f}^2-2 M_{H^\pm}^2)
+
Q_{f}
(m_{f}^2-m_{f'}^2)
+
2 Q_{f'}
M_{H^\pm}^2
-
\big(
m_{f'}^2
-
m_{f}^2
\big)
\Delta Q_{ff'}
\Big]
\nonumber \\
&=&
0,
\\
c_{R}
&=&
\Delta Q_{ff'}
\big(
m_{f'}^2
-
m_{f}^2
\big)
\big( M_{H^\pm}^2-M_{W}^2 \big)
+
M_{H^\pm}^2
\Big[
\Delta Q_{ff'}
\big(
m_{f'}^2-m_{f}^2+M_{H^\pm}^2
\big)
+
2 Q_{f}
M_{W}^2
\Big]
\nonumber \\
&&
+
Q_{f}
\Big[
M_{H^\pm}^2
\big(
2 m_{f'}^2
-
2 m_{f}^2
+
M_{H^\pm}^2
-
2 M_{W}^2
\big)
+
M_{W}^2
\big(
m_{f}^2
-
m_{f'}^2
\big)
\Big]
\nonumber \\
&&
+
Q_{f'}
\Big[
M_{H^\pm}^2
\big(
2 m_{f}^2
-
2 m_{f'}^2
-
M_{H^\pm}^2
\big)
+
M_{W}^2
\big(
m_{f'}^2
-
m_{f}^2
\big)
\Big]
\nonumber \\
&=&
Q_{f}
\Big[
M_{H^\pm}^2
\big(
2 m_{f'}^2
-
2 m_{f}^2
+
M_{H^\pm}^2
-
2 M_{W}^2
\big)
+
M_{W}^2
\big(
m_{f}^2
-
m_{f'}^2
\big)
\nonumber \\
&&
-
\big(
m_{f'}^2
-
m_{f}^2
\big)
\big( M_{H^\pm}^2-M_{W}^2 \big)
-
M_{H^\pm}^2
\big(
m_{f'}^2-m_{f}^2+M_{H^\pm}^2
\big)
+
2 M_{W}^2 M_{H^\pm}^2
\Big]
\nonumber \\
&&
+
Q_{f'}
\Big[
M_{H^\pm}^2
\big(
2 m_{f}^2
-
2 m_{f'}^2
-
M_{H^\pm}^2
\big)
+
M_{W}^2
\big(
m_{f'}^2
-
m_{f}^2
\big)
\nonumber \\
&&
+
\big(
m_{f'}^2
-
m_{f}^2
\big)
\big( M_{H^\pm}^2-M_{W}^2 \big)
+
M_{H^\pm}^2
\big(
m_{f'}^2-m_{f}^2+M_{H^\pm}^2
\big)
\Big]
\nonumber \\
&=&
0.
\end{eqnarray}
In terms of $\Delta$ and $\ln(\mu^2)$,
the form factor $F^{W^\pm H^\pm}_{2}$
can also be expressed as
\begin{eqnarray}
F^{W^\pm H^\pm}_{2}|_{
\Delta, ~\mu^2 - \text{term}}
&=&
-
\dfrac{e}{(4\pi)^2}
\dfrac{1}{
M_{H^\pm}^2
M_{W}^2
\big(M_{H^\pm}^2-M_{W}^2\big)^2}
\times
\\
&&\times
\sum
\limits_{\phi = h, H}
g_{\phi \, H^- W^+}
\Big[
M_{\phi}^2
c^{1}_{\phi}
+
M_{H^\pm}^2
c^{2}_{\phi}
+
M_{W}^2
c^{3}_{\phi}
+
c^{4}_{\phi}
+
c^{5}_{\phi}
+
c^{6}_{\phi}
+
c^{7}_{\phi}
\Big]
\nonumber\\
&=&
-
\dfrac{e}{(4\pi)^2}
\dfrac{1}{
M_{H^\pm}^2
M_{W}^2
\big(M_{H^\pm}^2-M_{W}^2\big)^2}
\times
\\
&&
\times
\sum
\limits_{\phi = h, H}
\Big[
c_{W^\pm}
\,
g_{\phi \, H^- W^+}
\,
g_{\phi W^\pm W^\mp}
+
c_{H^\pm}
\,
g_{\phi \, H^- W^+}
\,
g_{\phi H^\pm H^\mp}
\Big]
\nonumber\\
&=&
M_{W}^2
\big(
M_{H^\pm}^2
-
M_{W}^2
\big)
\big(
M_{H^\pm}^2
+
M_{W}^2
\big)
\sum
\limits_{\phi = h, H}
g_{\phi \, H^- W^+}
\,
g_{\phi W^\pm W^\mp}
=
0.\nonumber
\end{eqnarray}
To arrive the result in the last line
of the above equation, we verified that
$c_{H^{\pm}}=0$ and
$c_{W^\pm}$ is given by
\begin{eqnarray}
c_{H^\pm}
&=&
2 M_{W}^2
\Big[
M_{\phi}^2
\big(M_{H^\pm}^2-M_{W}^2\big)
-
M_{H^\pm}^2
\big(M_{H^\pm}^2-M_{W}^2\big)
+
M_{\phi}^2
M_{W}^2
-
2
M_{H^\pm}^2
\big(
M_{\phi}^2
-
M_{H^\pm}^2
\big)
\nonumber \\
&&
+
M_{H^\pm}^2
\big(
M_{\phi}^2-M_{H^\pm}^2
-M_{W}^2 \big) \Big]
=0,
\\
c_{W^\pm}
&=&
\big(M_{H^\pm}^2-M_{W}^2\big)
\Big[
\big(
M_{H^\pm}^2
-
M_{W}^2
\big)^2
-
M_{\phi}^2
M_{H^\pm}^2
-
M_{H^\pm}^2
\big(
M_{H^\pm}^2
-
M_{\phi}^2
-
3 M_{W}^2
\big)
\Big]
\nonumber\\
&=&
M_{W}^2
\big(
M_{H^\pm}^2
-
M_{W}^2
\big)
\big(
M_{H^\pm}^2
+
M_{W}^2
\big).
\end{eqnarray}
In conclusion, we have verified
that all form factors are UV-finite
and independent of the renormalization
scale $\mu^2$.
\section{Pair Charged Higgs Bosons production
at muon--TeV Colliders}
In this section, we investigate
the production of a charged Higgs
boson pair at muon--TeV colliders through
the processes
$\mu^+ \mu^- \rightarrow H^{+} H^{-} \rightarrow
W^{+} W^{-} h \gamma$
and
$\mu^+ \mu^- \rightarrow \gamma \gamma
\rightarrow H^{+} H^{-} \rightarrow W^{+} W^{-} h \gamma$
in the scope of the THDM. As a representative
scenario, we focus on
the phenomenological investigations of
the Type-I THDM. We first perform parameter scans
of the model and present the branching ratios of the
charged Higgs decays as functions of the relevant
parameters within the viable parameter space.
Subsequently, we simulate the signal of charged Higgs
pair production at future muon--TeV colliders, including the
relevant Standard Model backgrounds.
\subsection{Scanning parameters} 
We first perform scans over the parameter space of the
Type-I THDM within the following ranges:
$M_{h}=125.09~\textrm{GeV}$, $M_{H} \in [130, 1000]~\textrm{GeV}$,
$s_{\beta-\alpha} \in [0.97, 1]$, $M_{A,H^{\pm}} \in [130, 1000]~\textrm{GeV}$,
$t_{\beta} \in [0.5, 45]$, and $m_{12}^2 \in [0, 10^6]~\textrm{GeV}^2$.
The parameter scan is carried out in several steps. First, theoretical constraints
are applied to eliminate unphysical regions. The surviving points are then tested against EWPOs. In the next step, the remaining
points are examined using \texttt{HiggsBounds-5.10.1}, which incorporates exclusion limits from Higgs searches at colliders, and \texttt{HiggsSignals-2.6.1}, which evaluates compatibility with Higgs precision measurements at the LHC. Finally, the parameter space is further constrained using \textbf{SuperIso v4.1} to account for flavor physics observables. After applying all these constraints,
the allowed parameter space is then systematically analyzed and discussed.

\begin{figure}[H]
\centering
\begin{tabular}{cc}
\includegraphics[width=8cm, height=6cm]
{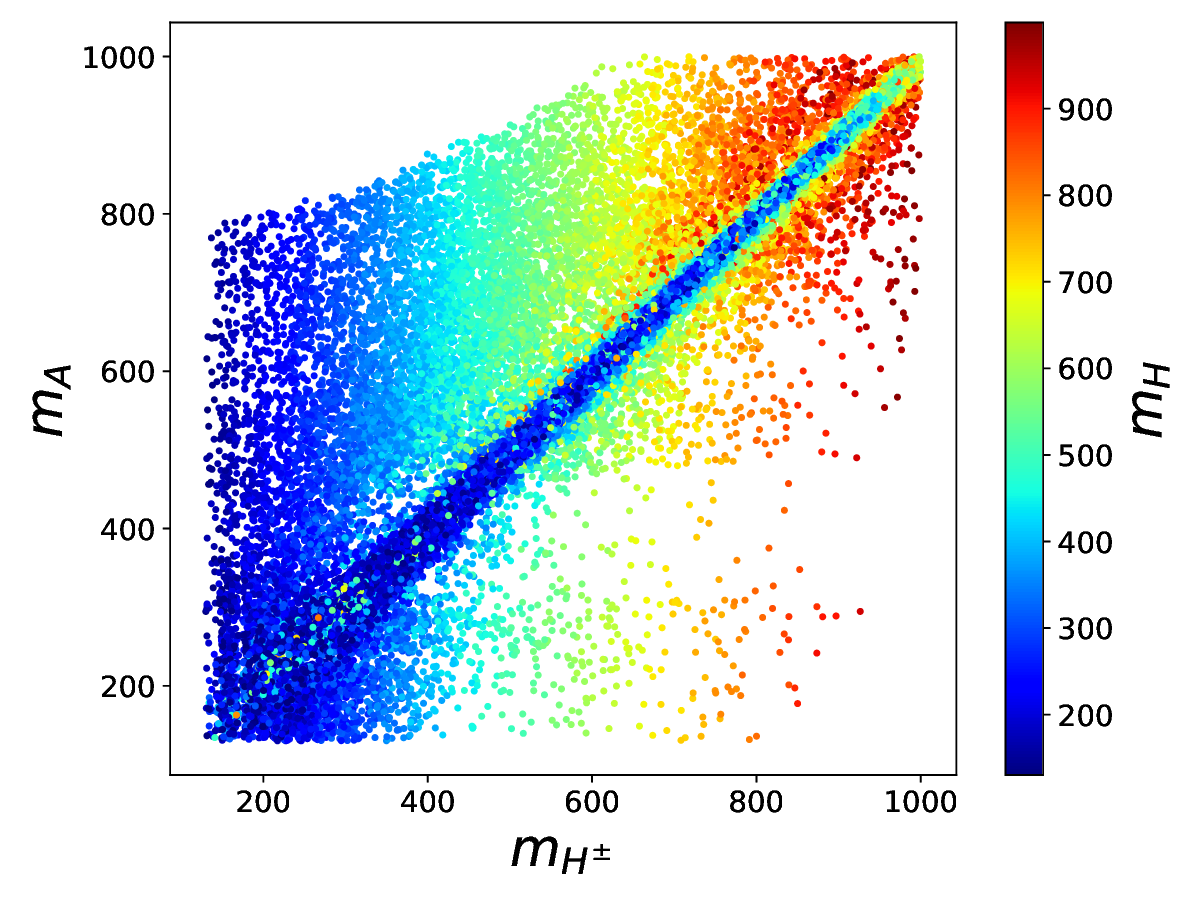}
&
\includegraphics[width=8cm, height=6cm]
{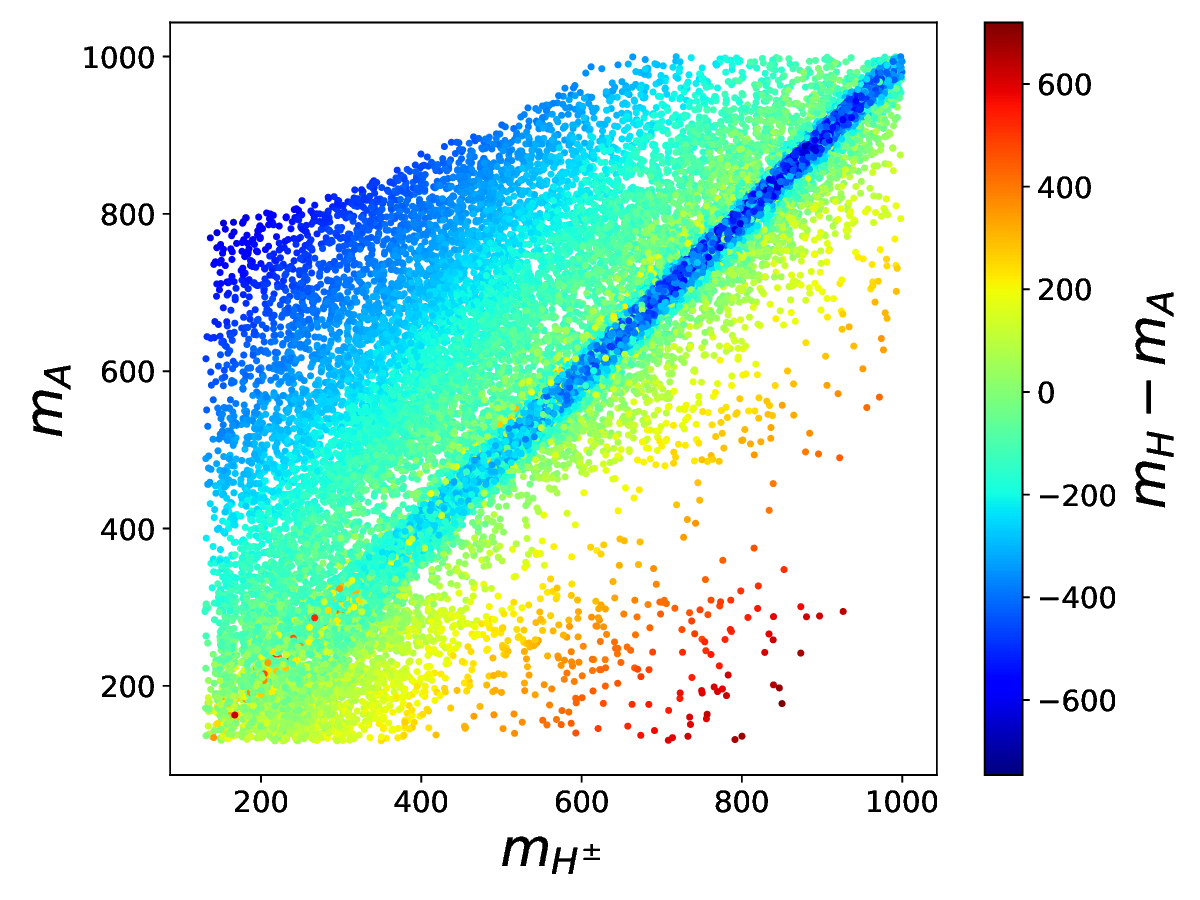}
\end{tabular}
\caption{\label{MATHDM1}
The allowed regions of the parameter
space for the Type-I THDMare illustrated
in scatter plots showing the correlations
among $M_{A}$, $M_{H^{\pm}}$,
and $M_{H}$, as well as between
$M_{A}$, $M_{H^{\pm}}$,
and $M_{H}-M_{A}$.
}
\end{figure}

\begin{figure}[H]
\centering
\begin{tabular}{cc}
\includegraphics[width=8cm, height=6cm]
{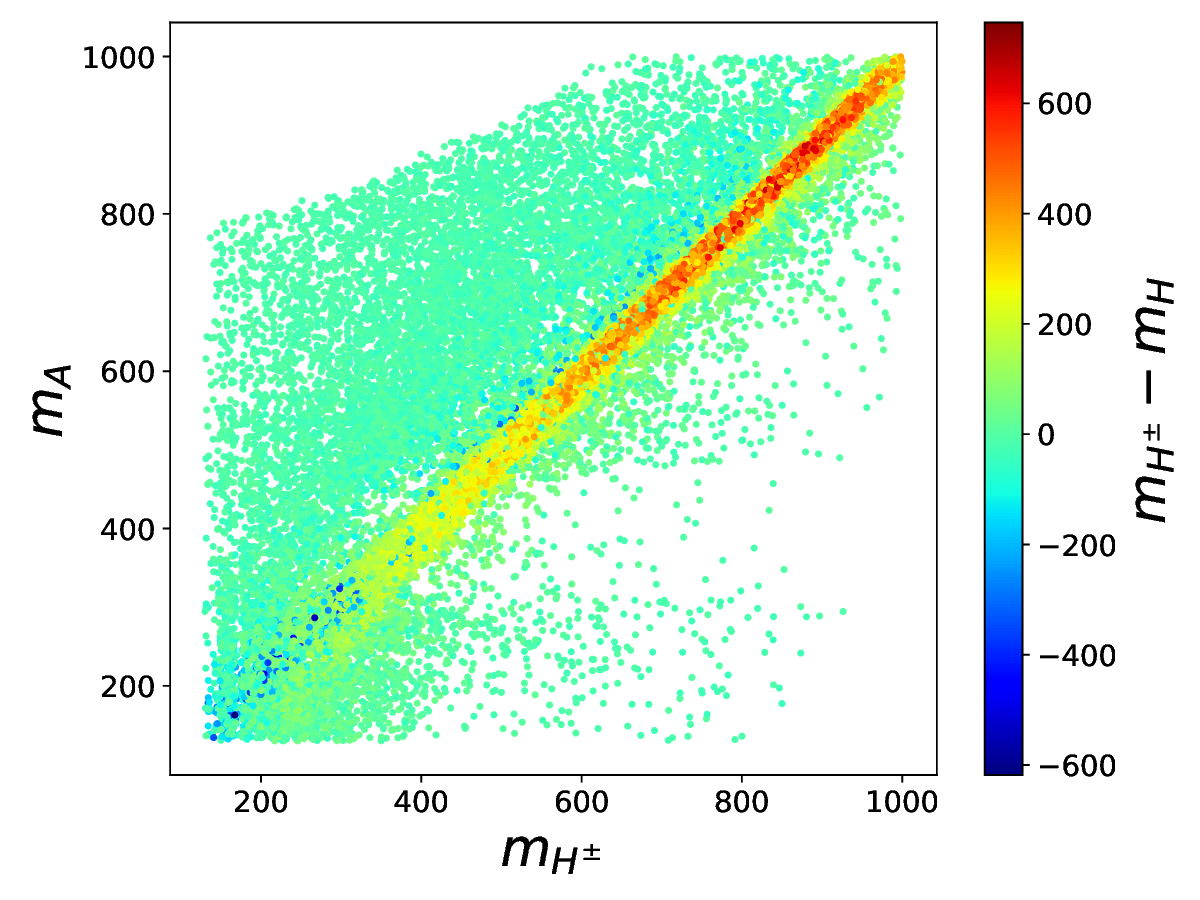}
&
\includegraphics[width=8cm, height=6cm]
{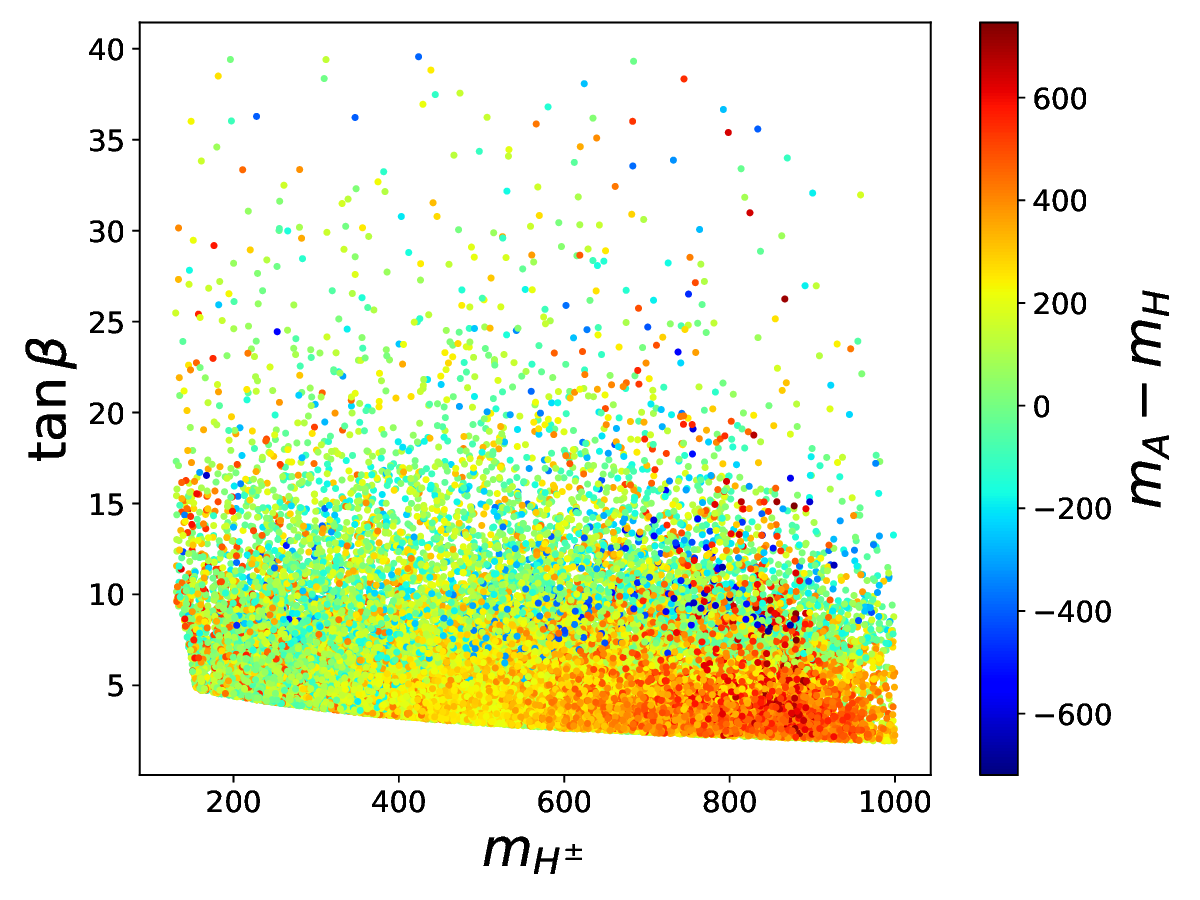}
\end{tabular}
\caption{\label{MATHDM2}
The allowed regions of the parameter space
for the Type-I THDM are shown in Fig.~\ref{MATHDM2}.
In the left panel, we display the correlations among
$M_A$, $M_{H^{\pm}}$, and $M_{H^{\pm}} - M_H$; in
the right panel, the correlations among $t_{\beta}$,
$M_{H^{\pm}}$, and $M_{A} - M_H$ are presented.}
\end{figure}

\begin{figure}[H]
\centering
\begin{tabular}{cc}
\includegraphics[width=8cm, height=6cm]
{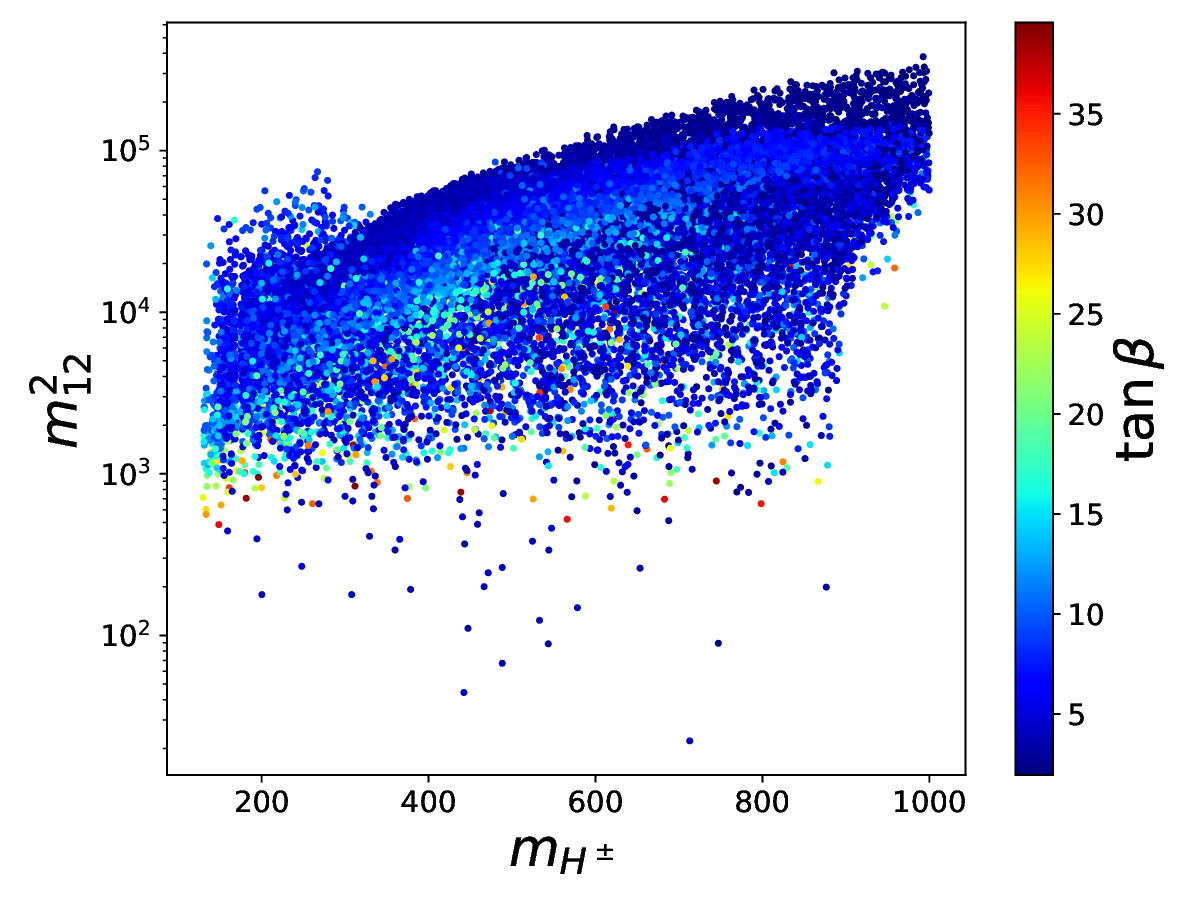}
&
\includegraphics[width=8cm, height=6cm]
{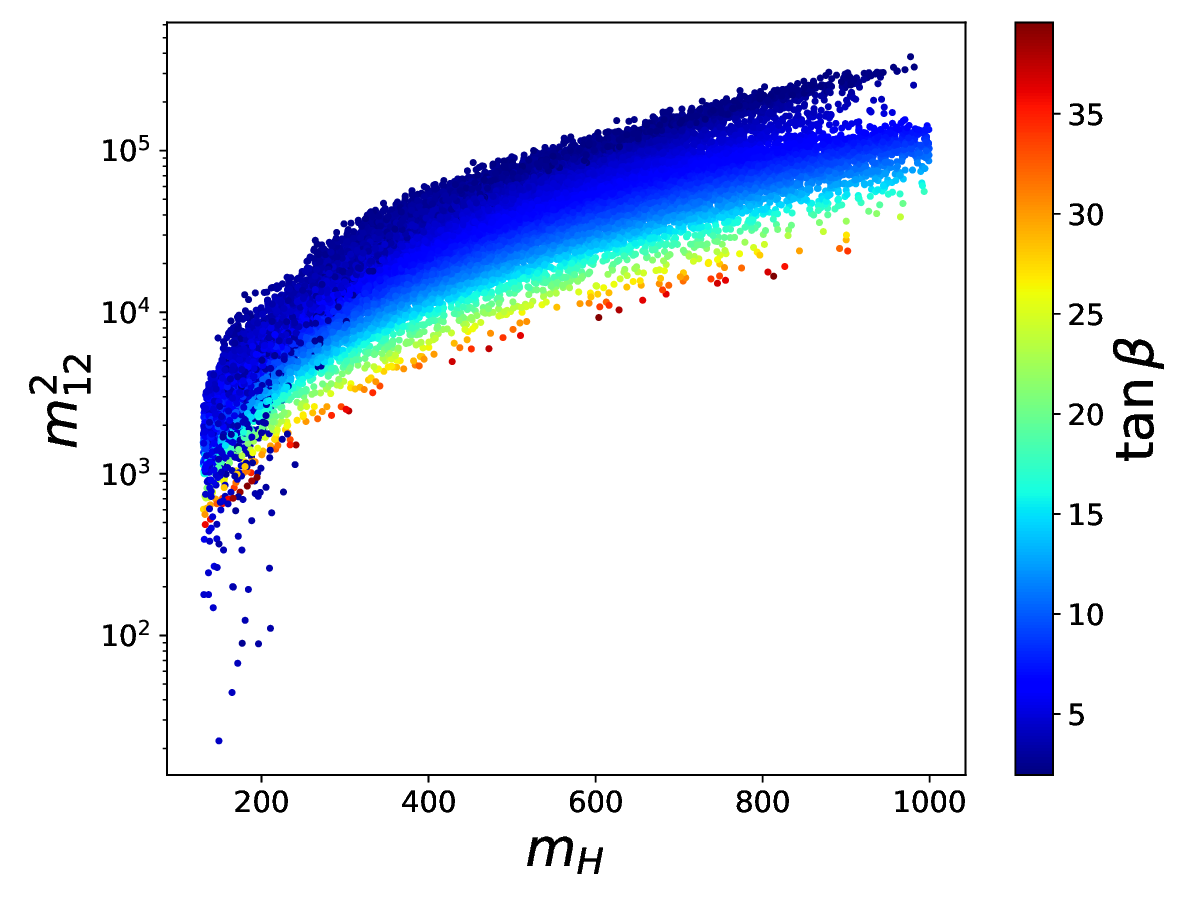}
\end{tabular}
\caption{\label{MATHDM3}
The correlations in the parameter space,
including $m_{12}^2$, $M_{H^{\pm}}^2$,
and $t_{\beta}$ in the left scatter plot,
and $m_{12}^2$, $M_H^2$, and $t_{\beta}$
in the right scatter plot, are shown in
these plots.}
\end{figure}

The viable parameter regions of the Type-I THDM are illustrated using scatter plots, showing the correlations among $M_{A}$, $M_{H^{\pm}}$, and $M_{H}$ in the left panel, and between $M_{A}$, $M_{H^{\pm}}$, and $M_{H} - M_{A}$ in the right panel, as depicted in Fig.~\ref{MATHDM1}. The results indicate that the regions with $M_{A} = M_{H^{\pm}}$ or $M_{H} = M_{H^{\pm}}$ are preferred by the EWPOs.
We next examine the viable regions of the parameter 
space for the Type-I THDM, as illustrated in Fig.~\ref{MATHDM2}. 
The left panel shows the correlations among $M_{A}$, $M_{H^{\pm}}$,
and $M_{H^{\pm}} - M_{H}$, while the right panel
presents the correlations among $t_{\beta}$, 
$M_{H^{\pm}}$, and $M_{A} - M_{H}$. 
In the left panel, the results indicate that the regions
with $M_{A} = M_{H^{\pm}}$ or $M_{H} = M_{H^{\pm}}$
are preferred by the EWPOs.
As shown in the right panel, the viable regions
are mostly favored in the range $2 \leq t_{\beta} \lesssim 20$.
We finally investigate the correlations in
the parameter space, including $m_{12}^2$,
$M_{H^{\pm}}^2$, and $t_{\beta}$ in the
left scatter plot, and $m_{12}^2$, $M_H^2$,
and $t_{\beta}$ in the right scatter plot,
as shown in Fig.~\ref{MATHDM3}. Across all
ranges of charged Higgs masses, the soft
$Z_2$-breaking parameter is found within
$10^3 \leq m_{12}^2 \leq 10^5$. The data
are generally preferred in regions with
$t_{\beta} \lesssim 20$. A similar 
preferred region of the parameter 
space is observed in the right 
plot as in the left.
\subsection{Branching ratios for
$H^{\pm} \rightarrow W^{\pm }\gamma$}
In this subsection, the
branching ratios for the decay
$H^{\pm} \rightarrow W^{\pm} \gamma$
are investigated 
within the Type-I THDM. For the total
decay width of the charged Higgs boson,
we consider all relevant decay modes,
including $H^{\pm} \rightarrow f\bar{f}'$,
$W^\pm \gamma$, $W^\pm Z$, and $W^\pm \phi_j$,
where $\phi_j = h, H, A$. The one-loop
expressions for the fermionic decays
$H^{\pm} \rightarrow f\bar{f}'$, with
$f = u, c, t, \ell$ and $f' = d, s, b, \nu_\ell$,
as well as for the decays
$H^{\pm} \rightarrow W^\pm \phi_j$,
are taken from Ref.~\cite{Aiko:2021can}.
The one--loop--induced process
$H^{\pm} \rightarrow W^\pm \gamma$
is computed in the present work,
while the expression for
$H^{\pm} \rightarrow W^\pm Z$
is adopted from Ref.~\cite{KHP}.

Based on the viable regions of the
parameter space for the Type-I THDM,
we consider two benchmark points.
In the first scenario, we set 
$M_{A} = M_{H} = 160~\mathrm{GeV}$, 
whereas in the second scenario, 
we take $M_{A} = M_{H} = 500~\mathrm{GeV}$. 
In both cases, we fix $s_{\beta-\alpha} = 0.98$ 
and $m_{12}^2 = 10^{4}~\mathrm{GeV}^2$.
The charged Higgs mass is varied within
the range $200~\mathrm{GeV} \leq
M_{H^{\pm}} \leq 600~\mathrm{GeV}$,
and the mixing angle is taken as
$t_{\beta} = 2, 8, 14, 20$ in the
subsequent plots.

\begin{figure}[H]
\centering
\begin{tabular}{cc}
\hspace{-4cm}
Br$\{H^{\pm} \rightarrow AB\}$
&
\hspace{-4cm}
Br$\{H^{\pm} \rightarrow AB\}$
\\
\includegraphics[width=8cm, height=6cm]
{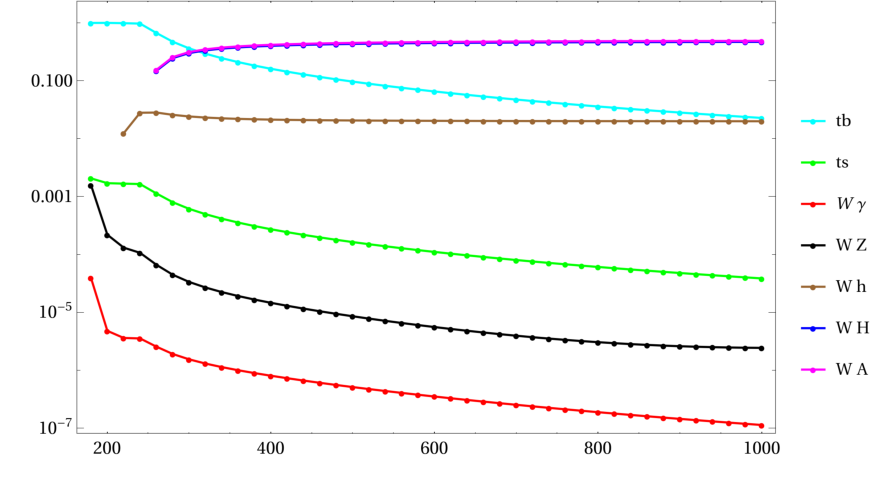}
 &
\includegraphics[width=8cm, height=6cm]
{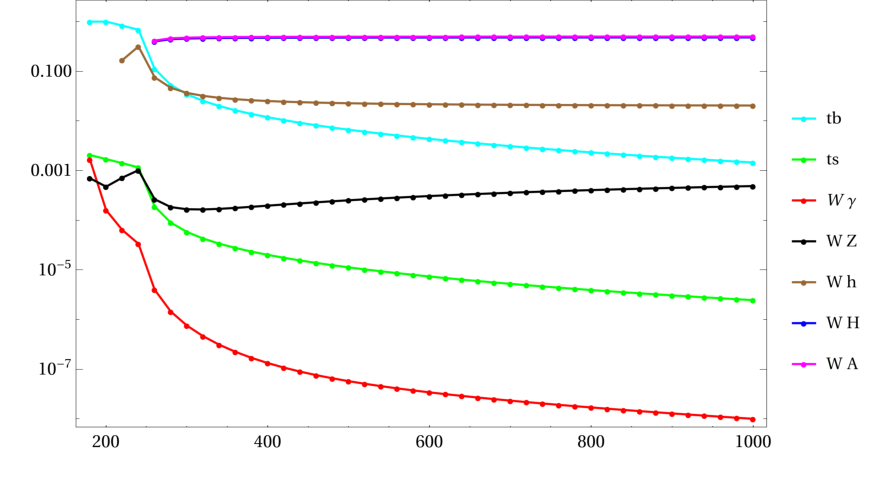}
\\
\includegraphics[width=8cm, height=6cm]
{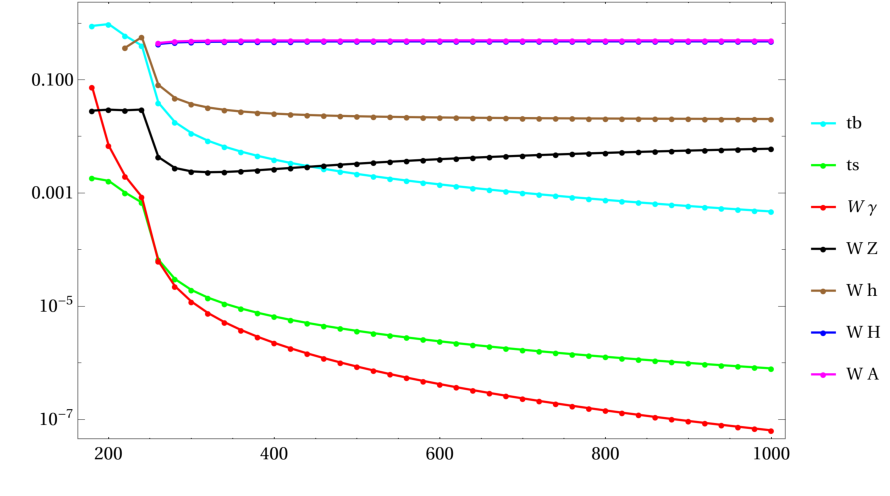}
 &
\includegraphics[width=8cm, height=6cm]
{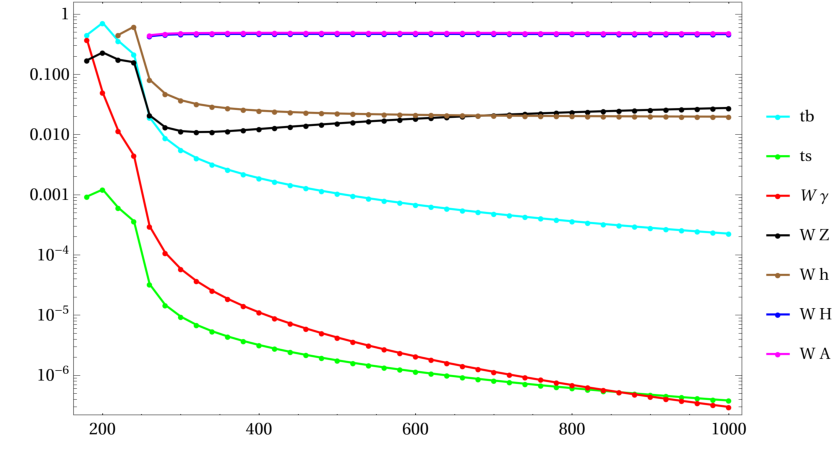}
\\
\hspace{4cm}
$M_{H^\pm}$ [GeV]
&
\hspace{4cm}
$M_{H^\pm}$ [GeV]
\end{tabular}
\caption{\label{BranchingHpm1}
The branching ratios of all charged Higgs
decay modes in the Type-I THDM are depicted
for the first benchmark scenario, corresponding
to $t_{\beta} = 2$ (top left), $t_{\beta} = 8$
(top right), $t_{\beta} = 14$ (bottom left),
and $t_{\beta} = 20$ (bottom right).
}
\end{figure}

\begin{figure}[H]
\centering
\begin{tabular}{cc}
\hspace{-4cm}
Br$\{H^{\pm} \rightarrow AB\}$
&
\hspace{-4cm}
Br$\{H^{\pm} \rightarrow AB\}$
\\
\includegraphics[width=8cm, height=6cm]
{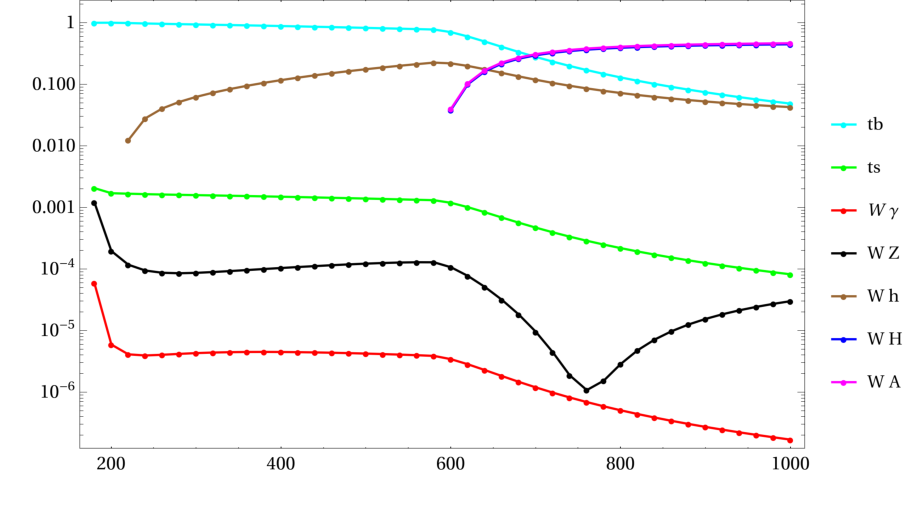}
 &
\includegraphics[width=8cm, height=6cm]
{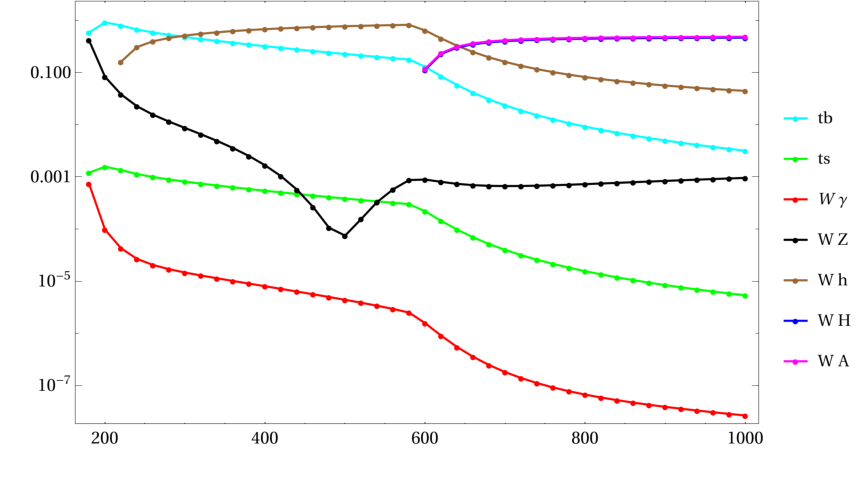}
\\
\includegraphics[width=8cm, height=6cm]
{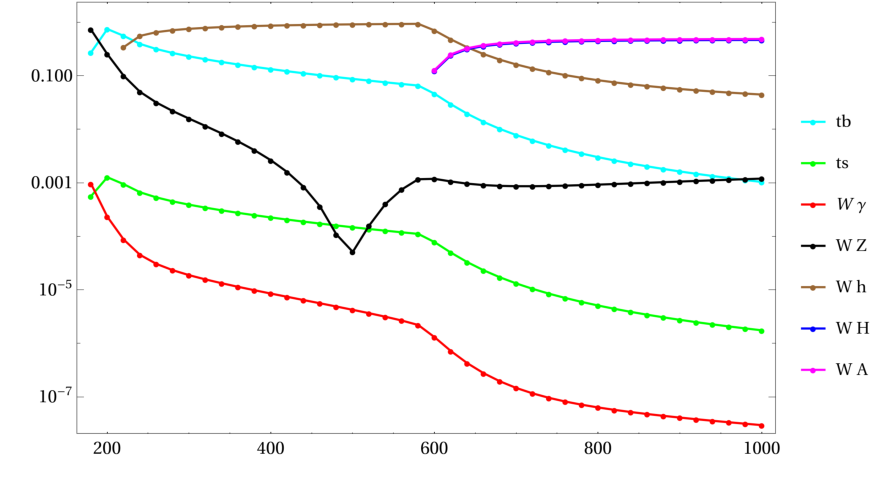}
 &
\includegraphics[width=8cm, height=6cm]
{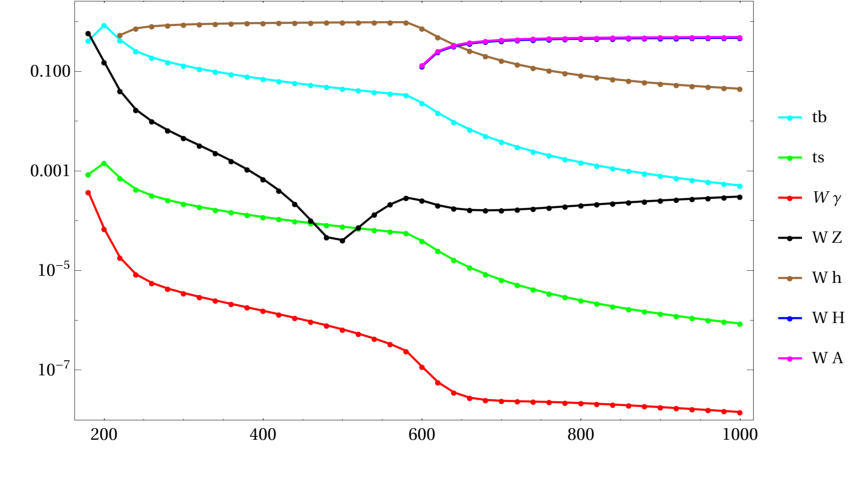}
\\
\hspace{4cm}
$M_{H^\pm}$ [GeV]
&
\hspace{4cm}
$M_{H^\pm}$ [GeV]
 \end{tabular}
\caption{\label{BranchingHpm2}
Branching ratios for all decay modes
of the charged Higgs boson in the Type-I THDM
for the second benchmark scenario, with
$t_{\beta} = 2$ (top left), $t_{\beta} = 8$
(top right), $t_{\beta} = 14$ (bottom left),
and $t_{\beta} = 20$ (bottom right), are
shown in Fig.~\ref{BranchingHpm2}.
}
\end{figure}

In Fig.~\ref{BranchingHpm1}, we present the decay fractions of all charged Higgs decay modes for the first benchmark scenario, with $t_{\beta} = 2$ (top left), $t_{\beta} = 8$ (top right), $t_{\beta} = 14$ (bottom left), and $t_{\beta} = 20$ (bottom right). For $t_{\beta}=2$, the decay $H^{\pm} \rightarrow tb$ dominates in the region $M_{H^{\pm}} \leq (M_{H/A}+M_W)$. Additionally, for $t_{\beta}=8, 14, 20$, other channels such as $H^{\pm} \rightarrow ts$, $H^{\pm} \rightarrow W\gamma$, and $H^{\pm} \rightarrow WZ$ also contribute significantly within the same mass region.
Once the threshold $M_{H^{\pm}} \geq (M_{H/A}+M_W)$ is crossed, the decays $H^{\pm} \rightarrow W\phi_j$ become dominant for all values of $t_{\beta}$. Specifically, for $t_{\beta}=2$, the channels $H^{\pm} \rightarrow tb$ and $H^{\pm} \rightarrow WH$ provide substantial contributions, whereas for $t_{\beta}=8, 14, 20$, the $H^{\pm} \rightarrow WZ$ channel also yields a sizable contribution.
Notably, the decay fractions of $H^{\pm} \rightarrow W\gamma$ can reach $\mathcal{O}(10^{-1})$ in the low-mass regime of
the charged Higgs for large values of $t_{\beta}$. 
This behavior arises because the fermionic and bosonic contributions interfere destructively: for large $t_{\beta}$, the fermionic contribution is suppressed, while in the low-mass region, the mixing between the charged Higgs and the $W$ boson in the loop becomes significant. Conversely, in the high-mass region of the charged Higgs, the decay $H^{\pm} \rightarrow W\gamma$ is strongly suppressed, with branching ratios of order $10^{-6}$ for all values of $t_{\beta}$.
The branching fractions for all decay modes of
the charged Higgs boson in the Type-I THDM
for the second benchmark scenario, corresponding
to $t_{\beta} = 2$ (top left),
$t_{\beta} = 8$ (top right), $t_{\beta} = 14$
(bottom left), and $t_{\beta} = 20$ (bottom right),
are presented in Fig.~\ref{BranchingHpm2}.
Overall, the behavior of the branching ratios
is similar to that observed in the first
benchmark scenario. In particular, the branching
ratio of $H^{\pm} \rightarrow W\gamma$ can
reach $\mathcal{O}(10^{-2})$ in the low-mass
regime of the charged Higgs for large values
of $t_{\beta}$. In other regions, it is of the
order of $10^{-7}$ and can be neglected.
\subsection{Processes $\mu^+\mu^- \rightarrow
H^{+}H^{-} \rightarrow W^+W^- h\gamma$}
As pointed out in the previous subsection,
the branching fractions
of $H^{\pm} \rightarrow W\gamma$
can reach $\mathcal{O}(10^{-1})$
in the charged Higgs low-mass regime
and for large values of $t_{\beta}$
in the first benchmark scenario. This represents
a highly probable case that should be considered
for collider signatures. Moreover, when the threshold
$M_{H^{\pm}} \geq (M_{H/A}+M_W)$ opens,
the decay modes $H^{\pm} \rightarrow W\phi_j$
become dominant for all values of $t_{\beta}$.
For these reasons, we examine the production
signals of $H^{+}H^{-}$ at muon colliders in the
TeV range via the process
$\mu^+\mu^- \rightarrow H^{+}H^{-}
\rightarrow W^+W^- h\gamma$. For this analysis,
the SM background process $\mu^+\mu^- \rightarrow
W^+W^- h\gamma$ is also taken into account.
Both signal and SM background, we apply
kinematic cuts on the final-state
photon as follows $E_{\gamma} \geq 1~\text{MeV}$.
Furthermore, in order to reduce SM background,
invariant-mass cuts are applied as 
$|M_{W^+h}-M_{H^{\pm}}| \leq 10$ GeV and
$|M_{W^-\gamma}-M_{H^{\pm}}| \leq 10$ GeV.
The SM background is simulated using
{\tt GRACE}~\cite{Belanger:2003sd}
to generate tree-level cross sections.

\begin{figure}[H]
\centering
\begin{tabular}{cc}
\includegraphics[width=9.3cm, height=8cm]
{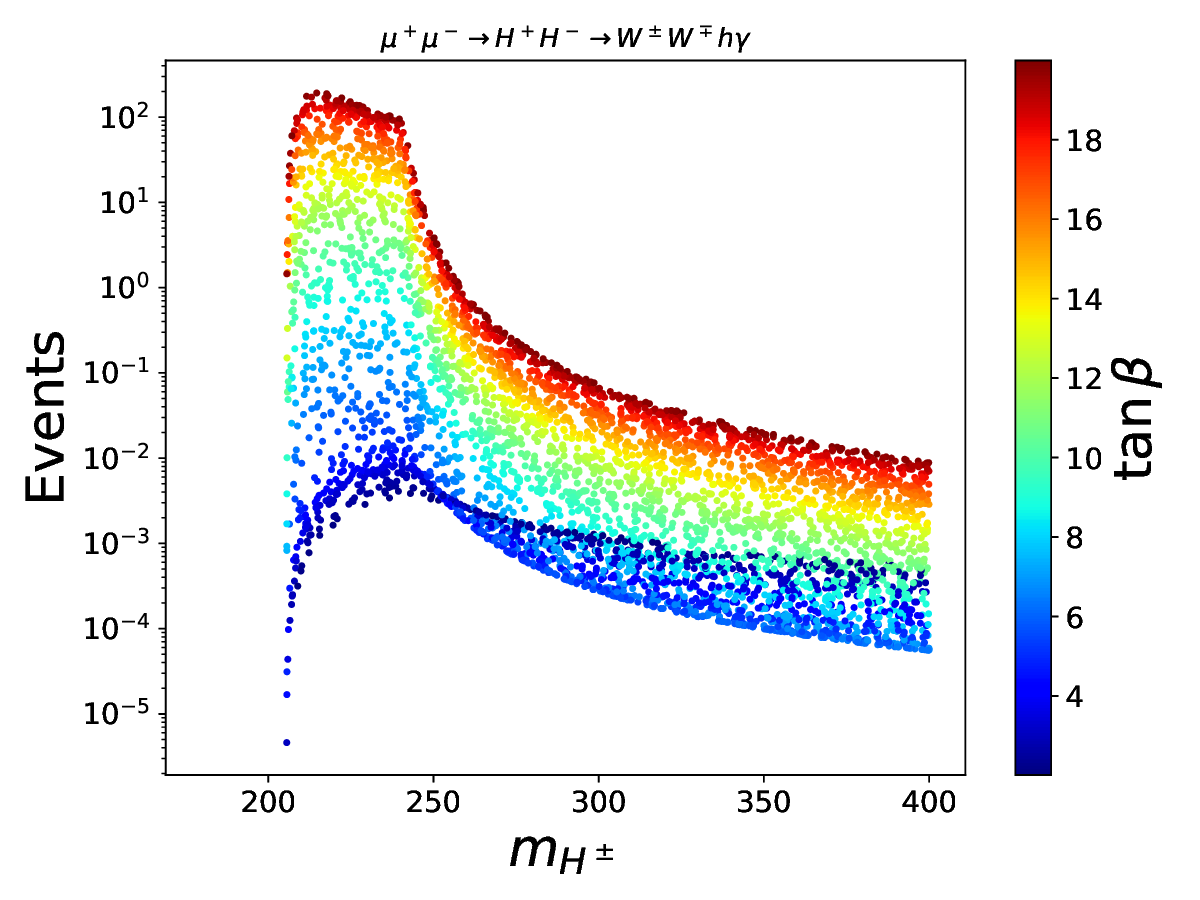}
&
\includegraphics[width=7.3cm, height=8cm]
{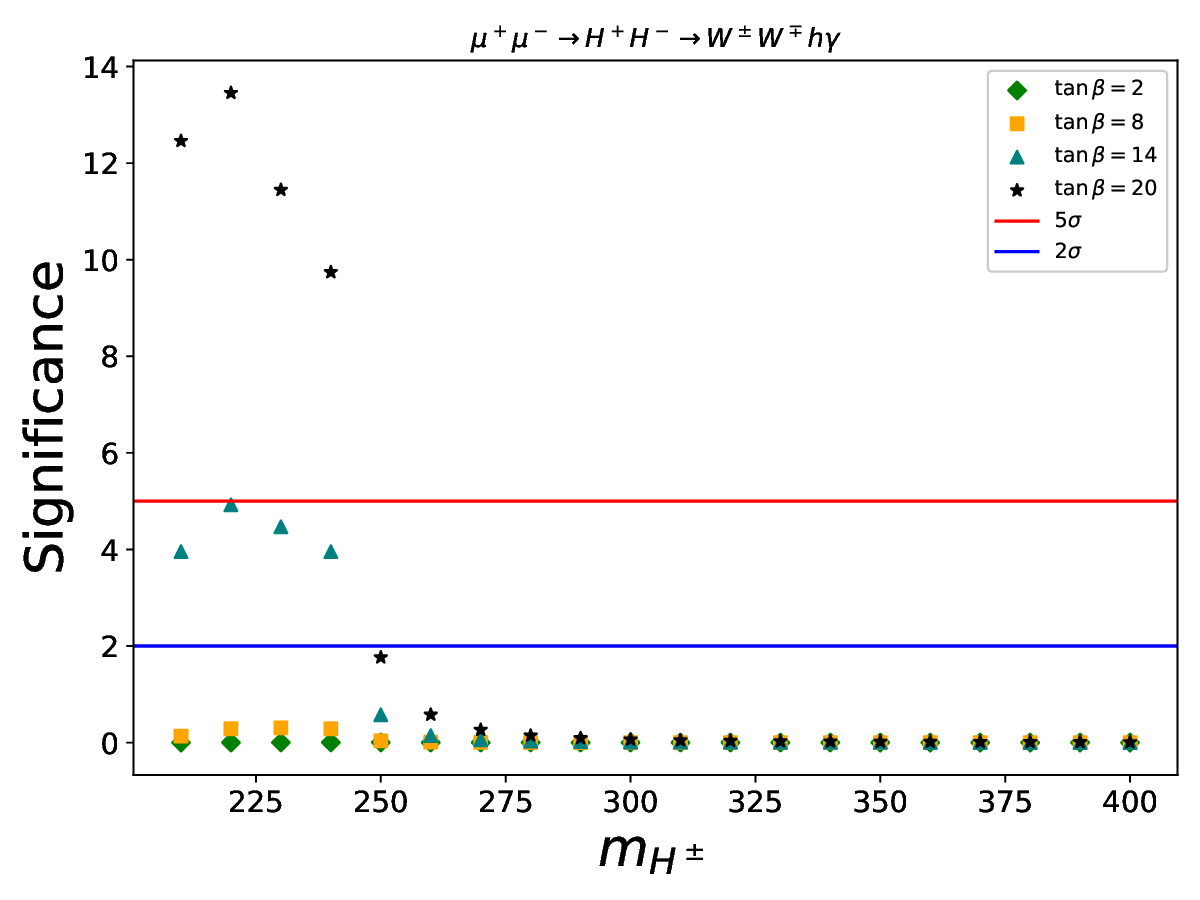}
\end{tabular}
\caption{\label{significantZ}
Events for the signal process are shown
in the parameter space of $(t_{\beta},
M_{H^{\pm}})$ at $\sqrt{s} = 3~\text{TeV}$
with an integrated luminosity of
$\mathcal{L} = 3000~\text{fb}^{-1}$ (left panel).
The corresponding statistical significance
of the signal is shown as a function
of the charged Higgs mass at $\sqrt{s} = 3~\text{TeV}$
with the same integrated luminosity (right panel).
}
\end{figure}

The number of events for the considered
processes is calculated as $N_{\rm Event}
= \mathcal{L} \, \sigma_S$, 
with $\mathcal{L}$ the integrated luminosity
and $\sigma_S$ the signal/background
cross section accordingly.
The statistical significance of the signal
is evaluated using
$\mathcal{S} = \frac{N_S}{\sqrt{N_S + N_B}}$,
where $N_S$ and $N_B$ correspond to the number
of signal and background events, respectively.
Figure~\ref{significantZ} illustrates the event rates
for the $W^+W^- h\gamma$ signal in the Type-I THDM
as a function of $t_{\beta}$ and $M_{H^{\pm}}$ for
the first benchmark scenario (left panel), along with
the corresponding statistical significance (right panel)
at $\sqrt{s} = 3$~TeV and
$\mathcal{L} = 3000~\text{fb}^{-1}$.
In the regions $210~\text{GeV} \leq M_{H^{\pm}} \leq 240~\text{GeV}$
and $t_{\beta} \geq 10$, the number of events is substantial.
Consequently, the statistical significance exceeds $2\sigma$ within
this region (in particular, $\mathcal{S}$
surpasses $5\sigma$ for $t_{\beta} = 20$),
whereas in other regions it remains negligible.
\subsection{Processes
$\mu^+\mu^- \rightarrow
\gamma \gamma \rightarrow
H^{+}H^{-} \rightarrow W^+W^- h\gamma$}
Another process considered as an
application of this work is the scattering
$\mu^+\mu^- \rightarrow \gamma\gamma
\rightarrow H^{+}H^{-} \rightarrow W^+W^- h\gamma$.
We first compute the partonic process
$\gamma\gamma \rightarrow H^{+}H^{-}$ within
the THDM framework
using {\tt FeynArts/FormCalc}~\cite{Hahn:2000kx}.
The total cross section is obtained by 
convolving the partonic cross section 
with the photon structure function $f_{\gamma/\mu}$:
\begin{equation}
\label{totalgamgam}
\sigma_{\text{tot}}(\mu^+ \mu^-
\to \gamma \gamma \to H^+ H^-, s)
=
\int_{\tfrac{2m_{H^\pm}}
{\sqrt{s}}}^{x_{\text{max}}}
dz \, \frac{dL_{\gamma\gamma}}{dz}
\, \hat{\sigma}(\hat{s} = z^2 s).
\end{equation}
The photon luminosity distribution 
function is given by
\begin{equation}
\label{luminosity}
\frac{dL_{\gamma\gamma}}{dz} = 2z
\int_{z^2/x_{\text{max}}}^{x_{\text{max}}}
\frac{dx}{x} \, f_{\gamma/\mu}(x)
\, f_{\gamma/\mu}\!
\left(\frac{z^2}{x}\right),
\end{equation}
where $f_{\gamma/\mu}(x)$ denotes
the photon structure function, 
with $x$ being the fraction of 
the incident muon's energy 
carried by the back-scattered 
photon. The explicit form 
of $f_{\gamma/\mu}(x)$ is 
given in~\cite{Zarnecki:2002qr}. 
In these formulas, 
we adopt $x_{\rm max} = 0.83$ 
following~\cite{Telnov:1989sd}.

\begin{figure}[H]
\centering
\begin{tabular}{cc}
\includegraphics[width=9cm, height=8cm]
{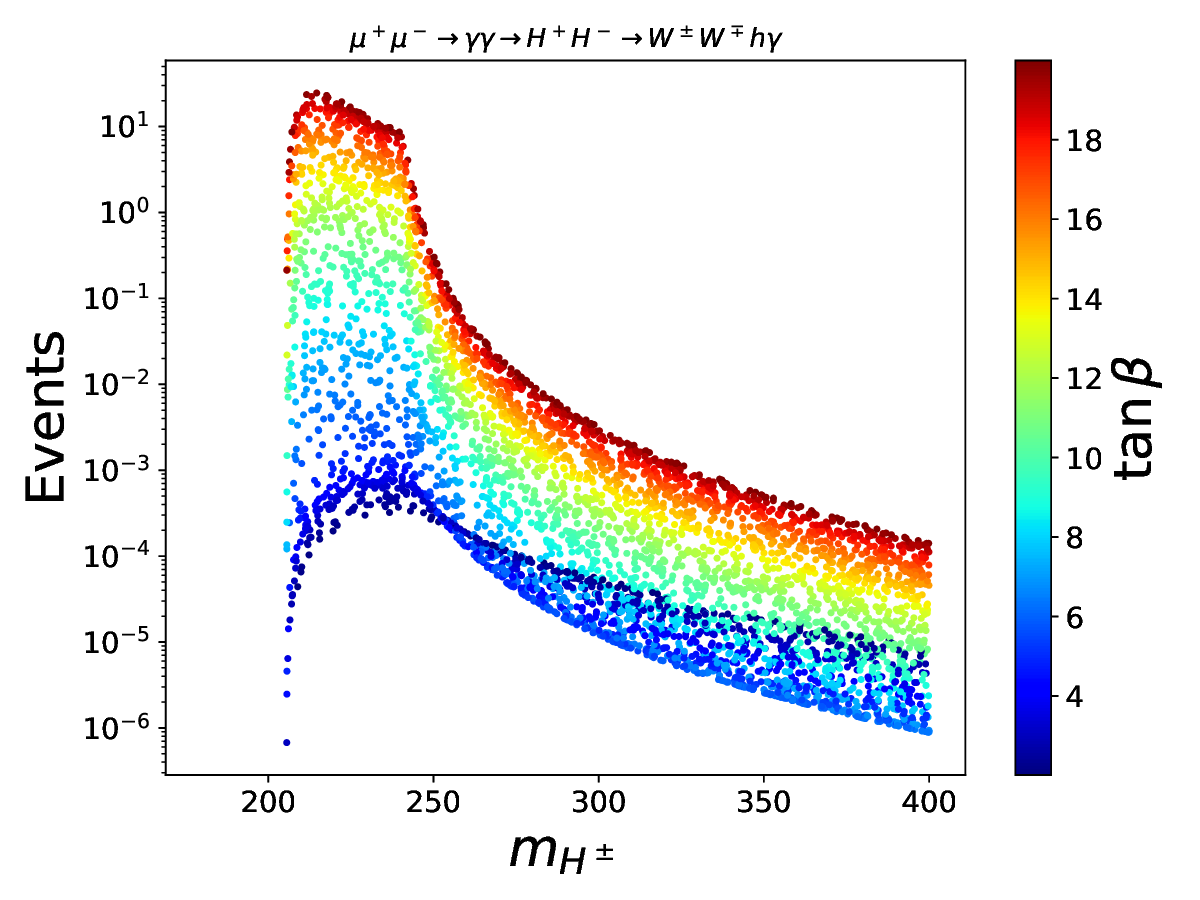}
 &
\includegraphics[width=7.5cm, height=8cm]
{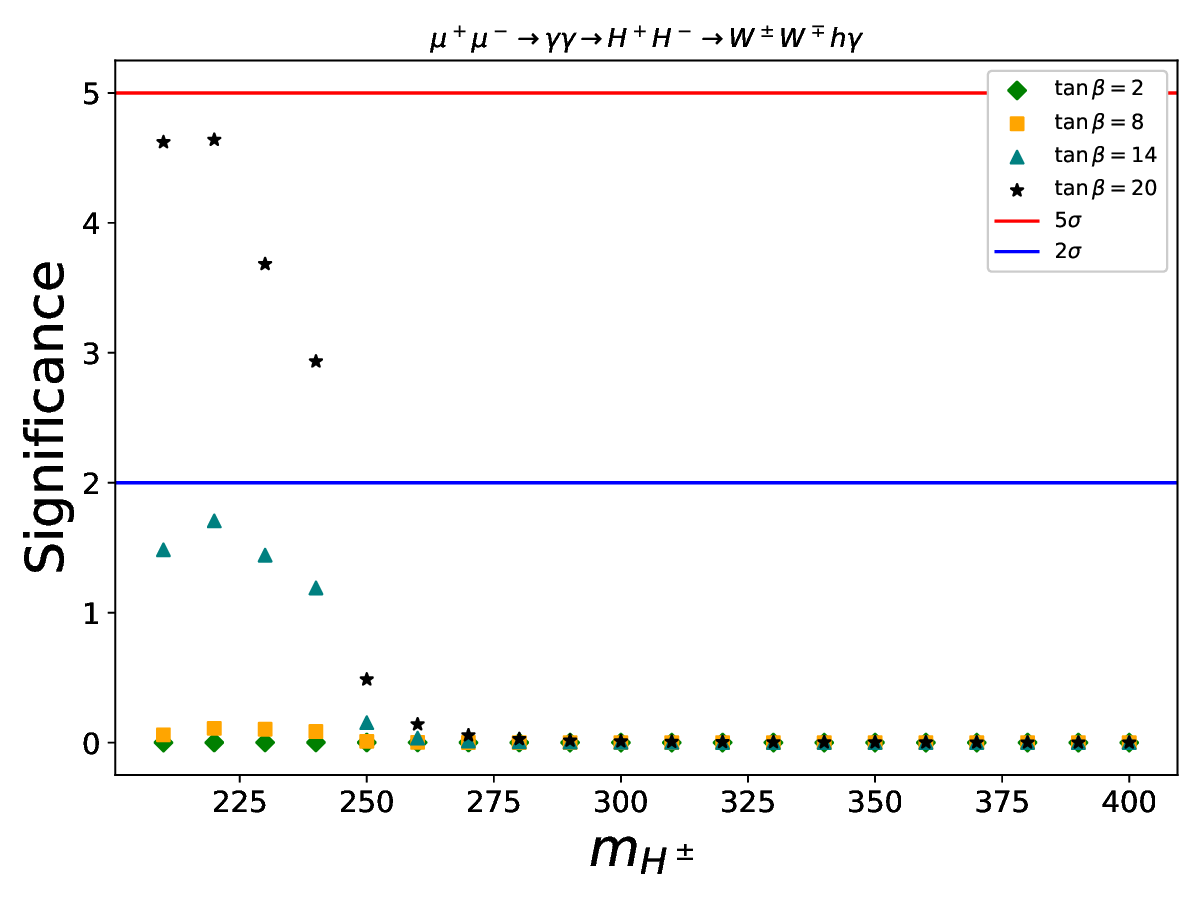}
 \end{tabular}
\caption{\label{significantZ2}
Events for the signal process are displayed 
in the $(t_{\beta}, M_{H^{\pm}})$ parameter 
space at $\sqrt{s} = 3$~TeV with the integrated 
luminosity of $\mathcal{L} = 3000~\text{fb}^{-1}$ 
(left panel). The statistical significance 
of the signal is shown as a function of the 
charged Higgs mass at $\sqrt{s} = 3$~TeV with 
the same integrated luminosity (right panel).
}
\end{figure}

For both the signal and the SM background, 
an energy cut on the final-state photon, 
$E_{\gamma} \geq 1~\text{MeV}$, is applied 
in this computation. Furthermore, to reduce
the SM background, we impose invariant mass cuts
$|M_{W^+h}-M_{H^{\pm}}| \leq 10~\text{GeV}$ and 
$|M_{W^-\gamma}-M_{H^{\pm}}| \leq 10~\text{GeV}$.
The SM background is simulated using 
{\tt GRACE}~\cite{Belanger:2003sd} to 
generate tree-level cross sections.
In Fig.~\ref{significantZ2}, the event rates
for the signal process are displayed in 
the parameter space of $(t_{\beta},
M_{H^{\pm}})$ at $\sqrt{s} = 3$~TeV 
with an integrated luminosity of 
$\mathcal{L} = 3000~\text{fb}^{-1}$ (left panel). 
We observe that the $WWh\gamma$ events 
are significant in the region 
$210~\text{GeV} \leq M_{H^{\pm}} 
\leq 240~\text{GeV}$ and $t_{\beta}
\geq 10$, while in other regions 
the event rates are very small. 
The corresponding statistical 
significance of the signal is 
shown as a function of the charged 
Higgs mass at $\sqrt{s} = 3$~TeV 
with $\mathcal{L} = 3000~\text{fb}^{-1}$ (right panel). 
Consistently, the statistical 
significance exceeds $2\sigma$ 
within $210~\text{GeV} \leq M_{H^{\pm}} 
\leq 240~\text{GeV}$ and $t_{\beta} \geq 10$, 
whereas in other regions 
it remains insignificant.
\section{Conclusions} 
In this work, we study the one--loop--induced decay channel $H^{\pm} \to W^{\pm}\gamma$ in the general $\mathcal{R}_\xi$ gauge. We explicitly verify the gauge invariance ($\xi$-independence and Ward identity checks), ultraviolet finiteness, and renormalization-scale independence of the one-loop form factors, thereby confirming the internal consistency of our calculations. For the phenomenological analysis, we perform a parameter scan of the Type-I THDM and, within the resulting viable parameter space, evaluate the branching fractions of this decay mode. 
Furthermore, we investigate charged Higgs pair 
production at muon--TeV colliders through the 
representative processes 
$\mu^+\mu^- \to H^{+}H^{-} \to W^+W^- h\gamma$ 
and 
$\mu^+\mu^- \to \gamma\gamma \to H^{+}H^{-} \to W^+W^- h\gamma$, as typical applications of our results. The total cross sections are computed within the viable parameter space of the Type-I THDM. The corresponding signal significances are simulated, including relevant Standard Model backgrounds, at a center-of-mass energy of 
$\sqrt{s} = 3~\text{TeV}$. With the high integrated luminosity expected at muon--TeV colliders, up to $\mathcal{L} = 3000~\text{fb}^{-1}$, we find that the signals can be observed with a $2\sigma$ significance for several benchmark points within the allowed parameter space of the THDM.
\\

\noindent
{\bf Acknowledgment:}~
This research is funded by Vietnam
National Foundation for Science and
Technology Development (NAFOSTED) under
the grant number $103.01$-$2023.16$.
\section*{Appendix A:  One-loop form factors 
for $H^{\pm}\rightarrow W^{\pm}\gamma$       
in the general $R_{\xi}$-gauge}              
One-loop form factors for $H^{\pm}\rightarrow
W^{\pm}\gamma$ in the general $R_{\xi}$-gauge are
presented in detail in this Appendix.
\subsection*{\underline{
Form factors
$F^{W^\pm H^\pm}_{i, \Delta}:$}
}
\begin{figure}[H]
\centering
\includegraphics[width=10cm, height=3.3cm]
{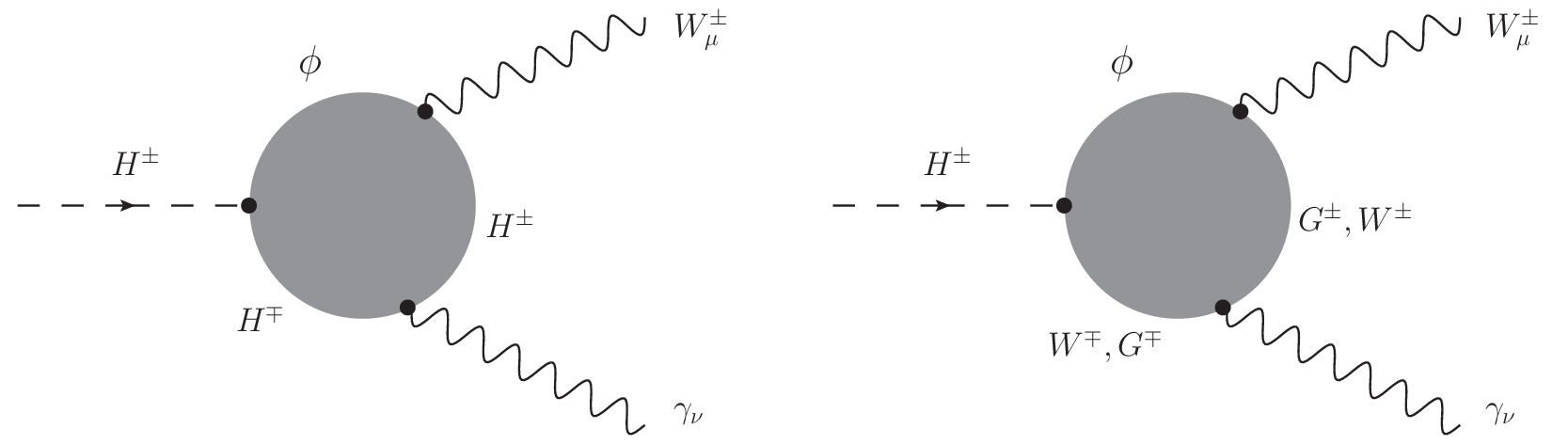}
\\
\includegraphics[width=10cm, height=3.3cm]
{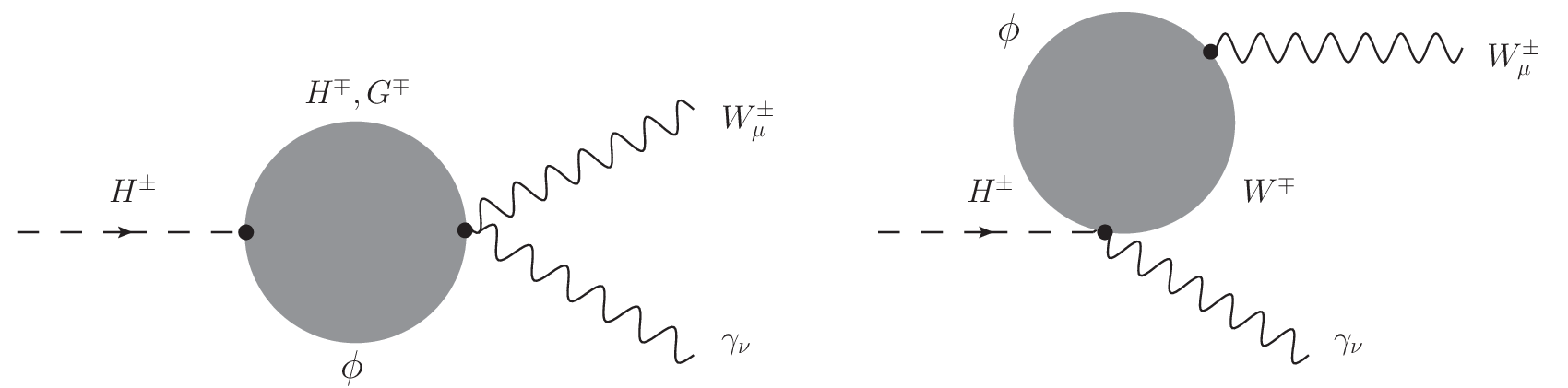}
\caption{
One-loop triangle diagrams contributing
to the decay process, in which charged
Higgs bosons, $W$ bosons and Goldstone
bosons are propagated in the loop.
}
\label{Fig:TrigsBoson}
\end{figure}

We first obtain the one-loop
bosonic form factors, namely
$F^{W^\pm H^\pm}_{i,\Delta}$ for $i = 1,2,3$,
arising from the one-loop diagrams shown in
Fig.~\ref{Fig:TrigsBoson}.
Two relevant contributions correspond
to the scalar Higgs bosons
$\phi \equiv h, H$,
together with $H^\pm$ and $W^\pm$
exchanged in the loop.
In particular, the form
factors can be written in
terms of their individual
contributions as
\begin{eqnarray}
F^{W^\pm H^\pm}_{i, \Delta}
&=&
\sum
\limits_{\phi = h, H}
\Big(
F^{\phi-H^\pm}_{i, \Delta}
+
F^{\phi-W^\pm}_{i, \Delta}
\Big).
\end{eqnarray}
Within the general $\mathcal{R}_{\xi}$ gauge,
the Goldstone bosons $G^\pm$ take part in
the one-loop diagrams. Their couplings can
be expressed in terms of the physical ones
using the relations listed below:
\begin{eqnarray}
g_{\gamma G^\pm G^\mp}
&=&
-
g_{\gamma W^\pm W^\mp},
\quad
g_{\gamma W^\pm G^\mp}
=
-
M_W \,
g_{\gamma W^\pm W^\mp},
\\
g_{\phi W^+ G^-}
&=&
-
\frac{1}{2 M_W}
g_{\phi W^\pm W^\mp}
,
\quad
g_{\phi W^- G^+}
=
\frac{1}{2 M_W}
g_{\phi W^\pm W^\mp}
,
\\
g_{\phi G^\mp W^\pm \gamma}
&=&
\frac{1}{2 M_W}
g_{\phi W^\pm W^\mp}
\,
g_{\gamma W^\pm W^\mp}
,
\quad
g_{\phi H^- G^+}
=
\frac{
M_{\phi}^2
-
M_{H^\pm}^2
}{M_W}
g_{\phi H^- W^+}
.
\end{eqnarray}
With the above relations,
we obtain the following results:
\begin{eqnarray}
F^{\phi-H^\pm}_{1, \Delta}
&=&
-
\dfrac{1}{(4\pi)^2}
\dfrac{g_{\phi H^\pm H^\mp}
}{
\big(M_{H^\pm}^2-M_{W}^2\big)}
\Big\{
g_{\gamma H^\pm H^\mp}
\,
g_{\phi H^- W^+}
\Big[
(M_{\phi}^2-M_{H^\pm}^2-M_{W}^2)
B_0(M_{W}^2,M_{\phi}^2,M_{H^\pm}^2)
\nonumber \\
&&
-
\Big[
g_{\gamma H^\pm H^\mp}
\,
g_{\phi H^- W^+}
(M_{\phi}^2-2 M_{H^\pm}^2)
+
g_{\phi H^- W^+ \gamma}
(M_{H^\pm}^2-M_{W}^2)
\Big]
B_0(M_{H^\pm}^2,M_{\phi}^2,M_{H^\pm}^2)
\nonumber \\
&&
+
(M_{H^\pm}^2
-M_{W}^2)
\Big(
1
+
2 M_{H^\pm}^2
C_0(M_{W}^2,0,M_{H^\pm}^2,
M_{\phi}^2,
M_{H^\pm}^2,M_{H^\pm}^2)
\Big)
\Big]
\Big\},
\\
&&
\nonumber
\\
F^{\phi-H^\pm}_{2, \Delta}
&=&
\dfrac{
2
}{(4\pi)^2}
\dfrac{
g_{\gamma H^\pm H^\mp}
\,
g_{\phi H^- W^+}
\,
g_{\phi H^\pm H^\mp}
}{
M_{H^\pm}^2
\big(M_{H^\pm}^2-M_{W}^2\big)^2}
\Big\{
(M_{H^\pm}^2-M_{W}^2)
\Big[
A_0(M_{\phi}^2)
-
A_0(M_{H^\pm}^2)
\Big]
\nonumber\\
&&
+
M_{H^\pm}^2
(M_{\phi}^2-M_{H^\pm}^2-M_{W}^2)
B_0(M_{W}^2,M_{\phi}^2,M_{H^\pm}^2)
\nonumber\\
&&
-
\Big[
2 M_{H^\pm}^2 (M_{\phi}^2-M_{H^\pm}^2)
-M_{\phi}^2 M_{W}^2
\Big]
B_0(M_{H^\pm}^2,M_{\phi}^2,M_{H^\pm}^2)
\nonumber \\
&&
+
M_{H^\pm}^2
(M_{H^\pm}^2-M_{W}^2)
\Big[
1
+
2 M_{H^\pm}^2
C_0(M_{W}^2,0,M_{H^\pm}^2,
M_{\phi}^2,M_{H^\pm}^2,M_{H^\pm}^2)
\Big]
\Big\}.
\end{eqnarray}
The other form factors with $\phi$
and $W^\pm, G^\pm$ propagating in
the loop are expressed as follows:
\begin{eqnarray}
\label{Eq:B1mTrigsPhiWpm}
F^{\phi-W^\pm}_{1, \Delta}
&=&
-
\dfrac{1}{(4\pi)^2}
\dfrac{g_{\phi W^\pm W^\mp}
}{
12 M_{W}^4
\big(M_{H^\pm}^2-M_{W}^2\big)^2}
\times
\\
&&
\times
\Bigg\{
(2 M_{W}^2)
(M_{H^\pm}^2-M_{W}^2)^2
\Big[
\big(
3 M_{\phi}^2
-
3 M_{H^\pm}^2
-
M_{W}^2
\big)
g_{\gamma W^\pm W^\mp}
\,
g_{\phi H^- W^+}
\nonumber \\
&&
\hspace{2cm}
-
M_{W}^2
\,
g_{\phi H^- W^+ \gamma}
+
(\xi M_{W}^2)
\Big(
g_{\phi H^- W^+ \gamma}
-
2
g_{\gamma W^\pm W^\mp}
\,
g_{\phi H^- W^+}
\Big)
\Big]
\nonumber \\
&&
+
\Big\{
g_{\gamma W^\pm W^\mp}
\,
g_{\phi H^- W^+}
\Big[
M_{\phi}^2
(4 M_{H^\pm}^2 M_{W}^2
+M_{W}^4-M_{H^\pm}^4)
\nonumber \\
&&
+
M_{W}^2
(M_{H^\pm}^2-M_{W}^2)
\Big(
M_{H^\pm}^2 (11 + \xi)
-
(7 + \xi)
M_{W}^2
\Big)
\Big]
\nonumber \\
&&
\hspace{2.2cm}
+
g_{\phi H^- W^+ \gamma}
\,
(M_{H^\pm}^2-M_{W}^2)^2
\Big(
M_{W}^2 (1 + \xi)
-M_{\phi}^2
\Big)
\Big\}
A_0(\xi M_{W}^2)
\nonumber \\
&&
+
\Big\{
g_{\gamma W^\pm W^\mp}
\,
g_{\phi H^- W^+}
\Big[
M_{W}^4
(20 M_{H^\pm}^2-M_{\phi}^2-8 M_{W}^2)
\nonumber \\
&&
-
4 M_{H^\pm}^2 M_{W}^2
(M_{\phi}^2+3 M_{H^\pm}^2)
+
M_{\phi}^2 M_{H^\pm}^4
\Big]
\nonumber \\
&&
\hspace{3.4cm}
+
g_{\phi H^- W^+ \gamma}
\,
(M_{\phi}^2-2 M_{W}^2)
(M_{H^\pm}^2-M_{W}^2)^2
\Big\}
A_0(M_{W}^2)
\nonumber \\
&&
+
\Big[
(1 - \xi)
M_{W}^2
(M_{H^\pm}^2-M_{W}^2)^2
\big(g_{\gamma W^\pm W^\mp}
\,
g_{\phi H^- W^+}
+
g_{\phi H^- W^+ \gamma}\big)
\Big]
A_0(M_{\phi}^2)
\nonumber \\
&&
+
\Big[
(M_{H^\pm}^2-M_{W}^2)^2
\big(
g_{\gamma W^\pm W^\mp}
\,
g_{\phi H^- W^+}
+
g_{\phi H^- W^+ \gamma}
\big)
\times
\nonumber \\
&&\hspace{1.2cm} \times
\Big(
2 M_{\phi}^2 M_{W}^2 (1 + \xi)
-
M_{W}^4 (1 - \xi)^2
-
M_{\phi}^4
\Big)
\Big]
B_0(M_{W}^2,M_{\phi}^2, \xi M_{W}^2)
\nonumber \\
&&
+
\Big\{
g_{\gamma W^\pm W^\mp}
\,
g_{\phi H^- W^+}
\times
\nonumber\\
&&
\hspace{1.5cm}
\times
\Big[
M_{\phi}^4 (M_{H^\pm}^2+5 M_{W}^2)
-
2 M_{\phi}^2 M_{W}^2
(5 M_{H^\pm}^2+7 M_{W}^2)
+
24 M_{W}^6
\Big]
\nonumber\\
&&
+
g_{\phi H^- W^+ \gamma}
(M_{H^\pm}^2-M_{W}^2)
\times
\nonumber \\
&&\hspace{1.5cm} \times
\big(
M_{\phi}^4
-4 M_{\phi}^2 M_{W}^2+12 M_{W}^4
\big)
\Big]
(M_{H^\pm}^2-M_{W}^2)
B_0(M_{W}^2,M_{\phi}^2,M_{W}^2)
\nonumber \\
&&
+
g_{\gamma W^\pm W^\mp}
\,
g_{\phi H^- W^+}
(2 M_{W}^4)
\Big[
(1 - \xi)
M_{\phi}^2
(M_{H^\pm}^2+M_{W}^2)
\nonumber \\
&&
\hspace{1.4cm}
+
(M_{H^\pm}^2-M_{W}^2)
\Big(
M_{H^\pm}^2 (5 - \xi)
-
M_{W}^2 (3 + \xi)
\Big)
\Big\}
B_0(0,M_{W}^2, \xi M_{W}^2)
\nonumber \\
&&
+
\Big[
g_{\gamma W^\pm W^\mp}
\,
g_{\phi H^- W^+}
(6 M_{W}^2)
(M_{\phi}^2-M_{H^\pm}^2)
(M_{H^\pm}^2-M_{W}^2)^2
\Big]
\times
\nonumber\\
&&
\hspace{8.5cm}
\times
B_0(M_{H^\pm}^2,
M_{\phi}^2, \xi M_{W}^2)
\nonumber \\
&&
+
\Big[
g_{\gamma W^\pm W^\mp}
\,
g_{\phi H^- W^+}
(6 M_{W}^2)
(M_{H^\pm}^2-M_{W}^2)
\times
\nonumber \\
&&\hspace{1.5cm} \times
(M_{H^\pm}^2-M_{\phi}^2+M_{W}^2)
(M_{\phi}^2+M_{H^\pm}^2-3 M_{W}^2)
\Big]
B_0(M_{H^\pm}^2,M_{\phi}^2,M_{W}^2)
\nonumber \\
&&
-
\Big[
g_{\gamma W^\pm W^\mp}
\,
g_{\phi H^- W^+}
(4 M_{W}^4 \xi)
(M_{H^\pm}^2-M_{W}^2)^2
\Big]
B_0(0, \xi M_{W}^2, \xi M_{W}^2)
\nonumber \\
&&
+
\Big[
g_{\gamma W^\pm W^\mp}
\,
g_{\phi H^- W^+}
(8 M_{W}^4)
(M_{H^\pm}^2-M_{W}^2)^2
\Big]
B_0(0,M_{W}^2,M_{W}^2)
\nonumber \\
&&
+
\Big[
g_{\gamma W^\pm W^\mp}
\,
g_{\phi H^- W^+}
(12 M_{W}^4)
(M_{H^\pm}^2-M_{W}^2)^2
\times
\nonumber \\
&&\hspace{2.5cm} \times
(M_{\phi}^2+M_{H^\pm}^2-3 M_{W}^2)
\Big]
C_0(M_{W}^2,0,M_{H^\pm}^2,
M_{\phi}^2,M_{W}^2,M_{W}^2)
\Bigg\},
\nonumber \\
&&
\nonumber\\
\label{Eq:B2mTrigsPhiWpm}
F^{\phi-W^\pm}_{2, \Delta}
&=&
\dfrac{1}{(4\pi)^2}
\dfrac{
g_{\gamma W^\pm W^\mp}
\,
g_{\phi H^- W^+}
\,
g_{\phi W^\pm W^\mp}
}{
3 M_{H^\pm}^2 M_{W}^2
\big(M_{H^\pm}^2-M_{W}^2\big)^4}
\times
\\
&&
\times
\Bigg\{
M_{H^\pm}^2
(M_{H^\pm}^2-M_{W}^2)^2
\Big[
3 M_{\phi}^2 (M_{H^\pm}^2-M_{W}^2)
-3 M_{H^\pm}^4+4 M_{H^\pm}^2 M_{W}^2+3 M_{W}^4
\Big]
\nonumber \\
&&
+
M_{H^\pm}^2
(M_{H^\pm}^2+M_{W}^2)
\Big[
4 M_{\phi}^2 (M_{H^\pm}^2-M_{W}^2)+6 M_{H^\pm}^4
\nonumber \\
&&
\hspace{5cm}
-
3 M_{H^\pm}^2 M_{W}^2
(\xi+5) - 3 M_{W}^4 (\xi-5)
\Big]
A_0(\xi M_{W}^2)
\nonumber \\
&&
+
\Big[
-
M_{\phi}^2
(M_{H^\pm}^2-M_{W}^2)
\Big(
7 M_{H^\pm}^4-2 M_{H^\pm}^2 M_{W}^2+3 M_{W}^4
\Big)
-
7 M_{H^\pm}^8
-
3 M_{W}^8
\nonumber \\
&&
+
M_{H^\pm}^2
M_{W}^2
\Big(
11 M_{H^\pm}^4
-
4 M_{H^\pm}^2 M_{W}^2
-
9 M_{W}^4
+
3 \xi
(M_{H^\pm}^2+M_{W}^2)^2
\Big)
\Big]
A_0(M_{W}^2)
\nonumber \\
&&
+
\Big[
3 (M_{H^\pm}^2-M_{W}^2)^3
(M_{\phi}^2-M_{H^\pm}^2-M_{W}^2)
\Big]
A_0(M_{\phi}^2)
\nonumber \\
&&
+
3 (M_{H^\pm}^2
-M_{W}^2)^2
\Big[
M_{W}^2
\Big(
M_{\phi}^4
+
4 M_{\phi}^2 M_{H^\pm}^2
+
M_{H^\pm}^4
+
M_{W}^4
\Big)
\nonumber \\
&&
\hspace{1cm}
-
2 M_{W}^4
(M_{\phi}^2+3 M_{H^\pm}^2)
+
2 M_{\phi}^2 M_{H^\pm}^2
(M_{H^\pm}^2-M_{\phi}^2)
\Big]
B_0(M_{H^\pm}^2,M_{\phi}^2,M_{W}^2)
\nonumber \\
&&
+
3 M_{H^\pm}^2
(M_{H^\pm}^2-M_{W}^2)^2
\Big[
M_{\phi}^2
(M_{\phi}^2-M_{H^\pm}^2-3 M_{W}^2)
-
2 M_{W}^2
\times
\nonumber \\
&&\hspace{6.5cm} \times
(M_{H^\pm}^2-3 M_{W}^2)
\Big]
B_0(M_{W}^2,M_{\phi}^2,M_{W}^2)
\nonumber \\
&&
+
M_{H^\pm}^2 M_{W}^2 (\xi-1)
(M_{H^\pm}^2+M_{W}^2)
\Big[
4 M_{\phi}^2
(M_{W}^2-M_{H^\pm}^2)
\nonumber \\
&&
-
6 M_{H^\pm}^4
+
3 M_{W}^2
\Big(
M_{H^\pm}^2
(\xi+5)
+
(\xi-5)
M_{W}^2
\Big)
\Big]
B_0(0,M_{W}^2, \xi M_{W}^2)
\nonumber \\
&&
+
\Big[
4 M_{H^\pm}^4 M_{W}^2
(M_{H^\pm}^2-M_{W}^2)^2
\Big]
B_0(0,M_{W}^2,M_{W}^2)
\nonumber \\
&&
+
\Big[
6 M_{H^\pm}^2 M_{W}^2
(M_{H^\pm}^2-M_{W}^2)^3
\times
\nonumber \\
&&\hspace{2.5cm} \times
(M_{\phi}^2+M_{H^\pm}^2-3 M_{W}^2)
\Big]
C_0(M_{W}^2,0,M_{H^\pm}^2,
M_{\phi}^2,M_{W}^2,M_{W}^2)
\Bigg\}.
\nonumber
\end{eqnarray}
The remaining one-loop form factors
for $i = 3$, which are factored out
of the Levi-Civita symbol
$\epsilon^{\mu\nu\rho\sigma}$,
give zero contribution in the
bosonic loop:
\begin{eqnarray}
F^{\phi-H^\pm}_{3, \Delta}
=
F^{\phi-W^\pm}_{3, \Delta}
=
0.
\end{eqnarray}
\subsection*{
\underline{
Form factors
$F^{W^\pm H^\pm}_{i, \Pi}$:}
}
\begin{figure}[H]
\centering
\includegraphics[width=14cm, height=4cm]
{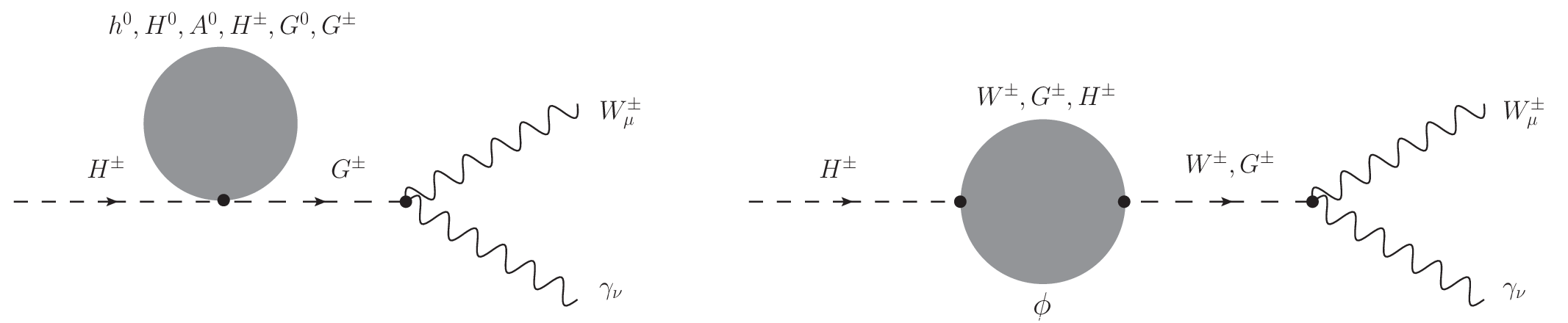}
\caption{
Self-energy contributions on the external
$H^\pm$ leg, which are also included in
the calculation of the decay process.
}
\label{Fig:SelfBoson}
\end{figure}

The one-loop form factors
$F^{W^\pm H^\pm}_{i,\Pi}$ for $i=1,2,3$,
originating from the Feynman
topologies shown in
Fig.~\ref{Fig:SelfBoson},
can be decomposed as follows:
\begin{eqnarray}
F^{W^\pm H^\pm}_{i, \Pi}
&=&
F_{i, \Pi - 1P}
+
\sum
\limits_{\phi = h, H}
\Big(
F^{\phi-H^\pm}_{i, \Pi - 2P}
+
F^{\phi-W^\pm}_{i, \Pi - 2P}
\Big).
\end{eqnarray}
Each contribution is expressed
in terms of scalar one-loop integrals.
The first term, obtained from
the left topology in
Fig.~\ref{Fig:SelfBoson},
takes the form
\begin{eqnarray}
\label{Eq:B1mSE1P}
F_{1, \Pi - 1P}
&=&
\frac{-
g_{\gamma W^\pm W^\mp}
}{(4\pi)^2}
\frac{
M_{W}
}{
2 \big(M_{H^\pm}^2
- \xi M_{W}^2\big)
}
\Big\{
g_{G^0 G^0 H^- G^+}
\,
A_0(\xi M_{Z}^2)
+
2 g_{G^+ G^- H^- G^+}
\,
A_0(\xi M_{W}^2)
\nonumber
\\
&&
+
g_{A A H^- G^+}
\,
A_0(M_{A}^2)
+
2 g_{H^+ H^- H^- G^+}
\,
A_0(M_{H^\pm}^2)
+
\sum
\limits_{\phi = h, H}
g_{\phi \phi \, H^- G^+}
\,
A_0(M_{\phi}^2)
\Big\}.
\end{eqnarray}
The second contribution is
calculated
from the right topology of
Fig.~\ref{Fig:SelfBoson}.
The factors
are obtained as follows:
\begin{eqnarray}
F^{\phi-H^\pm}_{1, \Pi - 2P}
&=&
\frac{1}{(4\pi)^2}
\frac{
g_{\gamma W^\pm W^\mp}
\,
g_{\phi H^- W^+}
\,
g_{\phi H^\pm H^\mp}
}{
M_{H^\pm}^2
\big(M_{H^\pm}^2
- \xi M_{W}^2 \big)}
\Big\{
(\xi M_{W}^2)
\Big[
A_0(M_{H^\pm}^2)
-
A_0(M_{\phi}^2)
\Big]
\nonumber \\
&&
+
\Big[
(M_{H^\pm}^2 - M_{\phi}^2)
(M_{H^\pm}^2 - \xi M_{W}^2)
\Big]
B_0(M_{H^\pm}^2,
M_{\phi}^2,M_{H^\pm}^2)
\Big\},
\\
F^{\phi-W^\pm}_{1, \Pi - 2P}
&=&
-
\frac{1}{(4\pi)^2}
\frac{
g_{\gamma W^\pm W^\mp}
\,
g_{\phi H^- W^+}
\,
g_{\phi W^\pm W^\mp}
}{
2 M_{H^\pm}^2 M_{W}^2
\big(M_{H^\pm}^2
-\xi M_{W}^2\big)}
\times
\\
&&
\times
\Big\{
M_{W}^2
\Big[
\xi
\big(
M_{H^\pm}^2
-M_{\phi}^2
+M_{W}^2
\big)
-
M_{H^\pm}^2
\Big]
A_0(M_{\phi}^2)
\nonumber \\
&&
+
(M_{H^\pm}^2
-M_{\phi}^2
+M_{W}^2)
(M_{H^\pm}^2 - \xi M_{W}^2)
A_0(M_{W}^2)
\nonumber \\
&&
+
M_{H^\pm}^2
(M_{\phi}^2
-M_{H^\pm}^2
+ \xi M_{W}^2)
A_0(\xi M_{W}^2)
\nonumber \\
&&
-
\big(
M_{H^\pm}^2
- \xi M_{W}^2
\big)
\Big[
\big(
M_{\phi}^2
-
M_{H^\pm}^2
-
M_{W}^2
\big)^2
-
4
M_{H^\pm}^2
M_{W}^2
\Big]
\times
\nonumber
\\
&&
\hspace{6cm}
\times
B_0(M_{H^\pm}^2,M_{\phi}^2,M_{W}^2)
\nonumber \\
&&
-
\big(
M_{H^\pm}^2
- \xi M_{W}^2
\big)
M_{H^\pm}^2 (M_{\phi}^2-M_{H^\pm}^2)
B_0(M_{H^\pm}^2,M_{\phi}^2, \xi M_{W}^2)
\Big\}.
\nonumber
\end{eqnarray}
Remaining factors
for $i = 2, 3$ have
no contribution,
\begin{eqnarray}
F_{i, \Pi - 1P}
=
F^{\phi-H^\pm}_{i, \Pi - 2P}
=
F^{\phi-W^\pm}_{i, \Pi - 2P}
=
0
\end{eqnarray}
in which the general couplings
involved in these Goldstone bosons
$G^\pm$ are rewritten in terms of
physical particles such as
$W^\pm$-bosons as follows:
\begin{eqnarray}
g_{\phi G^+ G^-}
&=&
-
\frac{M_{\phi}^2}{2 M_W^2}
\,
g_{\phi W^\pm W^\mp},
\quad
g_{\phi G^- G^+}
=
-
\frac{M_{\phi}^2}{2 M_W^2}
\,
g_{\phi W^\pm W^\mp}.
\end{eqnarray}
\subsection*{
\underline{
Form factors
$F^{W^\pm H^\pm}_{i, T}$:}
}
\begin{center}
\begin{figure}[H]
\centering
\includegraphics[width=16cm, height=6cm]
{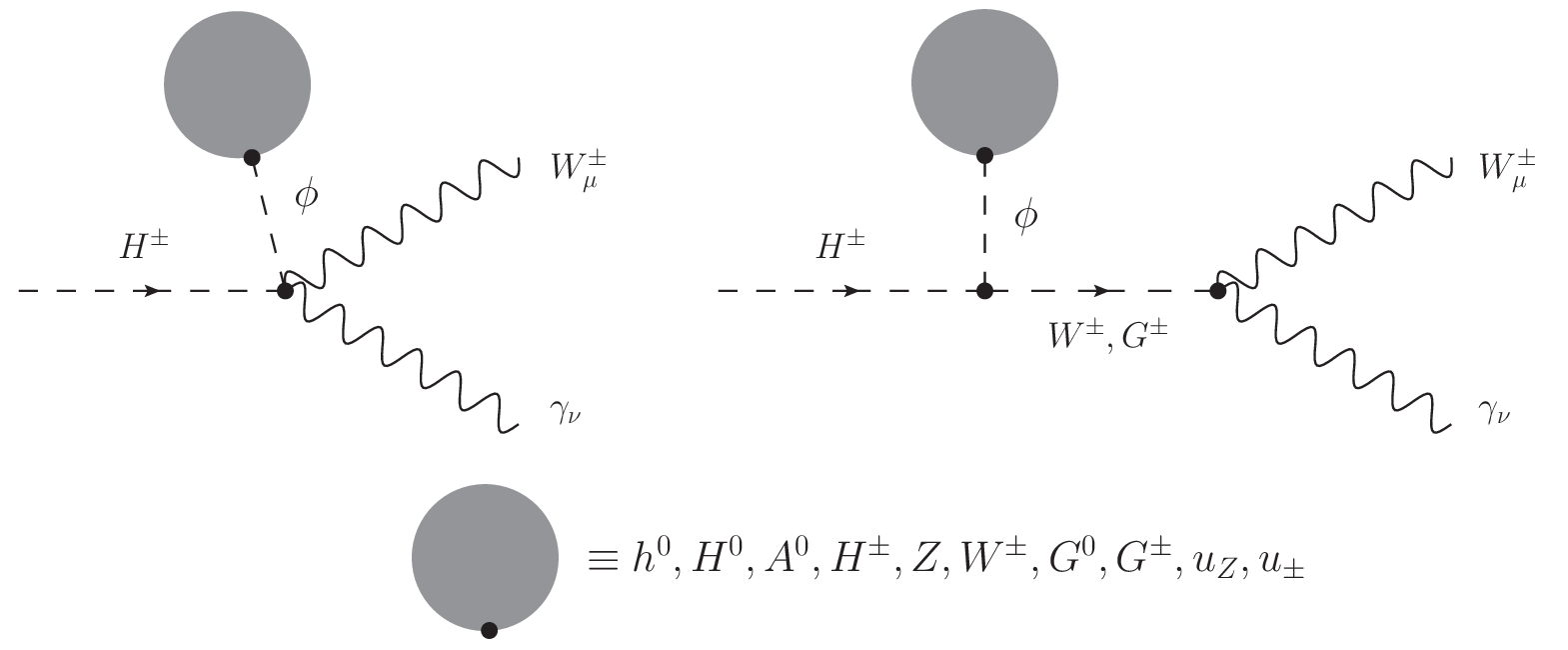}
\caption{
Tadpole diagrams associated with
bosonic loops involving the external
scalar pole $\phi \equiv h, H$
are also included in the computation
of the process.
}
\label{Fig:TadBoson}
\end{figure}
\end{center}

The form factors $F^{W^\pm H^\pm}_{i,T}$
($i = 1, 2, 3$; see Fig.~\ref{Fig:TadBoson})
correspond to scalar Higgs poles
$\phi \equiv h, H$ attached to the
bubble diagrams, with loop contributions
from $h, H, A, H^\pm, Z, W^\pm, G^0, G^\pm$,
as well as the corresponding ghost particles
$u_Z$ and $u_\pm$.
\begin{eqnarray}
\label{Eq:B1mTPhi}
F^{W^\pm H^\pm}_{i, T}
&=&\frac{1}{(4\pi)^2}
\sum
\limits_{\phi = h, H}
\Big[
\frac{
g_{\gamma W^\pm W^\mp}
\,
g_{\phi H^- W^+}
+
g_{\phi H^- W^+ \gamma}
}{
4 M_{W}^2
M_{\phi}^2}
-
\frac{
g_{\gamma W^\pm W^\mp}
\,
g_{\phi H^- W^+}
}{
4 M_{W}^2
\big(M_{H^\pm}^2
- \xi M_{W}^2 \big)}
\Big]
\times
\nonumber\\
&& \times
\Big\{
g_{\phi ZZ}
\,
(M_{\phi}^2 c_W^2)
A_0(\xi M_{Z}^2)
+
g_{\phi W^\pm W^\mp}
\,
(2 M_{\phi}^2)
A_0(\xi M_{W}^2)
\nonumber \\
&&
-
2 M_{W}^2
\Big(
g_{\phi A A}
\,
A_0(M_{A}^2)
+
g_{\phi h h}
\,
A_0(M_{h}^2)
\nonumber \\
&&
+
g_{\phi H H}
\,
A_0(M_{H}^2)
+
2 g_{\phi H^\pm H^\mp}
\,
A_0(M_{H^\pm}^2)
\nonumber
\\
&&
+
2 g_{\phi W^\pm W^\mp}
\,
[
-
3
A_0(M_{W}^2)
+
2 M_{W}^2
]
+
g_{\phi ZZ}
\,
[
-
3
A_0(M_{Z}^2)
+
2 M_{Z}^2
]
\Big)
\Big\},
\\
F^{\phi-W^\pm H^\pm}_{2, T}
&=&
F^{\phi-W^\pm H^\pm}_{3, T}
=
0.
\end{eqnarray}
Where the general couplings
involving the Goldstone
bosons $G^\pm, G^0$ and the
ghost fields $u_\pm, u_Z$
are expressed in terms of
the physical particles,
namely the $W^\pm$ and $Z$
bosons, as follows:
\begin{eqnarray}
g_{\phi G^0 G^0}
&=&
-
\frac{c_W^2}{2 M_{W}^2}
M_{\phi}^2
\,
g_{\phi ZZ},
\quad
g_{\phi u_Z \bar{u}_Z}
=
-
\frac{
g_{\phi ZZ}
}{2},
\quad
g_{\phi u_{\pm} \bar{u}_\pm}
=
-
\frac{
g_{\phi W^\pm W^\mp}
}{2}.
\end{eqnarray}
\subsection*{
\underline{
Form factors
$F^{f}_{i, \Delta}$:}
}
\begin{center}
\begin{figure}[H]
\centering
\includegraphics[width=12cm, height=3cm]
{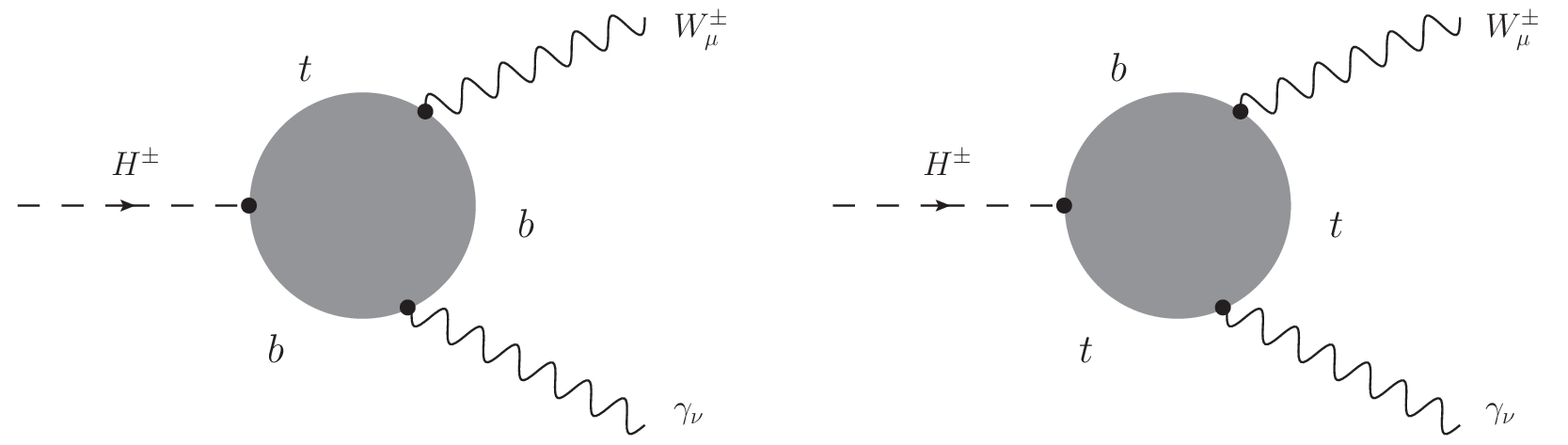}
\caption{
One-loop triangle diagrams with
internal fermion loops are also
taken into account in the process
under the evaluation.}
\label{Fig:TrigsFermion}
\end{figure}
\end{center}

The form factors $F^{f}_{i,\Delta}$
for $i = 1, 2, 3$, with fermion
contributions from the mixing of
both top ($t$) and bottom ($b$)
quarks in Fig.~\ref{Fig:TrigsFermion},
are expressed as follows:
\begin{eqnarray}
F^{f}_{1, \Delta}
&=&
\frac{
g_{W^+ \bar{t} \, b}
}{(4\pi)^2}
N^C_Q
\Big\{
g^{R}_{H^- t \bar{b}}
(2 \; eQ_b \cdot
m_{t}^2)
\,
B_{0}(0,m_{b}^2,m_{b}^2)
+
g^{L}_{H^- t \bar{b}}
(2\; eQ_t
\cdot m_{b}^2)
\,
B_{0}(0,m_{t}^2,m_{t}^2)
\nonumber \\
&&
+
g^{R}_{H^- t \bar{b}}
\,
(
eQ_b
\cdot
m_{t}^2)
\Big[
2 \big(
m_{t}^2
-m_{b}^2
-M_{W}^2
\big)
C_{0}
-
4 C_{00}
\\
&&
\hspace{1.7cm}
+
2 \big(M_{H^\pm}^2-M_{W}^2\big)
C_{1}
-
\big(M_{H^\pm}^2+3 M_{W}^2\big)
C_{2}
\Big]
(0,M_{H^\pm}^2,M_{W}^2,
m_{b}^2,m_{b}^2,m_{t}^2)
\nonumber\\
&&
+
g^{L}_{H^- t \bar{b}}
\,
(
eQ_b
\cdot
m_{b}^2)
\Big[
\big(M_{H^\pm}^2-M_{W}^2\big)
\big(
C_{0}
+
C_{2}
\big)
-
4 C_{00}
\Big]
(0,M_{H^\pm}^2,M_{W}^2,
m_{b}^2,m_{b}^2,m_{t}^2)
\nonumber \\
&&
+
g^{L}_{H^- t \bar{b}}
\,
(
eQ_t
\cdot
m_{b}^2)
\Big[
\big(
2 m_{b}^2
+M_{H^\pm}^2
-2 m_{t}^2
+M_{W}^2
\big)
C_{0}
\nonumber \\
&&
+
\big(M_{H^\pm}^2+3 M_{W}^2\big)
C_{1}
+
\big(3 M_{H^\pm}^2+M_{W}^2\big)
C_{2}
-
4 C_{00}
\Big]
(M_{W}^2,0,M_{H^\pm}^2,
m_{b}^2,m_{t}^2,m_{t}^2)
\nonumber\\
&&
+
g^{R}_{H^- t \bar{b}}
\,
(
eQ_t
\cdot
m_{t}^2)
\Big[
\big(M_{W}^2 - M_{H^\pm}^2\big)
\big(
C_{1}
+
C_{2}
\big)
-
4 C_{00}
\Big]
(M_{W}^2,0,M_{H^\pm}^2,
m_{b}^2,m_{t}^2,m_{t}^2)
\Big\},
\nonumber \\
F^{f}_{2, \Delta}
&=&
\frac{2\;
g_{W^+ \bar{t} \, b}
}{(4\pi)^2}
N^C_Q
\Big\{
-
g^{L}_{H^- t \bar{b}}
\,
(
eQ_t
\cdot
m_{b}^2)
\times
\\
&&
\times
\Big[
C_{0}
+
C_{1}
+
3 C_{2}
+
2 C_{12}
+
2 C_{22}
\Big](M_{W}^2,0,M_{H^\pm}^2,
m_{b}^2,m_{t}^2,m_{t}^2)
\nonumber \\
&&
+
g^{R}_{H^- t \bar{b}}
\,
(
eQ_t
\cdot
m_{t}^2)
\Big[
C_{1}
-
C_{2}
-
2 C_{12}
-
2 C_{22}
\Big]
(M_{W}^2,0,M_{H^\pm}^2,
m_{b}^2,m_{t}^2,m_{t}^2)
\nonumber \\
&&
-
g^{L}_{H^- t \bar{b}}
\,
(
eQ_b
\cdot
m_{b}^2)
\Big[
C_{0}
+
C_{2}
-
2 C_{12}
\Big](0,M_{H^\pm}^2,M_{W}^2,
m_{b}^2,m_{b}^2,m_{t}^2)
\nonumber \\
&&
+
g^{R}_{H^- t \bar{b}}
\,
(
eQ_b
\cdot
m_{t}^2)
\Big[
C_{2}
+
2 C_{12}
\Big]
(0,M_{H^\pm}^2,M_{W}^2,
m_{b}^2,m_{b}^2,m_{t}^2)
\Big\}.
\nonumber
\end{eqnarray}
The form factor
$F^{f}_{3, \Delta}$ for
triangle fermion loop group
contributing to total
form factor $F_{3}$
is reading as follows
\begin{eqnarray}
F^{f}_{3, \Delta}
&=&-2
\frac{
g_{W^+ \bar{t} b}
}{(4\pi)^2}
N^C_Q
\Big\{
eQ_b
\Big[
g^{LR}_{H^- t \bar{b}}
C_{2}
+
g^{L}_{H^- t \bar{b}}
\,
(m_{b}^2)
\,
C_{0}
\Big]
(0,M_{H^\pm}^2,M_{W}^2,
m_{b}^2,m_{b}^2,m_{t}^2)
\nonumber \\
&&
+
eQ_t
\Big[
g^{LR}_{H^- t \bar{b}}
\big(
C_{1}
+
C_{2}
\big)
+
g^{L}_{H^- t \bar{b}}
\,
(m_{b}^2)
\,
C_{0}
\Big]
(M_{W}^2,0,M_{H^\pm}^2,
m_{b}^2,m_{t}^2,m_{t}^2)
\Bigg\}.
\nonumber
\end{eqnarray}
In this case, the color factor $N^C_Q$ for quarks,
such as the top quark $t$ and bottom quark $b$
exchanged in the loop, has a value of 3.
General coupling
$g^{LR}_{H^- t \bar{b}}
=
\Big(
g^{L}_{H^- t \bar{b}}
\,
m_{b}^2
+
g^{R}_{H^- t \bar{b}}
\,
m_{t}^2
\Big)$ is used.
The corresponding general couplings for the vector
boson--fermion vertices are
$g^{L}_{\gamma f \bar{f}} =
g^{R}_{\gamma f \bar{f}}
\equiv g_{\gamma f \bar{f}}$
and
$g^{R}_{W^\pm f \bar{f'}} = 0, \;
g^{L}_{W^\pm f \bar{f'}} \equiv
g_{W^\pm f \bar{f'}}
= \frac{e}{\sqrt{2} s_W}$.
\subsection*{
\underline{
Form factors
$F^{f}_{i, \Pi}$:
}}
\begin{center}
\begin{figure}[H]
\centering
\includegraphics[width=8cm, height=3cm]
{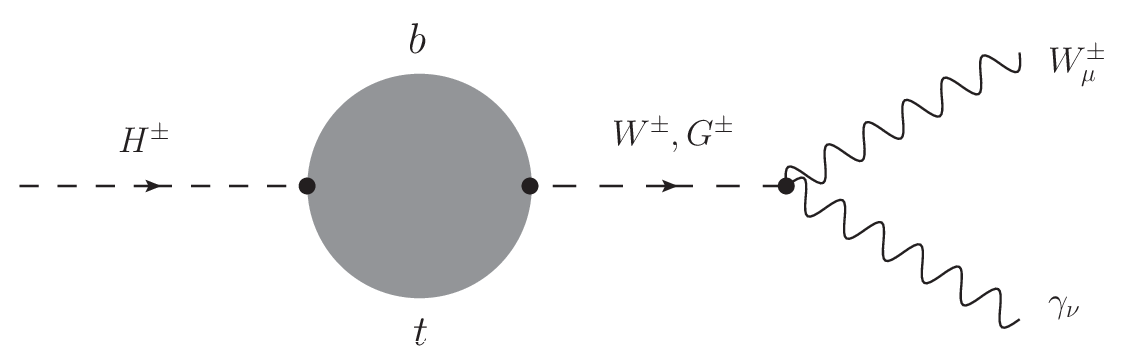}
\caption{Self-energy contributions on the
external $H^\pm$ leg arising from fermion
loops are also included in the calculation.
}
\label{Fig:SelfFermion}
\end{figure}
\end{center}

The form factor $F^{f}_{i, \Pi}$
for $i = 1, 2, 3$ from
the Figure.~\ref{Fig:SelfFermion}
is expressed as follows
\begin{eqnarray}
F^{f}_{1, \Pi}
&=&
\frac{1}{(4\pi)^2}
\frac{2
\cdot
g_{\gamma W^\pm W^\mp}
\cdot
g_{W^+ \bar{t} b}
}{
\big(M_{H^\pm}^2-M_{W}^2\big)}
N^C_Q
\times
\\
&&
\times
\Big\{
g^{L}_{H^- t \bar{b}}
\,
(m_{b}^2)
A_{0}(m_{b}^2)
+
g^{L}_{H^- t \bar{b}}
\,
(m_{b}^2)
\big(M_{W}^2-M_{H^\pm}^2\big)
\Big[
B_{0}
+
B_{1}
\Big]
(M_{H^\pm}^2,m_{b}^2,m_{t}^2)
\nonumber \\
&&
-
g^{R}_{H^- t \bar{b}}
\,
(m_{t}^2)
A_{0}(m_{b}^2)
\nonumber \\
&&
+
g^{R}_{H^- t \bar{b}}
\,
(m_{t}^2)
\Big[
\big(M_{H^\pm}^2-m_{t}^2+m_{b}^2\big)
B_{0}
+
\big(M_{H^\pm}^2+M_{W}^2\big)
B_{1}
\Big]
(M_{H^\pm}^2,m_{b}^2,m_{t}^2)
\Big\},
\nonumber \\
F^{f}_{2, \Pi}
&=&
F^{f}_{3, \Pi}
=
0.
\end{eqnarray}
Where the general couplings
involved in these Goldstone bosons
$G^\pm$ and quarks $f, f'$
exchanging in loop are
parameterized
as $ g_{G^- f \bar{f'}}
= m_{f'}
g^{L}_{G^- f \bar{f'}}
\,
P_L
+
m_{f}
\,
g^{R}_{G^- f \bar{f'}}
\,
P_R$
and
$g_{G^+ f' \bar{f}}
= m_{f}
\,
g^{L}_{G^+ f' \bar{f}}
\,
P_L
+
m_{f'}
\,
g^{R}_{G^+ f' \bar{f}}
\,
P_R$. They are expressed
in terms of physical couplings
following relations:
\begin{eqnarray}
g^{L}_{G^+ f' \bar{f}}
&=&
g^{R}_{G^- f \bar{f'}}
=
-
\dfrac{1}{M_W}
\,
g_{W^\pm f \bar{f'}},
\quad
g^{R}_{G^+ f' \bar{f}}
=
g^{L}_{G^- f \bar{f'}}
=
+
\dfrac{1}{M_W}
\,
g_{W^\pm f \bar{f'}}.
\end{eqnarray}
\subsection*{
\underline{
Form factors
$F^{f}_{i, T}$:}
}
\begin{center}
\begin{figure}[H]
\centering
\includegraphics[width=16cm, height=6cm]
{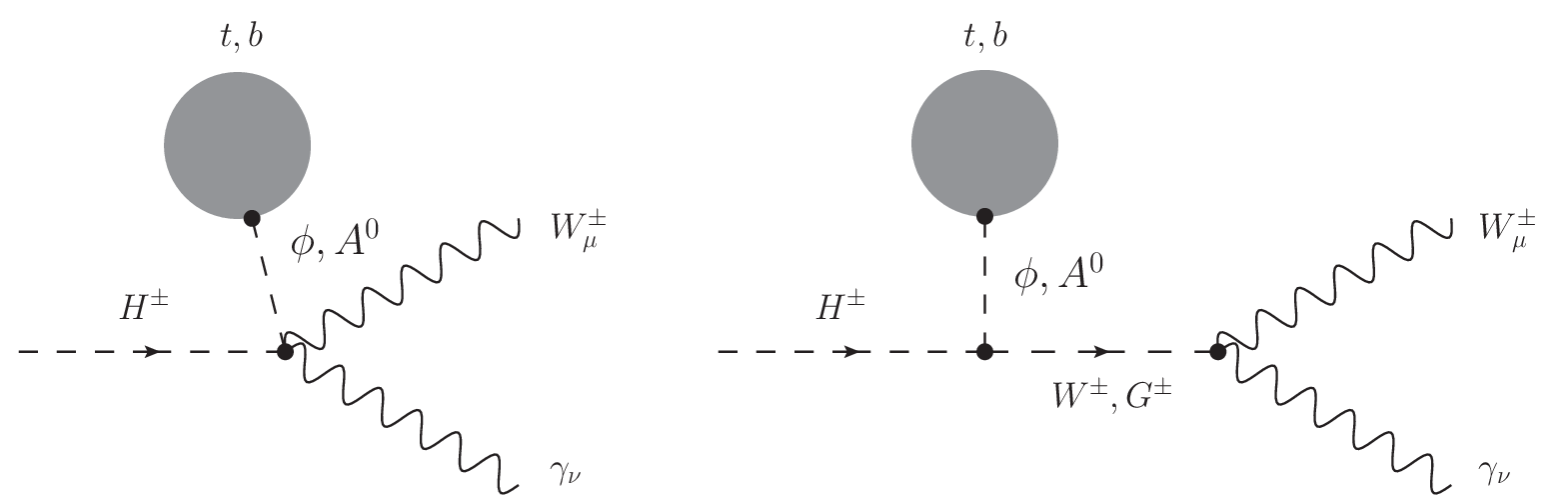}
\caption{Tadpole diagrams induced by fermion
loops with external scalar poles $\phi \equiv h, H$,
and $A$ are taken into account in the computation.
}
\label{Fig:TadFermion}
\end{figure}
\end{center}

The form factor $F^{f}_{i, T}$ for
$i = 1, 2, 3$ is expressed in terms
of the one-loop contributions with
poles $\phi \equiv h, H$ and $A$
as shown in Fig.~\ref{Fig:TadFermion}
as follows:
\begin{eqnarray}
F^{f}_{i, T}
&=&
F^{A - f}_{i, T}
+
\sum
\limits_{\phi = h, H}
F^{\phi - f}_{i, T}.
\end{eqnarray}
where the decomposed
form factors are presented as
\begin{eqnarray}
F^{A - f}_{1, T}
&=&
\frac{1}{(4\pi)^2}
\frac{2}{M_{A}^2
\big(M_{H^\pm}^2-M_{W}^2\big)}
N^C_Q
\times
\nonumber\\
&&
\times
\Big[
M_{A}^2
\,
g_{A H^- W^+}
\,
g_{\gamma W^\pm W^\mp}
+
\big(M_{W}^2-M_{H^\pm}^2\big)
(
g_{A H^- W^+}
\,
g_{\gamma W^\pm W^\mp}
+
g_{A H^- W^+ \gamma}
)
\Big]
\times
\nonumber \\
&& \times
\Big[
\Big(
g^{L}_{A b \bar{b}}
+
g^{R}_{A b \bar{b}}
\Big)
\,
m_{b}^2
\,
A_{0}(m_{b}^2)
+
\Big(
g^{L}_{A t \bar{t}}
+
g^{R}_{A t \bar{t}}
\Big)
\,
m_{t}^2
\,
A_{0}(m_{t}^2)
\Big]
\\
F^{\phi - f}_{1, T}
&=&
-
\frac{1}{(4\pi)^2}
\frac{4}{M_{\phi}^2
\big(M_{H^\pm}^2-M_{W}^2\big)}
N^C_Q
\times
\\
&&
\times
\Big[
\big(M_{W}^2-M_{H^\pm}^2\big)
\,
g_{\phi H^- W^+ \gamma}
+
\big(M_{\phi}^2-M_{H^\pm}^2+M_{W}^2\big)
\,
g_{\gamma W^\pm W^\mp}
\,
g_{\phi H^- W^+}
\Big]
\times
\nonumber\\
&&
\times
\Big[
g_{\phi b \bar{b}}
\,
m_{b}^2
\,
A_{0}(m_{b}^2)
+
g_{\phi t \bar{t}}
\,
m_{t}^2
\,
A_{0}(m_{t}^2)
\Big],
\nonumber\\
F^{A / \phi - f}_{2, T}
&=&
F^{A / \phi - f}_{3, T}
=
0.
\end{eqnarray}
The general couplings
involved in these Goldstone bosons
$G^\pm$ are rewritten in terms of
physical particles such as
$W^\pm$-bosons as follows
\begin{eqnarray}
g_{A H^- G^+}
&=&
\dfrac{1}{M_W}
\big(
M_{A}^2
-
M_{H^\pm}^2
\big)
\,
g_{A H^- W^+}.
\end{eqnarray}
The relevant general couplings of
the scalar Higgs bosons $\phi$ and
the pseudoscalar Higgs $A$ to
fermion vertices are given by
$g^{L}_{\phi f \bar{f}} =
g^{R}_{\phi f \bar{f}}
\equiv g_{\phi f \bar{f}}$
and
$g^{R}_{A f \bar{f}}
= -\,g^{L}_{A f \bar{f}}
\equiv g_{A f \bar{f}}$.
%
%
%
%
\section*{Appendix B:
Analytic Checks of $\xi$-Independence
}
Analytic checks for the $\xi$-independence
of the calculations are presented in this
Appendix. Without loss of generality, we
consider the one-loop
form factor with the bosonic contribution
as a representative
example. In particular, the one-loop form
factors
$F^{(W^\pm,H^\pm)}_{i}$ for $i = 1, 2, 3$,
originating from the mixing between the
vector boson
and the charged Higgs, can be decomposed
as follows:
\begin{eqnarray}
F^{W^\pm H^\pm}_{i}
&=&
F^{W^\pm H^\pm}_{i, \Delta}
+
F^{W^\pm H^\pm}_{i, \Pi}
+
F^{W^\pm H^\pm}_{i, T}.
\end{eqnarray}
It is reminded that the form factor
$F^{W^\pm H^\pm}_{3}$ equals zero,
as shown explicitly in the above sections.
The remaining form factors are expressed
in terms of the scalar one-loop
integrals $A_0$, $B_0$, and $C_0$
as follows:
\begin{eqnarray}
\label{Eq:Bim}
(4\pi)^2
\times
F^{W^\pm H^\pm}_{i}
&=&
c^{0}_{i}
+
\Big\{
c^{1}_{i}
A_0(M_{h}^2)
+
c^{2}_{i}
A_0(M_{H}^2)
+
c^{3}_{i}
A_0(M_{H^\pm}^2)
+
c^{4}_{i}
A_0(M_{A}^2)
\nonumber \\
&&
+
c^{5}_{i}
A_0(M_{Z}^2)
+
c^{6}_{i}
A_0( \xi M_{Z}^2)
+
c^{7}_{i}
A_0(M_{W}^2)
+
c^{8}_{i}
A_0( \xi M_{W}^2)
\nonumber \\
&&
+
c^{9}_{i}
B_0(0,M_{W}^2,
\xi M_{W}^2)
+
c^{10}_{i}
B_0(0,M_{W}^2, M_{W}^2)
+
c^{11}_{i}
B_0(0, \xi M_{W}^2,
\xi M_{W}^2)
\Big\},
\nonumber\\
&&
+
\sum
\limits_{\phi = h, H}
\Big\{
c^{1, \phi}_{i}
B_0(M_{H^\pm}^2,M_{\phi}^2,M_{H^\pm}^2)
+
c^{2, \phi}_{i}
B_0(M_{H^\pm}^2,M_{\phi}^2,M_{W}^2)
\nonumber \\
&&
+
c^{3, \phi}_{i}
B_0(M_{W}^2,M_{\phi}^2,M_{H^\pm}^2)
+
c^{4, \phi}_{i}
B_0(M_{W}^2,M_{\phi}^2,M_{W}^2)
\nonumber \\
&&
+
c^{5, \phi}_{i}
B_0(M_{H^\pm}^2,M_{\phi}^2, \xi M_{W}^2)
+
c^{6, \phi}_{i}
B_0(M_{W}^2,M_{\phi}^2, \xi M_{W}^2)
\nonumber \\
&&
+
c^{7, \phi}_{i}
C_0(M_{W}^2,0,
M_{H^\pm}^2,M_{\phi}^2,M_{W}^2,M_{W}^2)
\nonumber \\
&&
+
c^{8, \phi}_{i}
C_0(M_{W}^2,0,M_{H^\pm}^2,
M_{\phi}^2,M_{H^\pm}^2,M_{H^\pm}^2)
\Big\}.
\end{eqnarray}
All coefficients $c_j^k$ and $c_j^{k,\phi}$
are presented in the following
paragraphs. We note that the relations
given below are used to demonstrate
the $\xi$-independence of the coefficients
and to replace unphysical couplings with
the corresponding physical ones. These
relations are shown as
\begin{eqnarray}
\sum
\limits_{\phi = h, H}
g_{\phi W^\pm W^\mp}
\,
g_{\phi H^- W^+ \gamma}
&=&
\sum
\limits_{\phi = h, H}
g_{\phi ZZ}
\,
g_{\phi H^- W^+ \gamma}
=0,
\\
\sum
\limits_{\phi = h, H}
g_{\phi W^\pm W^\mp}
\,
g_{\phi H^- W^+}
&=&
\sum
\limits_{\phi = h, H}
g_{\phi ZZ}
\,
g_{\phi H^- W^+}
=
0.
\end{eqnarray}
Furthermore, one uses
\begin{eqnarray}
g_{h h H^- G^+}
&=&
\sum
\limits_{\phi = h, H}
\frac{
g_{\phi H^- W^+}
\,
g_{\phi h h}
}{M_{W}}
+
\dfrac{g_{h H^- W^+}}{M_{W}^3}
\Big[
\big(
M_{h}^2
-
M_{H^\pm}^2
\big)
\,
g_{h W^\pm W^\mp}
-
2 M_{W}^2
\,
g_{h H^\pm H^\mp}
\Big],
\\
g_{H H H^- G^+}
&=&
\sum
\limits_{\phi = h, H}
\frac{
g_{\phi H^- W^+}
\,
g_{\phi H H}
}{M_{W}}
+
\frac{g_{H H^- W^+}}{M_{W}^3}
\Big[
\big(
M_{H}^2
-
M_{H^\pm}^2
\big)
\,
g_{H W^\pm W^\mp}
-
2 M_{W}^2
\,
g_{H H^\pm H^\mp}
\Big],
\nonumber\\
\\
g_{A A H^- G^+}
&=&
\sum
\limits_{\phi = h, H}
\frac{
g_{\phi H^- W^+}
\,
g_{\phi A A}
}{M_{W}},
\\
g_{H^+ H^- H^- G^+}
&=&
\sum
\limits_{\phi = h, H}
\frac{2 \,
g_{\phi H^- W^+}
\,
g_{\phi H^\pm H^\mp}
}{M_{W}},
\\
g_{G^+ G^- H^- G^+}
&=&
-
\sum
\limits_{\phi = h, H}
\frac{
	M_{\phi}^2
	\,
	g_{\phi H^- W^+}
	\,
	g_{\phi W^\pm W^\mp}
}{M_{W}^3},
\\
g_{G^0 G^0 H^- G^+}
&=&
-
\frac{c_W^2}{2 M_{W}^3}
\sum
\limits_{\phi = h, H}
M_{\phi}^2
\,
g_{\phi H^- W^+}
\,
g_{\phi ZZ}.
\end{eqnarray}
As a result, all coefficients $c_i$
are taken the form of:
\begin{eqnarray}
c^{0}_{1}
&=&
-
\frac{e}{2 M_{W}^2}
\sum
\limits_{\phi = h, H}
\Big[
M_{\phi}^2
\,
g_{\phi W^\pm W^\mp}
\,
g_{\phi H^- W^+}
-
(2 M_{W}^2)
\,
g_{\phi H^\pm H^\mp}
\,
g_{\phi H^- W^+}
\Big],
\\
c^{1}_{1}
&=&
\frac{e}{
2 M_{H^\pm}^2 M_{W}^2 }
g_{h H^- W^+}
\,
\Big[
\big(2 M_{W}^2\big)
\,
g_{h H^\pm H^\mp}
+
\big(
M_{H^\pm}^2
-
M_{h}^2
+
M_{W}^2
\big)
\,
g_{h W^\pm W^\mp}
\Big]
\nonumber \\
&&
+
\frac{e M_{H^\pm}^2 }{
2 M_{H^\pm}^2 M_{W}^2
\big(M_{H^\pm}^2
- \xi M_{W}^2\big)}
\times
\nonumber\\
&&
\times
\Big\{
g_{h H^- W^+}
\,
\Big[
\big(
M_{h}^2
-
M_{H^\pm}^2
\big)
\,
g_{hW^\pm W^\mp}
-
2 M_{W}^2
\,
g_{h H^\pm H^\mp}
+
M_{W}^2
\,
g_{h h h}
\Big]
\nonumber \\
&&
+
M_{W}^2
\,
g_{H H^- W^+}
\,
g_{H h h}
-
M_{W}^3
\,
g_{hhH^- G^+}
\Big\}.
\end{eqnarray}
We verify that the last term
of $c_1^1$ is vanished as
\begin{eqnarray}
&&
g_{h H^- W^+}
\,
\Big[
\big(
M_{h}^2
-
M_{H^\pm}^2
\big)
\,
g_{h W^\pm W^\mp}
-
2 M_{W}^2
\,
g_{h H^\pm H^\mp}
+
M_{W}^2
\,
g_{h h h}
\Big]
\nonumber\\
&&
+
M_{W}^2
\,
g_{H H^- W^+}
\,
g_{H h h}
-
M_{W}^3
\,
g_{h h H^- G^+}
\nonumber \\
&&
=
-\dfrac{3 \lambda_{5} M_{W}^3 s_{\beta - \alpha}
	c_{\beta - \alpha}}{2 s_{2 \beta}}
\Big(
s_{2 \beta}
-
s_{2 \alpha}
-
2 \,
c_{\alpha + \beta}
\,
s_{\beta - \alpha}
\Big)
=0.
\end{eqnarray}
Subsequently, we arrive at
\begin{eqnarray}
c_1^1
&=&
\dfrac{e}{
2 M_{H^\pm}^2 M_{W}^2}
\,
g_{h H^- W^+}
\,
\Big[
\big(2 M_{W}^2\big)
\,
g_{h H^\pm H^\mp}
-
\big(
M_{h}^2
-
M_{H^\pm}^2
-
M_{W}^2
\big)
\,
g_{h W^\pm W^\mp}
\Big],
\\
c^{2}_{1}
&=&
\frac{e}{
2 M_{H^\pm}^2 M_{W}^2
}
g_{H H^- W^+}
\,
\Big[
\big(2 M_{W}^2\big)
\,
g_{H H^\pm H^\mp}
+
\big(
M_{H^\pm}^2
-
M_{H}^2
+
M_{W}^2
\big)
\,
g_{H W^\pm W^\mp}
\Big]
\nonumber \\
&&
+
\frac{e\; M_{H^\pm}^2}
{
2 M_{H^\pm}^2 M_{W}^2
\big(M_{H^\pm}^2
- \xi M_{W}^2\big)}
\times
\\
&& \times
\Big\{
g_{H H^- W^+}
\,
\Big[
\big(
M_{H}^2
-
M_{H^\pm}^2
\big)
\,
g_{H W^\pm W^\mp}
-
2 M_{W}^2
\,
g_{H H^\pm H^\mp}
+
M_{W}^2
\,
g_{H H H}
\Big]
\nonumber \\
&&
+
M_{W}^2
\,
g_{h H^- W^+}
\,
g_{h H H}
-
M_{W}^3
\,
g_{H H H^- G^+}
\Big\}.
\nonumber
\end{eqnarray}
We also have
\begin{eqnarray}
&&
g_{H H^- W^+}
\,
\Big[
\big(
M_{H}^2
-
M_{H^\pm}^2
\big)
\,
g_{H W^\pm W^\mp}
-
2 M_{W}^2
\,
g_{H H^\pm H^\mp}
+
M_{W}^2
\,
g_{H H H}
\Big]
\nonumber \\
&&
+
M_{W}^2
\,
g_{h H^- W^+}
\,
g_{h H H}
-
M_{W}^3
\,
g_{H H H^- G^+}
\nonumber \\
&&
=
\dfrac{3 \lambda_{5} M_{W}^3
s_{\beta - \alpha}
c_{\beta - \alpha}}{2 s_{2 \beta}}
\Big(
s_{2 \alpha}
+
s_{2 \beta}
-
2 \,
s_{\alpha + \beta}
\,
c_{\beta - \alpha}
\Big)=
0.
\end{eqnarray}
One then arrives at
\begin{eqnarray}
c_1^2
&=&
\dfrac{e}{
2 M_{H^\pm}^2 M_{W}^2}
\,
g_{H H^- W^+}
\,
\Big[
\big(2 M_{W}^2\big)
\,
g_{H H^\pm H^\mp}
-
\big(
M_{H}^2
-
M_{H^\pm}^2
-
M_{W}^2
\big)
\,
g_{H W^\pm W^\mp}
\Big],
\\
c^{3}_{1}
&=&
-
\dfrac{g_{\gamma W^\pm W^\mp}
}{
M_{H^\pm}^2
\big(M_{H^\pm}^2
- \xi M_{W}^2\big)}
\Big\{
\sum
\limits_{\phi = h, H}
g_{\phi H^- W^+}
\,
g_{\phi H^\pm H^\mp}
\,
\big(
M_{H^\pm}^2
-
\xi M_{W}^2
\big)
\nonumber \\
\nonumber \\
&&
+
M_{H^\pm}^2
\Big[
M_{W}
\,
g_{H^+ H^- H^- G^+}
-
2
\sum
\limits_{\phi = h, H}
g_{\phi H^- W^+}
\,
g_{\phi H^\pm H^\mp}
\Big]
\Big\}.
\end{eqnarray}
The second term in this coefficient
is vanished due to the following
identity
\begin{eqnarray}
&&
M_{W}
\,
g_{H^+ H^- H^- G^+}
-
2
\sum
\limits_{\phi = h, H}
g_{\phi H^- W^+}
\,
g_{\phi H^\pm H^\mp}
=
- \dfrac{2 \lambda_5 M_{W}}{s_{2 \beta}}
\big(
c_{2 \beta}
-
c_{\alpha + \beta}
\,
c_{\beta - \alpha}
+
s_{\alpha + \beta}
\,
s_{\beta - \alpha}
\big)=
0. \nonumber\\
\end{eqnarray}
As a result, the coefficient
$c_1^3$ gets
\begin{eqnarray}
c_1^3
&=&
-
\dfrac{e}{M_{H^\pm}^2}
\sum
\limits_{\phi = h, H}
g_{\phi H^- W^+}
\,
g_{\phi H^\pm H^\mp},
\\
c^{4}_{1}
&=&
-
\dfrac{g_{\gamma W^\pm W^\mp}}
{2 \big(M_{H^\pm}^2
- \xi M_{W}^2\big)}
\Big[
M_{W}
\,
g_{A A H^- G^+}
-
\sum
\limits_{\phi = h, H}
g_{\phi H^- W^+}
\,
g_{\phi A A}
\Big] = 0.
\end{eqnarray}
In fact, from this coefficient,
we can express the term in the
bracket as follows:
\begin{eqnarray}
&&
M_{W}
\,
g_{A A H^- G^+}
-
\sum
\limits_{\phi = h, H}
g_{\phi H^- W^+}
\,
g_{\phi A A}
=
-\frac{\lambda_5 M_{W}}{s_{2 \beta}}
\big(
c_{2 \beta}
-
c_{\alpha + \beta}
\,
c_{\beta - \alpha}
+
s_{\alpha + \beta}
\,
s_{\beta - \alpha}
\big)=
0.
\end{eqnarray}
Other coefficients are given by
\begin{eqnarray}
c^{5}_{1}
&=&
\sum
\limits_{\phi = h, H}
\dfrac{3 \, g_{\phi ZZ}}{2}
\Big[
\dfrac{1}{M_{\phi}^2}
\Big(
g_{\gamma W^\pm W^\mp}
\,
g_{\phi H^- W^+}
+
g_{\phi H^- W^+ \gamma}
\Big)
-
\dfrac{
g_{\gamma W^\pm W^\mp}
\,
g_{\phi H^- W^+}
}{
\big(M_{H^\pm}^2
- \xi M_{W}^2 \big)}
\Big] =0, \\
c^{6}_{1}
&=&
-
\dfrac{g_{\gamma W^\pm W^\mp}}{
4 M_{W}^2
\big(M_{H^\pm}^2
- \xi M_{W}^2\big)}
\Big[
\big(2 M_{W}^3\big)
\,
g_{G^0 G^0 H^- G^+}
+
\sum
\limits_{\phi = h, H}
\big(M_{\phi}^2 c_W^2\big)
\,
g_{\phi H^- W^+}
\,
g_{\phi ZZ}
\Big] =0. \nonumber\\
\end{eqnarray}
This can be verified by
\begin{eqnarray}
\big(2 M_{W}^3\big)
\,
g_{G^0 G^0 H^- G^+}
+
\sum
\limits_{\phi = h, H}
\big(M_{\phi}^2 c_W^2\big)
\,
g_{\phi H^- W^+}
\,
g_{\phi ZZ}
&=&
0.
\end{eqnarray}
For $c_1^7$, one presents
\begin{eqnarray}
c^{7}_{1}
&=&
\sum
\limits_{\phi = h, H}
\dfrac{
g_{\gamma W^\pm W^\mp}
\,
g_{\phi H^- W^+}
\,
g_{\phi W^\pm W^\mp}
}{
6 M_{H^\pm}^2 M_{W}^2
\big(M_{H^\pm}^2
- M_{W}^2\big)^2}
\Big[
M_{\phi}^2
\,
\Big(
4
M_{H^\pm}^4
-
5
M_{H^\pm}^2
M_{W}^2
+
3
M_{W}^4
\Big)
\Big],
\\
c^{8}_{1}
&=&
-
\sum
\limits_{\phi = h, H}
\dfrac{g_{\phi W^\pm W^\mp}
}{
12 M_{W}^4
\big(M_{H^\pm}^2
- M_{W}^2\big)^2}
\,
M_{\phi}^2
\times
\nonumber \\
\nonumber \\
&& \times
\Big[
g_{\gamma W^\pm W^\mp}
\,
g_{\phi H^- W^+}
\Big(
4 M_{H^\pm}^2 M_{W}^2
+
M_{W}^4
-
M_{H^\pm}^4
\Big)
-
g_{\phi H^- W^+ \gamma}
\,
\big(M_{H^\pm}^2-M_{W}^2\big)^2
\Big]
\nonumber \\
\nonumber \\
&&
-
\dfrac{g_{\gamma W^\pm W^\mp}}{
M_{W}^2
\big(M_{H^\pm}^2
- \xi M_{W}^2\big)}
\Big[
M_{W}^3
\,
g_{G^+ G^- H^- G^+}
+
\sum
\limits_{\phi = h, H}
M_{\phi}^2
\,
g_{\phi H^- W^+}
\,
g_{\phi W^\pm W^\mp}
\Big]
\nonumber \\
&=&
-
\sum
\limits_{\phi = h, H}
\dfrac{
g_{\gamma W^\pm W^\mp}
\,
g_{\phi H^- W^+}
\,
g_{\phi W^\pm W^\mp}
}{
6 M_{W}^2
\big(M_{H^\pm}^2
- M_{W}^2\big)^2}
M_{\phi}^2
\,
\big(
M_{H^\pm}^2
+
M_{W}^2
\big).
\end{eqnarray}
In order to arrive the last line
result, we have already applied
the following relation
\begin{eqnarray}
g_{G^+ G^- H^- G^+}
&=&
\dfrac{e^2
}{
	2 M_W^2 s_W^2}
s_{\beta - \alpha}
c_{\beta - \alpha}
\big(
M_{h}^2
-
M_{H}^2
\big)
\end{eqnarray}
and confirm subsequently
\begin{eqnarray}
M_{W}^3
\,
g_{G^+ G^- H^- G^+}
+
\sum
\limits_{\phi = h, H}
M_{\phi}^2
\,
g_{\phi H^- W^+}
\,
g_{\phi W^\pm W^\mp}
=0.
\end{eqnarray}
The remaining coefficients are listed as
\begin{eqnarray}
c^{9}_{1}
&=&
\dfrac{g_{\gamma W^\pm W^\mp}
}{
\big(M_{H^\pm}^2
- M_{W}^2\big)^2}
\,
\dfrac{M_{H^\pm}^2+M_{W}^2}{6}
\big(\xi - 1\big)
\sum
\limits_{\phi = h, H}
M_{\phi}^2
\,
g_{\phi W^\pm W^\mp}
\,
g_{\phi H^- W^+}
,\\
c^{10}_{1}
&=&
-
\dfrac{2 \, g_{\gamma W^\pm W^\mp}}{3}
\sum
\limits_{\phi = h, H}
g_{\phi W^\pm W^\mp}
\,
g_{\phi H^- W^+} = 0,
\\
c^{11}_{1}
&=&
\dfrac{
g_{\gamma W^\pm W^\mp}
}{3}
\xi
\sum
\limits_{\phi = h, H}
g_{\phi W^\pm W^\mp}
\,
g_{\phi H^- W^+} = 0.
\end{eqnarray}
The other coefficients
$c^{i, \phi}_{1}$ for
$i = 1, \ldots, 8$
are also expressed as follows:
\begin{eqnarray}
c^{1, \phi}_{1}
&=&
\dfrac{e}{M_{H^\pm}^2}
\dfrac{
g_{\phi H^\pm H^\mp}
\,
g_{\phi H^- W^+}
}{
\big(M_{H^\pm}^2
- M_{W}^2\big)}
\Big[
M_{\phi}^2
M_{W}^2
-
2
M_{H^\pm}^2
\big(
M_{\phi}^2
-
M_{H^\pm}^2
\big)
\Big],
\\
c^{2, \phi}_{1}
&=&
\dfrac{e}{2 M_{W}^2 M_{H^\pm}^2}
\dfrac{
g_{\phi H^- W^+}
\,
g_{\phi W^\pm W^\mp}
}{
\big(M_{H^\pm}^2
- M_{W}^2\big)}
\times
\nonumber\\
&&
\times
\Big\{
\big(M_{H^\pm}^2
- M_{W}^2\big)
\Big[
\big(
M_{\phi}^2
-
M_{H^\pm}^2
-
M_{W}^2
\big)^2
-
4
M_{H^\pm}^2
M_{W}^2
\Big]
\nonumber \\
&&
-
M_{H^\pm}^2
(M_{H^\pm}^2-M_{\phi}^2+M_{W}^2)
(M_{\phi}^2+M_{H^\pm}^2-3 M_{W}^2)
\Big\}, \\
c^{3, \phi}_{1}
&=&
\dfrac{e}{\big(M_{H^\pm}^2-M_{W}^2\big)}
\,
g_{\phi H^- W^+}
\,
g_{\phi H^\pm H^\mp}
\,
\Big(
M_{\phi}^2
-M_{H^\pm}^2
-M_{W}^2
\Big),     \\
c^{4, \phi}_{1}
&=&
-
\dfrac{e}{2 M_{W}^2}
\dfrac{
g_{\phi H^- W^+}
\,
g_{\phi W^\pm W^\mp}
}{
\big(M_{H^\pm}^2-M_{W}^2\big)}
\Big[
M_{\phi}^2
(M_{\phi}^2-M_{H^\pm}^2-3 M_{W}^2)
-
2 M_{W}^2
(M_{H^\pm}^2-3 M_{W}^2)
\Big],
\\
c^{5, \phi}_{1}
&=&
-
\dfrac{g_{\phi W^\pm W^\mp}
}{
12 M_{W}^4
\big(M_{H^\pm}^2
-M_{W}^2\big)^2}
\Big[
g_{\gamma W^\pm W^\mp}
\,
g_{\phi H^- W^+}
\big(
6 M_{W}^2
\big)
\big(
M_{\phi}^2
-M_{H^\pm}^2
\big)
\big(
M_{H^\pm}^2
-M_{W}^2
\big)^2
\Big]
\nonumber \\
\nonumber \\
&&
+
\dfrac{
g_{\gamma W^\pm W^\mp}
\,
g_{\phi H^- W^+}
\,
g_{\phi W^\pm W^\mp}
}{
2 M_{H^\pm}^2 M_{W}^2
\big(M_{H^\pm}^2
- \xi M_{W}^2\big)}
\Big[
M_{H^\pm}^2
\big(
M_{H^\pm}^2
- \xi M_{W}^2
\big)
\big(
M_{\phi}^2
-M_{H^\pm}^2
\big)
\Big]=
0,
\\
c^{6, \phi}_{1}
&=&
-
\dfrac{g_{\phi W^\pm W^\mp}
}{
12 M_{W}^4}
\Big(
g_{\gamma W^\pm W^\mp}
\,
g_{\phi H^- W^+}
+
g_{\phi H^- W^+ \gamma}
\Big)
\Big[
2 M_{\phi}^2 M_{W}^2
\big(
1 + \xi
\big)
-
M_{W}^4
\big(
1 - \xi
\big)^2
-
M_{\phi}^4
\Big]
\nonumber \\
&=&
0,
\\
c^{7, \phi}_{1}
&=&
-
g_{\phi W^\pm W^\mp}
\,
g_{\gamma W^\pm W^\mp}
\,
g_{\phi H^- W^+}
\,
\Big(
M_{\phi}^2
+M_{H^\pm}^2
-3 M_{W}^2
\Big),
\\
c^{8, \phi}_{1}
&=&
-
g_{\phi H^\pm H^\mp}
\,
g_{\gamma H^\pm H^\mp}
\,
g_{\phi H^- W^+}
\,
(2 M_{H^\pm}^2).
\end{eqnarray}
Similarly, other coefficients are also
given by
\begin{eqnarray}
c^{0}_{2}
&=&
\frac{e}{M_{W}^2}
\frac{1}{
\big(M_{H^\pm}^2-M_{W}^2\big)}
\sum
\limits_{\phi = h, H}
\Big[
M_{\phi}^2
\,
g_{\phi H^- W^+}
\,
g_{\phi W^\pm W^\mp}
-
(2 M_{W}^2)
\,
g_{\phi H^- W^+}
\,
g_{\phi H^\pm H^\mp}
\Big]
\nonumber \\
&=&
\dfrac{-2}{\big(M_{H^\pm}^2-M_{W}^2\big)}
\times
c^{0}_{1},
\\
c^{1}_{2}
&=&
-\dfrac{
g_{h H^- W^+}
}{\big(M_{H^\pm}^2-M_{W}^2\big)}
\dfrac{e}{
M_{H^\pm}^2 M_{W}^2}
\Big[
\big(2 M_{W}^2\big)
\,
g_{h H^\pm H^\mp}
-
\big(M_{h}^2
-M_{H^\pm}^2
-M_{W}^2\big)
\,
g_{h W^\pm W^\mp}
\Big]
\nonumber \\
&=&
\dfrac{-2}{\big(M_{H^\pm}^2-M_{W}^2\big)}
\times
c^{1}_{1},
\\
c^{2}_{2}
&=&
\dfrac{-
g_{H H^- W^+}
}{\big(M_{H^\pm}^2-M_{W}^2\big)}
\dfrac{e}{
M_{H^\pm}^2 M_{W}^2}
\Big[
\big(2 M_{W}^2\big)
\,
g_{H H^\pm H^\mp}
-
\big(M_{H}^2
-M_{H^\pm}^2
-M_{W}^2\big)
\,
g_{H W^\pm W^\mp}
\Big]
\nonumber \\
&=&
\dfrac{-2}{
\big(M_{H^\pm}^2-M_{W}^2\big)}
\times
c^{2}_{1},
\\
c^{3}_{2}
&=&
\dfrac{2\; e}{
\big(M_{H^\pm}^2-M_{W}^2\big)}
\dfrac{1}{M_{H^\pm}^2}
\sum
\limits_{\phi = h, H}
g_{\phi H^- W^+}
\,
g_{\phi H^\pm H^\mp}
=
\dfrac{-2}{
\big(M_{H^\pm}^2-M_{W}^2\big)}
\times
c^{3}_{1},
\\
c^{7}_{2}
&=&
-
\sum
\limits_{\phi = h, H}
\dfrac{
g_{\gamma W^\pm W^\mp}
\,
g_{\phi H^- W^+}
\,
g_{\phi W^\pm W^\mp}
}{
3 M_{H^\pm}^2 M_{W}^2
\big(M_{H^\pm}^2
-M_{W}^2\big)^3}
\Big[
M_{\phi}^2
\,
\Big(
7 M_{H^\pm}^4
-
2
M_{H^\pm}^2
M_{W}^2
+
3 M_{W}^4
\Big)
\Big],
\\
c^{8}_{2}
&=&
\sum
\limits_{\phi = h, H}
\dfrac{
g_{\gamma W^\pm W^\mp}
\,
g_{\phi H^- W^+}
\,
g_{\phi W^\pm W^\mp}
}{
3 M_{W}^2
\big(M_{H^\pm}^2
-M_{W}^2\big)^3}
\Big[
4 M_{\phi}^2
\,
\big(M_{H^\pm}^2+M_{W}^2\big)
\Big],
\\
c^{9}_{2}
&=&
-
\dfrac{
4 \,
g_{\gamma W^\pm W^\mp}
}{
\big(M_{H^\pm}^2
-M_{W}^2\big)^3}
\,
\dfrac{M_{H^\pm}^2+M_{W}^2}{3}
\big(\xi-1\big)
\sum
\limits_{\phi = h, H}
M_{\phi}^2
\,
g_{\phi H^- W^+}
\,
g_{\phi W^\pm W^\mp},
\\
c^{10}_{2}
&=&
\dfrac{
4 M_{H^\pm}^2
}{
3
\big(M_{H^\pm}^2
-M_{W}^2\big)^2}
\,
g_{\gamma W^\pm W^\mp}
\sum
\limits_{\phi = h, H}
g_{\phi H^- W^+}
\,
g_{\phi W^\pm W^\mp} = 0.
\end{eqnarray}
We also confirm that
$c^{4}_{2} = c^{5}_{2}
= c^{6}_{2} = c^{11}_{2} = 0$.
Furthermore, the remaining
coefficients
$c^{i, \phi}_{2}$ for
$i = 1, \ldots, 8$
are presented as follows:
\begin{eqnarray}
c^{1, \phi}_{2}
&=&
\dfrac{-2\; e}{\big(M_{H^\pm}^2-M_{W}^2\big)}
\dfrac{1}{M_{H^\pm}^2}
\dfrac{
g_{\phi H^- W^+}
\,
g_{\phi H^\pm H^\mp}
}{
\big(M_{H^\pm}^2
-M_{W}^2\big)}
\Big[
M_{\phi}^2 M_{W}^2
-
2 M_{H^\pm}^2
(M_{\phi}^2-M_{H^\pm}^2)
\Big]
\nonumber \\
&=&
\dfrac{-2}{
\big(M_{H^\pm}^2
-M_{W}^2\big)}
\times
c^{1, \phi}_{1},
\\
c^{2, \phi}_{2}
&=&
\dfrac{-2}{\big(M_{H^\pm}^2-M_{W}^2\big)}
\dfrac{
g_{\gamma W^\pm W^\mp}
\,
g_{\phi H^- W^+}
\,
g_{\phi W^\pm W^\mp}
}{
2 M_{W}^2 M_{H^\pm}^2
\big(M_{H^\pm}^2-M_{W}^2\big)}
\Big[
2 M_{W}^4
(M_{\phi}^2+3 M_{H^\pm}^2)
\nonumber \\
&&
-
M_{W}^2
\Big(
M_{\phi}^4
+
4 M_{\phi}^2 M_{H^\pm}^2
+
M_{H^\pm}^4
+
M_{W}^4
\Big)
-
2 M_{\phi}^2 M_{H^\pm}^2
(M_{H^\pm}^2-M_{\phi}^2)
\Big]
\nonumber \\
&=&
\dfrac{-2}{\big(M_{H^\pm}^2-M_{W}^2\big)}
\times
c^{2, \phi}_{1},
\\
c^{3, \phi}_{2}
&=&
\dfrac{-2}{\big(M_{H^\pm}^2-M_{W}^2\big)}
\Big\{
\dfrac{e}{\big(M_{H^\pm}^2-M_{W}^2\big)}
\,
g_{\phi H^- W^+}
\,
g_{\phi H^\pm H^\mp}
\,
\big(M_{\phi}^2-M_{H^\pm}^2-M_{W}^2\big)
\Big\}
\nonumber \\
&=&
\dfrac{-2}{\big(M_{H^\pm}^2-M_{W}^2\big)}
\times
c^{3, \phi}_{1},
\\
c^{4, \phi}_{2}
&=&
\dfrac{
g_{\gamma W^\pm W^\mp}
\,
g_{\phi H^- W^+}
\,
g_{\phi W^\pm W^\mp}
}{
M_{W}^2
\big(M_{H^\pm}^2-M_{W}^2\big)^2}
\Big[
M_{\phi}^2
(M_{\phi}^2-M_{H^\pm}^2-3 M_{W}^2)
-
2 M_{W}^2
(M_{H^\pm}^2-3 M_{W}^2)
\Big]
\nonumber \\
&=&
\dfrac{-2}{\big(M_{H^\pm}^2-M_{W}^2\big)}
\times
c^{4, \phi}_{1},
\\
c^{7, \phi}_{2}
&=&
\dfrac{-2}{\big(M_{H^\pm}^2-M_{W}^2\big)}
\Big[
(-e)
\,
(M_{\phi}^2+M_{H^\pm}^2-3 M_{W}^2)
\,
g_{\phi H^- W^+}
\,
g_{\phi W^\pm W^\mp}
\Big]
\nonumber \\
\nonumber \\
&=&
\dfrac{-2}{\big(M_{H^\pm}^2-M_{W}^2\big)}
\times
c^{7, \phi}_{1},
\\
c^{8, \phi}_{2}
&=&
\dfrac{-2}{\big(M_{H^\pm}^2-M_{W}^2\big)}
\Big[
-
(2 M_{H^\pm}^2)
\,
g_{\gamma H^\pm H^\mp}
\,
g_{\phi H^- W^+}
\,
g_{\phi H^\pm H^\mp}
\Big]
=
\frac{-2\times
c^{8, \phi}_{1}}{\big(M_{H^\pm}^2-M_{W}^2\big)}.
\end{eqnarray}
In addition, the coefficients
$c^{5, \phi}_{2}$ and $c^{6, \phi}_{2}$
vanish, i.e.,
$c^{5, \phi}_{2} = c^{6, \phi}_{2} = 0$.

Remaining terms dependence of $\xi$
is included the term $c^{8}_{i}
A_0( \xi M_{W}^2)
+
c^{9}_{i}
B_0(0,M_{W}^2, \xi M_{W}^2)$.
This term is expressed as
\begin{eqnarray}
\label{lastXI}
c^{8}_{i}
A_0( \xi M_{W}^2)
+
c^{9}_{i}
B_0(0,M_{W}^2, \xi M_{W}^2)
&=&
\Big[
c^{8}_{i}
-
\dfrac{c^{9}_{i}
}{
M_{W}^2 \big(1 - \xi\big) }
\Big]
A_0(\xi M_{W}^2)
+
\dfrac{c^{9}_{i}
}{
M_{W}^2 \big(1 - \xi\big) }
A_0(M_{W}^2).\nonumber\\
\end{eqnarray}
We note that the $B_0$ function has already
been expanded in terms of $A_0$ to obtain
the right-hand side of the above equation.
The expansion is given by:
\begin{eqnarray}
B_0(0,M_{1}^2, M_{2}^2)
&=&
\dfrac{1}{M_{1}^2 - M_{2}^2}
\Big[
A_0(M_{1}^2)
-
A_0(M_{2}^2)
\Big].
\end{eqnarray}
Because we find that
the coefficients $c_i^{9}/(1-\xi)$
are independent of $\xi$,
the last term in Eq.~\ref{lastXI}
is therefore $\xi$-independent.
The remaining task is to prove that
the term $A_0(\xi M_{W}^2)$
vanishes analytically, which can
be done as follows:
\begin{eqnarray}
c^{8}_{1}
-
\dfrac{1}{
M_{W}^2 \big(1 - \xi\big) }
\times
c^{9}_{1}
&=&
-
\sum
\limits_{\phi = h, H}
\dfrac{
g_{\gamma W^\pm W^\mp}
\,
g_{\phi H^- W^+}
\,
g_{\phi W^\pm W^\mp}
}{
6 M_{W}^2
\big(M_{H^\pm}^2
-M_{W}^2\big)^2}
\,
\Big[
M_{\phi}^2
\,
\big(
M_{H^\pm}^2
+
M_{W}^2
\big)
\Big]
\nonumber \\
&&
-
\dfrac{1}{
M_{W}^2 \big(1 - \xi\big) }
\Big[
\big(\xi - 1\big)
\dfrac{g_{\gamma W^\pm W^\mp}
}{
\big(M_{H^\pm}^2-M_{W}^2\big)^2}
\,
\dfrac{M_{H^\pm}^2+M_{W}^2}{6}
\times
\nonumber \\
&& \times
\sum
\limits_{\phi = h, H}
M_{\phi}^2
\,
g_{\phi W^\pm W^\mp}
\,
g_{\phi H^- W^+}
\Big]=
0.
\end{eqnarray}
Another terms is also vanished
in the same way
\begin{eqnarray}
c^{8}_{2}
-
\dfrac{1}{
M_{W}^2 \big(1 - \xi\big) }
\times
c^{9}_{2}
&=&
\sum
\limits_{\phi = h, H}
\dfrac{
g_{\gamma W^\pm W^\mp}
\,
g_{\phi H^- W^+}
\,
g_{\phi W^\pm W^\mp}
}{
3 M_{W}^2
\big(M_{H^\pm}^2-M_{W}^2\big)^3}
\Big[
4 M_{\phi}^2
\,
\big(M_{H^\pm}^2+M_{W}^2\big)
\Big]
\nonumber \\
&&\hspace{0.25cm}
-
\dfrac{1}{
M_{W}^2 \big(1 - \xi\big) }
\Big[
-
\big(\xi-1\big)
\dfrac{
4
\,
g_{\gamma W^\pm W^\mp}
}{
\big(M_{H^\pm}^2-M_{W}^2\big)^3}
\,
\dfrac{M_{H^\pm}^2+M_{W}^2}{3}
\times
\nonumber \\
&& \times
\sum
\limits_{\phi = h, H}
M_{\phi}^2
\,
g_{\phi H^- W^+}
\,
g_{\phi W^\pm W^\mp}
\Big]=
0.
\end{eqnarray}
As a result, the form factors are independent
of $\xi$.
\section*{Appendix C:
Analytic Ward Identity Checks}
After confirming the $\xi$-independence of
the results, the remaining $\xi$-independent
terms in the form factors of Eq.~\ref{Eq:Bim}
are found to satisfy the Ward identity.
In particular, the extra terms
$\frac{c^{9}_{i}}{M_{W}^2 \big(1 - \xi\big)}$
arising from the reduction in Eq.~(\ref{lastXI})
are combined with $c^{7}_{i}$ in Eq.~\ref{Eq:Bim}.
We thus analytically confirm the Ward identity.
The detailed verification is presented as follows:

\begin{itemize}
\item \underline{In the form
factor $F^{W^\pm H^\pm}_{1}$}:
 we obtain that
\begin{eqnarray}
&& c^{7}_{1}
\longrightarrow
c^{7}_{1}
+
\dfrac{c^{9}_{1}
}{
M_{W}^2 \big(1 - \xi\big) }
= \\
&&=
\sum
\limits_{\phi = h, H}
\dfrac{
g_{\gamma W^\pm W^\mp}
\,
g_{\phi H^- W^+}
\,
g_{\phi W^\pm W^\mp}
}{
6 M_{H^\pm}^2 M_{W}^2
\big(M_{H^\pm}^2-M_{W}^2\big)^2}
\Big[
M_{\phi}^2
\,
\Big(
4
M_{H^\pm}^4
-
5
M_{H^\pm}^2
M_{W}^2
+
3
M_{W}^4
\Big)
\Big]
\nonumber \\
&&
+
\dfrac{1}{
M_{W}^2 \big(1 - \xi\big) }
\Big[
\big(\xi - 1\big)
\dfrac{g_{\gamma W^\pm W^\mp}
}{
\big(M_{H^\pm}^2-M_{W}^2\big)^2}
\,
\dfrac{M_{H^\pm}^2+M_{W}^2}{6}
\sum
\limits_{\phi = h, H}
M_{\phi}^2
\,
g_{\phi W^\pm W^\mp}
\,
g_{\phi H^- W^+}
\Big]
\nonumber\\
&&=
\dfrac{
g_{\gamma W^\pm W^\mp}
}{
2 M_{H^\pm}^2 M_{W}^2}
\sum
\limits_{\phi = h, H}
M_{\phi}^2
\,
g_{\phi H^- W^+}
\,
g_{\phi W^\pm W^\mp}.
\nonumber
\end{eqnarray}
\item \underline{In the form factor $F^{W^\pm H^\pm}_{2}$}:
We also have
\begin{eqnarray}
&& c^{7}_{2}
\longrightarrow
c^{7}_{2}
+
\dfrac{c^{9}_{2}
}{
M_{W}^2 \big(1 - \xi\big) }
=\\
&&
= -
\sum
\limits_{\phi = h, H}
\dfrac{
g_{\gamma W^\pm W^\mp}
\,
g_{\phi H^- W^+}
\,
g_{\phi W^\pm W^\mp}
}{
3 M_{H^\pm}^2 M_{W}^2
\big(M_{H^\pm}^2-M_{W}^2\big)^3}
\Big[
M_{\phi}^2
\,
\Big(
7 M_{H^\pm}^4
-
2
M_{H^\pm}^2
M_{W}^2
+
3 M_{W}^4
\Big)
\Big]
\nonumber \\
&&-
\dfrac{1}{
M_{W}^2 \big(1 - \xi\big) }
\Big[
\big(\xi-1\big)
\dfrac{
4 \,
g_{\gamma W^\pm W^\mp}
}{
\big(M_{H^\pm}^2-M_{W}^2\big)^3}
\,
\dfrac{M_{H^\pm}^2+M_{W}^2}{3}
\sum
\limits_{\phi = h, H}
M_{\phi}^2
\,
g_{\phi H^- W^+}
\,
g_{\phi W^\pm W^\mp}
\Big]
\nonumber \\
&&=
\dfrac{-2}{\big(M_{H^\pm}^2-M_{W}^2\big)}
\Big[
\dfrac{
	g_{\gamma W^\pm W^\mp}
}{
	2 M_{H^\pm}^2 M_{W}^2}
\sum
\limits_{\phi = h, H}
M_{\phi}^2
\,
g_{\phi H^- W^+}
\,
g_{\phi W^\pm W^\mp}
\Big]
\nonumber \\
&&=
\dfrac{-2}{\big(M_{H^\pm}^2-M_{W}^2\big)}
\times
c^{7}_{1}. \nonumber
\end{eqnarray}
\end{itemize}
Finally, two form factors of
the mixing of W boson and charged Higgs
contribution with the
$\xi$-independence,
are expressed by Ward identity
as follows:
\begin{eqnarray}
F^{W^\pm H^\pm}_{2}
&=&
\dfrac{-2}{\big(M_{H^\pm}^2-M_{W}^2\big)}
\times
F^{W^\pm H^\pm}_{1}.
\end{eqnarray}
Where analytic formulas
for these form factors
are given in Eq.~\ref{masterFWH}.
\section*{Appendix D: Feynman diagrams  
for $H^{\pm}\rightarrow W^{\pm} \gamma$ 
in general $R_{\xi}$-gauge}             
A complete set of one-loop Feynman diagrams
relevant to the decay process $H^{\pm}
\rightarrow W^{\pm} \gamma$ in the general
$R_{\xi}$ gauge is provided in the
Appendix.

\end{document}